\documentclass[12pt]{article}

\usepackage{newtxtext,newtxmath}

\usepackage{graphicx}

\usepackage[letterpaper,margin=1in]{geometry}

\renewenvironment{abstract}
	{\quotation}
	{\endquotation}

\date{}

\makeatletter
\renewcommand{\fnum@figure}{\textbf{Figure \thefigure}}
\renewcommand{\fnum@table}{\textbf{Table \thetable}}
\makeatother

\usepackage{scicite}

\usepackage{url}

\usepackage{xspace}
\usepackage{longtable}

\newcommand{\farcs}{\mbox{\ensuremath{.\!\!^{\prime\prime}}}}

\newcommand{\ion}[2]{\text{#1\,\textsc{#2}}}
\newcommand{\sersic}{S\'{e}rsic}

\newcommand{\oiii}{\mbox{[\ion{O}{iii}]}}
\newcommand{\hi}{\mbox{\ion{H}{i}}}

\newcommand{\hb}{\mbox{H$\beta$}}
\newcommand{\ha}{\mbox{H$\alpha$}}

\newcommand{\lya}{\mbox{Ly$\alpha$}}

\newcommand{\zph}{\mbox{$z_\mathrm{phot}$}}

\newcommand{\venus}{\mbox{\texttt{VENUS}}}
\newcommand\msun{\mbox{$\mathrm{M}_\odot$}}
\newcommand\lsun{\mbox{$\mathrm{L}_\odot$}}

\newcommand{\micron}{\mbox{$\mu\mathrm{m}$}}
\newcommand{\rxc}{RXC\,J2211--0350\xspace}
\newcommand{\tgta}{R2211-RX1\xspace}
\newcommand{\tgtb}{R2211-RX2\xspace}
\newcommand{\arcsec}{\mbox{$^{\prime\prime}$}}

\usepackage{hyperref}
\hypersetup{colorlinks,
    linkcolor={red!50!black},
    citecolor={blue!50!black},
    urlcolor={blue!80!black}
}
\usepackage[dvipsnames]{xcolor}

\newcommand{\pending}[1]{{#1}}

\def\scititle{
Little red dot variability over a century reveals black hole envelope via a giant Einstein cross
}

\title{\bfseries \boldmath \scititle}

\author{
    Zijian~Zhang$^{1,2}$,
    Mingyu~Li$^{3}$,
    Masamune~Oguri$^{4,5}$,
    Xiaojing~Lin$^{3}$,
    Kohei~Inayoshi$^{1}$,  \and
    Catherine~Cerny$^{6}$,
    Dan~Coe$^{7,8,9}$,
    Jose~M.~Diego$^{10}$,
    Seiji~Fujimoto$^{11,12}$,
    Linhua~Jiang$^{1,2,\ast}$,  \and
    Guillaume~Mahler$^{13}$,
    Jorryt~Matthee$^{14}$,
    Rohan~P.~Naidu$^{15}$,
    Keren~Sharon$^{6}$,
    Yue~Shen$^{16,17}$,  \and
    Adi~Zitrin$^{18}$,
    Abdurro'uf$^{19}$,
    Hollis~Akins$^{20}$,
    Joseph~F.~V.~Allingham$^{18}$,
    Ricardo~Amor\'in$^{21}$, \and
    Yoshihisa~Asada$^{11,12}$, 
    Hakim~Atek$^{22}$,
    Franz~E.~Bauer$^{23}$,
    Maru\v{s}a~Brada\v{c}$^{24}$, \and
    Larry~D.~Bradley$^{7}$,
    Zheng~Cai$^{3}$, 
    Sebastiano~Cantalupo$^{25}$,
    Christopher~Conselice$^{26}$,\and
    Liang~Dai$^{27}$,  
    Pratika~Dayal$^{28,11,29}$,
    Eiichi~Egami$^{30}$,
    Daniel~J.~Eisenstein$^{31}$,\and
    Andreas~L.~Faisst$^{32}$,  
    Xiaohui~Fan$^{30}$, 
    Qinyue~Fei$^{11}$,
    Brenda~L.~Frye$^{30}$,\and
    Yoshinobu~Fudamoto$^{4}$, 
    Lukas~J.~Furtak$^{20,33}$,
    Miriam~Golubchik$^{18}$,\and
    Mauro~Gonz\'{a}lez-Otero$^{21}$,
    Yuichi~Harikane$^{34}$,
    {Tiger~Yu-Yang~Hsiao$^{20,33}$}, \and
    Yolanda~Jim\'enez-Teja$^{21,35}$, 
    Jeyhan~S.~Kartaltepe$^{36}$,
    Tomokazu~Kiyota$^{37,38}$, \and
    Anton~M.~Koekemoer$^{7}$, 
    {Kotaro~Kohno$^{39,40}$}, 
    Vasily~Kokorev$^{20,33}$,
    {Nimisha~Kumari$^{9}$}, \and
    Ivo~Labbe$^{41}$,
    Claudia~D.~P.~Lagos$^{42,43}$,
    Conor~Larison$^{7}$,
    Yongming~Liang$^{34,38}$, \and
    Ray~A.~Lucas$^{7}$,
    Jianwei~Lyu$^{30}$,
    Nicholas~S.~Martis$^{24}$,
    Georgios~E.~Magdis$^{43,44}$,\and
    Matteo~Messa$^{45}$,
    Minami~Nakane$^{34,46}$,
    Ga\"el~Noirot$^{7}$,
    Rafael~Ortiz~III$^{47}$, \and
    Masami Ouchi$^{38,34,37,48}$,
    Justin~D.~R.~Pierel$^{7}$,
    Marc~Postman$^{7}$,
    Naveen~Reddy$^{49}$,\and
    Massimo~Ricotti$^{50}$,
    Daniel~Schaerer$^{51,52}$,
    Raffaella~Schneider$^{53}$,
    Charles~C.~Steidel$^{54}$, \and
    Wei~Leong~Tee$^{55}$,
    Roberta~Tripodi$^{56,24,57}$,
    James~A.~A.~Trussler$^{31}$,
    Hiroya~Umeda$^{34,46}$,\and
    Francesco~Valentino$^{43,44}$,
    Eros~Vanzella$^{45}$,
    Feige~Wang$^{6}$,
    Rogier~Windhorst$^{47}$,\and
    Yunjing~Wu$^{3}$,
    Zihao~Wu$^{31}$,
    Hiroto~Yanagisawa$^{34,46}$,
    Jinyi~Yang$^{6}$,
    Fengwu~Sun$^{31,\ast}$ \and
	\footnotesize$^{1}$Kavli Institute for Astronomy and Astrophysics, Peking University, Beijing 100871, China\and
    \footnotesize$^{2}$Department of Astronomy, School of Physics, Peking University, Beijing 100871, China\and
    \footnotesize$^{3}$Department of Astronomy, Tsinghua University, Beijing 100084, People’s Republic of China\and
    \footnotesize$^{4}$Center for Frontier Science, Chiba University, 1-33 Yayoi-cho, Inage-ku, Chiba 263-8522, Japan\and
    \footnotesize$^{5}$Department of Physics, Graduate School of Science, Chiba University, 1-33 Yayoi-Cho, Inage-Ku, Chiba 263-8522, Japan\and
    \footnotesize$^{6}$Department of Astronomy, University of Michigan 1085 South University Avenue Ann Arbor, MI 48109, USA\and
    \footnotesize$^{7}$Space Telescope Science Institute (STScI), 3700 San Martin Drive, Baltimore, MD 21218, USA\and
    \footnotesize$^{8}$Center for Astrophysical Sciences, Department of Physics and Astronomy, The Johns Hopkins University, 3400 N Charles St. Baltimore, MD 21218, USA\and
    \footnotesize$^{9}$Association of Universities for Research in Astronomy (AURA), Inc.~for the European Space Agency (ESA)\and
    \footnotesize$^{10}$Instituto de F\'isica de Cantabria (CSIC-UC). Avda. Los Castros s/n. 39005 Santander, Spain\and
    \footnotesize$^{11}$David A. Dunlap Department of Astronomy and Astrophysics, University of Toronto, 50 St. George Street, Toronto, Ontario, M5S 3H4, Canada\and
    \footnotesize$^{12}$Dunlap Institute for Astronomy and Astrophysics, 50 St. George Street, Toronto, Ontario, M5S 3H4, Canada\and
    \footnotesize$^{13}$STAR Institute, Quartier Agora - All\'ee du six Ao\^ut, 19c B-4000 Li\`ege, Belgium\and
    \footnotesize$^{14}$Institute of Science and Technology Austria (ISTA), Am Campus 1, 3400 Klosterneuburg, Austria\and
    \footnotesize$^{15}$MIT Kavli Institute for Astrophysics and Space Research, 70 Vassar Street, Cambridge, MA 02139, USA\and
    \footnotesize$^{16}$Department of Astronomy, University of Illinois Urbana-Champaign, Urbana, IL 61801, USA\and
    \footnotesize$^{17}$National Center for Supercomputing Applications, University of Illinois Urbana-Champaign, Urbana, IL 61801, USA\and
    \footnotesize$^{18}$Department of Physics, Ben-Gurion University of the Negev, P.O. Box 653, Beer-Sheva 8410501, Israel\and
    \footnotesize$^{19}$Department of Astronomy, Indiana University, 727 East Third Street, Bloomington, IN 47405, USA\and
    \footnotesize$^{20}$Department of Astronomy, The University of Texas at Austin, 2515 Speedway Blvd Stop C1400, Austin, TX 78712, USA\and
    \footnotesize$^{21}$Instituto de Astrof\'{i}sica de Andaluc\'{i}a (CSIC), Apartado 3004, 18080 Granada, Spain\and
    \footnotesize$^{22}$Institut d'Astrophysique de Paris, CNRS, Sorbonne Universit\'e, 98bis Boulevard Arago, 75014, Paris, France\and
    \footnotesize$^{23}$ Instituto de Alta Investigaci{\'{o}}n,  Universidad de Tarapac{\'{a}}, Casilla 7D, Arica, 1010000, Chile \and
    \footnotesize$^{24}$University of Ljubljana FMF, Jadranska 19, 1000 Ljubljana, Slovenia  \and
    \footnotesize$^{25}$Dipartimento di Fisica G. Occhialini, Universit\`a degli Studi di Milano Bicocca, Piazza della Scienza 3, 20126 Milano, Italy\and
    \footnotesize$^{26}$Jodrell Bank Centre for Astrophysics, University of Manchester, Oxford Road, Manchester UK\and
    \footnotesize$^{27}$Department of Physics, University of California, 366 Physics North MC 7300, Berkeley, CA. 94720, USA\and
    \footnotesize$^{28}$Canadian Institute for Theoretical Astrophysics, 60 St George St, University of Toronto, Toronto, ON M5S 3H8, Canada\and
    \footnotesize$^{29}$Department of Physics, 60 St George St, University of Toronto, Toronto, ON M5S 3H8, Canada\and
    \footnotesize$^{30}$ Department of Astronomy/Steward Observatory, University of Arizona, 933 N. Cherry Ave, Tucson, AZ 85721, USA \and
    \footnotesize$^{31}$Center for Astrophysics $|$ Harvard \& Smithsonian, 60 Garden St., Cambridge, MA 02138, USA\and
    \footnotesize$^{32}$IPAC, California Institute of Technology, 1200 E. California Blvd. Pasadena, CA 91125, USA\and
    \footnotesize$^{33}$Cosmic Frontier Center, The University of Texas at Austin, Austin, TX 78712, USA\and
    \footnotesize$^{34}$Institute for Cosmic Ray Research, The University of Tokyo, 5-1-5 Kashiwanoha, Kashiwa, Chiba 277-8582, Japan\and
    \footnotesize$^{35}$Observat\'orio Nacional, Rua General Jos\'e Cristino, 77 - Bairro Imperial de S\~ao Crist\'ov\~ao, Rio de Janeiro, 20921-400, Brazil\and
    \footnotesize$^{36}$Laboratory for Multiwavelength Astrophysics, School of Physics and Astronomy, Rochester Institute of Technology, 84 Lomb Memorial Drive, Rochester, NY 14623, USA\and
    \footnotesize$^{37}$Department of Astronomical Science, The Graduate University for Advanced Studies, SOKENDAI, 2-21-1 Osawa, Mitaka, Tokyo, 181-8588, Japan\and
    \footnotesize$^{38}$National Astronomical Observatory of Japan, 2-21-1 Osawa, Mitaka, Tokyo, 181-8588, Japan\and
    \footnotesize$^{39}$Institute of Astronomy, Graduate School of Science, The University of Tokyo, 2-21-1 Osawa, Mitaka, Tokyo, 181-0015 Japan\and
    \footnotesize$^{40}$Research Center for the Early Universe, Graduate School of Science, The University of Tokyo, 7-3-1 Hongo, Bunkyo-ku, Tokyo 113-0033, Japan\and
    \footnotesize$^{41}$Centre for Astrophysics and Supercomputing, Swinburne University of Technology, Melbourne, VIC 3122, Australia\and
    \footnotesize$^{42}$International Centre for Radio Astronomy Research (ICRAR), M468, University of Western Australia, 35 Stirling Hwy, Crawley, WA 6009, Australia\and
    \footnotesize$^{43}$Cosmic Dawn Center (DAWN), Jagtvej 128, DK2200 Copenhagen N, Denmark\and
    \footnotesize$^{44}$DTU-Space, Technical University of Denmark, Elektrovej 327, 2800, Kgs. Lyngby, Denmark\and
    \footnotesize$^{45}$INAF -- OAS, Osservatorio di Astrofisica e Scienza dello Spazio di Bologna, via Gobetti 93/3, I-40129 Bologna, Italy\and
    \footnotesize$^{46}$Department of Physics, Graduate School of Science, The University of Tokyo, 7-3-1 Hongo, Bunkyo, Tokyo 113-0033, Japan\and
    \footnotesize$^{47}$School of Earth and Space Exploration, Arizona State University, Tempe, AZ 85287-6004, USA\and
    \footnotesize$^{48}$Kavli Institute for the Physics and Mathematics of the Universe (WPI), The University of Tokyo, 5-1-5 Kashiwanoha, Kashiwa, Chiba 277-8583, Japan\and
    \footnotesize$^{49}$Department of Physics and Astronomy, University of California, Riverside, 900 University Avenue, Riverside, CA 92521, USA\and
    \footnotesize$^{50}$Department of Astronomy, University of Maryland, College Park, 20742, USA \and
    \footnotesize$^{51}$Observatoire de Gen\`eve, Universit\'e de Gen\`eve, Chemin Pegasi 51, 1290 Versoix, Switzerland\and
    \footnotesize$^{52}$CNRS, IRAP, 14 Avenue E. Belin, 31400 Toulouse, France\and
    \footnotesize$^{53}$Department of Physics, Sapienza University of Rome, Pzz.le Aldo Moro 5, 00185 Rome, Italy\and
    \footnotesize$^{54}$Cahill Center for Astronomy and Astrophysics, California Institute of Technology, MS 249-17, Pasadena, CA 91125\and
    \footnotesize$^{55}$Department of Astronomy and Astrophysics, The Pennsylvania State University, 525 Davey Lab, University Park, PA 16802, USA\and
    \footnotesize$^{56}$INAF - Osservatorio Astronomico di Roma, Via Frascati 33, I-00078 Monte Porzio Catone, Italy\and
    \footnotesize$^{57}$IFPU - Institute for Fundamental Physics of the Universe, via Beirut 2, I-34151 Trieste, Italy\and
	\small$^\ast$Corresponding authors. Email: \href{mailto:fengwu.sun@cfa.harvard.edu}{fengwu.sun@cfa.harvard.edu} (Fengwu Sun) \& \href{mailto:jiangKIAA@pku.edu.cn}{jiangKIAA@pku.edu.cn} (Linhua Jiang)
}

\begin{document} 

\maketitle

\begin{abstract} \bfseries \boldmath
``Little red dots'' (LRDs) represent a new population of astronomical objects uncovered by JWST whose nature remains debated.
Although many LRDs are suspected as active galactic nuclei (AGN), they show little variability on days-years timescales.
We report the discovery of two gravitationally lensed LRDs at redshift\,$\sim 4.3$ behind the cluster RXCJ2211-0350, one of which (RX1) is quadruply imaged with time delays spanning $\sim$130\,years.
RX1 exhibits intrinsic color and brightness variations of up to 0.7\,magnitude among its images. 
These changes are consistent with blackbody-temperature variations of a photosphere, indicating long-term variability analogous to Cepheid-like pulsations but in a far more extended ($R \sim 2000$\,AU) and massive ($M\gtrsim10^6$\,\msun) systems. 
These results suggest LRDs as a distinct class of AGN with stellar-like envelopes.
\end{abstract}


JWST observations have unveiled a population of compact, high-redshift objects with unique V-shaped spectral energy distribution (SED), commonly referred to as ``little red dots'' (LRDs). 
Their compactness and broad Balmer emission lines point to the presence of accreting supermassive black holes (SMBHs) \cite{Harikane2023,Greene2024,Matthee2024,Kocevski2025}. 
However, LRDs also exhibit several properties that distinguish them from typical AGNs: they are unusually faint in the X-ray \cite{Yue2024,Maiolino2025}, radio \cite{Mazzolari2024}, and far-infrared bands \cite{Perez2024,Williams2024}, often show strong Balmer absorption and pronounced Balmer breaks \cite{Lin2024,Wang2024b,deGraaff2025_ruby,Juodzbalis2025,Naidu2025}.
These puzzling characteristics challenge our current understanding of AGN physics and galaxy-black hole coevolution, suggesting that LRDs may represent a transitional or previously unexplored phase of black hole growth in the early universe \cite{inayoshi2025_lrd1stagn}.

The variability of LRDs offers a powerful probe into their underlying physical nature. 
Measuring potential time lags between variations in the UV/optical continuum and broad-line emission can reveal the relative geometry of the SMBH, the broad-line region, and the dense gas envelope responsible for strong Balmer breaks and absorption features \cite{Inayoshi2025_densegas,Ji2025a,Kido2025,Lin2025_locallrd,Naidu2025}. 
Surprisingly, most LRDs display little to no photometric variability ($\lesssim5\%$) over an observed timescale of $\sim 3$ years \cite{Kokubo2024,Tee2025,Stone2025,Zhang2025a}. 
Strong gravitational lensing of LRDs by massive clusters provides a unique opportunity to probe variability over decades, leveraging time delays ($\Delta t_{\rm grav}$) that can span tens of years between multiple images. 
To date, only two multiply imaged LRDs, A2744-QSO1 ($z=7.04$, $\Delta t_{\rm grav}\sim20$\,years)\cite{Furtak2023,Furtak2025b} in Abell 2744 and A383-LRD1 ($z=6.02$, $\Delta t_{\rm grav}\sim5$\,years) in Abell 383\cite{Golubchik25}, has been reported. A2744-QSO1 hints at changes in emission-line equivalent width (EWs; but not in continuum luminosity or color) over a rest-frame time delay of $\sim$2.7 years\cite{Furtak2025b, Ji2025a}.
These timescales remain far shorter than the expected envelope dynamical timescale ($t_{\rm dyn}$, see Equation \ref{equ:tdyn} below) of several tens of years, which would be the characteristic variability timescale of LRD envelopes\cite{Burke2025,Inayoshi2025_SpectralUniformity,Kido2025}. More multiply imaged LRDs with longer time delays are needed to explore longer-term variability and establish its universality.

\subsection*{Discovery of two multiply imaged LRDs}
\begin{figure}[!t]
\centering
\includegraphics[width=\linewidth]{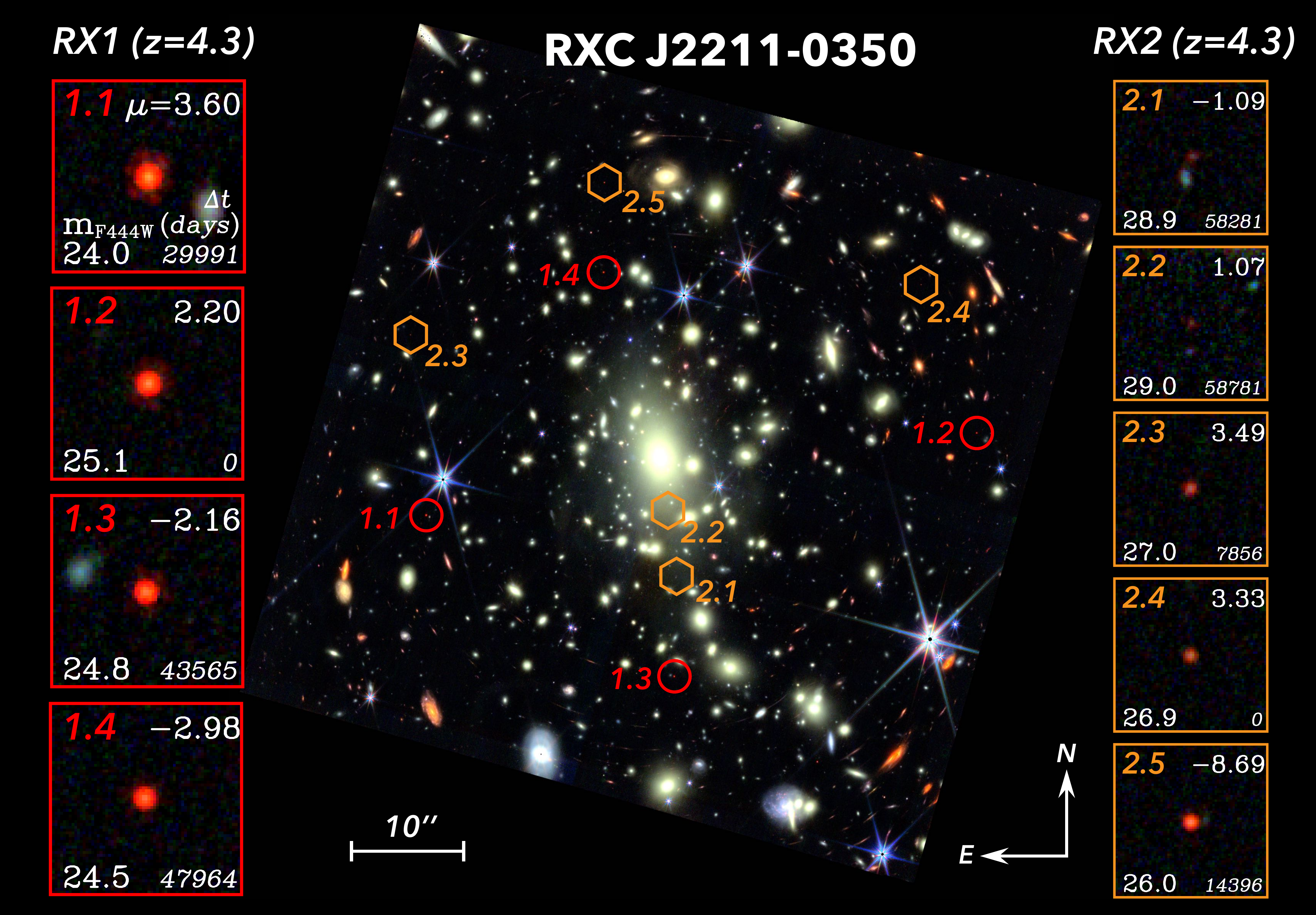}
\caption{\textbf{JWST NIRCam color composite image of the galaxy cluster \rxc.} 
Thanks to strong gravitational lensing, two multiply imaged LRDs, dubbed RX1 and RX2, are identified, both with photometric redshifts of $z=4.3$. 
The four images of RX1 are marked by red circles (1.1-1.4), and the five images of RX2 are marked by orange hexagons (2.1-2.5). The left (red) and right (orange) panels show zoom-in cutouts for each of these lensed images. In each cutout, the image name is given in the top left, the magnification ($\mu$; negative value indicates that the image parity is reversed and the image orientation is flipped) calculated from our cluster mass model is in the top right, the measured F444W-band magnitude (not corrected for magnification) is in the bottom left, and the inferred time delay ($\Delta t$ in days) is in the bottom right.
The uncertainties of $\mu$ and $\Delta t$ are typically $\lesssim$\,5\%.
}
\label{fig:overview}
\end{figure}

The two lensed LRDs in \rxc\ were first identified through our systematic search for strongly lensed LRDs using archival JWST/NIRCam imaging of over 140 galaxy clusters, primarily from the Strong LensIng and Cluster Evolution (SLICE; PID: 5594) program\cite{cerny25}.
Subsequent deep 10-band NIRCam imaging from the Vast Exploration for Nascent, Unexplored Sources (\venus; PID: 6882) program, taken on October 16, 2025, reveals their characteristic V-shape continuum and compact morphology (Figure \ref{fig:photz} and \ref{fig:Vshape_compact}), establishing both as bona fide LRD candidates awaiting spectroscopic confirmation. 
Our cluster mass model confirms the multiple images from both systems, which are gravitationally lensed into four and five distinct images, respectively, by the massive cluster potential of \rxc\ at $z=0.397$. One of them forms a perfect Einstein cross in the sky as shown in Figure \ref{fig:overview}. 
They are thus dubbed \tgta\ and \tgtb\ (RX for ``red cross'').
The source-plane positions (Figure \ref{fig:sourceplane_pos}) demonstrate that both sources reside in a potential galaxy overdensity encompassing nine objects at similar photometric redshift ($z_{\rm phot}$) and enclosed in a circle with radius of $\sim 10^{\prime\prime}$ (corresponding to $\sim 70$\,kpc at $z \sim 4.3$). We adopt the most probable redshift of $z_{\rm phot} \sim 4.3$ of this structure as the fiducial redshift of the two LRDs (see Materials and Methods). This value aligns with their \zph\ derived from the empirical LRD templates (Figure~\ref{fig:photz} and \ref{fig:lrd_temp}) and the geometric redshift from the cluster mass model. 
To date, \rxc\ hosts the largest known number of multiply imaged LRDs in any galaxy cluster observed by JWST. 

\medskip

Gravitational lensing not only magnifies these intrinsically faint sources but also introduces time delays of decades to over a century among their multiple images. 
We update the strong cluster mass model of \rxc\ constructed with \texttt{glafic} \cite{Oguri2010,Oguri2021}, taking advantage of the newly obtained JWST imaging data (Materials and Methods). 
The cluster mass model is tightly constrained with 146 multiple images from 43 sources, as shown in Figure~\ref{fig:lensing_model} and Table~\ref{tab:multi}. 
For \tgta\ and \tgtb, the magnifications and time delays derived by the cluster mass model are listed in Table \ref{tab:lens_param}. 
The magnifications for most images range from 2-4, reaching $\sim 8.7$ for \tgtb.5, with typical uncertainties of order $\sim$0.1-0.2.
The blackbody fitting to the intrinsic SEDs below yields bolometric luminosities for \tgta\ and \tgtb\ of $2.38_{-0.05}^{+0.04}\times10^{44}\rm ~erg~s^{-1}$ and $2.54_{-0.33}^{+0.32}\times10^{43}\rm ~erg~s^{-1}$, respectively. 
These values are slightly higher than those estimated from $L_{5100}$ (inferred using the intrinsic F300M photometry) using the $L_{5100}$-based bolometric correction in \cite{Greene2025Lbol}.
Assuming the Eddington ratio to be unity, the corresponding black hole masses ($M_{\rm BH}$) are estimated to be $1.9\times10^6~ M_\odot$ and $2\times10^5~ M_\odot$.
Thanks to a large cluster mass ($M \approx 8.1 \times10^{14}$\,\msun\ within 500\,kpc) and a large Einstein radius ($\Theta_\mathrm{E} \approx 47.5$\arcsec\ at $z \approx 4.3$), the lens model predicts that the relative time delays (observed-frame) between the multiple images span a uniquely long and broad range: $\sim$80-130 yr for \tgta, and $\sim$20-160 yr for \tgtb, substantially exceeding that of other multiply imaged LRDs\cite{Furtak2023,Ji2025a,Furtak2025b,Golubchik25}.

\subsection*{LRD variability observed on a century-scale}

We use magnification-corrected photometry to investigate their variability (Materials and Methods). 
We account for both photometric and magnification uncertainties in the error bars. In the short-wavelength (SW) bands, the signal-to-noise ratios (S/Ns) are low because the LRDs are faint in the rest-frame UV. To improve sensitivity, we stack the SW bands (F090W, F115W, and F150W) using a bandpass-weighted flux average for each image.
Figures~\ref{fig:RX1_vari}(A) and Figure~\ref{fig:RX2_lc} show the delensed photometric light curves of \tgta\ and \tgtb\ in the JWST filters, respectively. 
To quantify the degree of variability, we define the reduced chi-square $\chi^2_\nu$ relative to a constant mean flux, as indicated in the Figures.

\medskip

For \tgta, no significant variability is observed in most bands blueward F200W (i.e., rest-frame $<3800$\,\AA) with $\chi^2_\nu<1$, even after stacking. The exception is F150W2, which covers slightly longer wavelengths than F200W. 
In contrast, the long-wavelength (LW) bands show consistent and coordinated variability across different bands, with amplitudes up to $|\Delta m| \sim 0.7$ mag, S/N$\sim 4$, and $\chi^2_\nu=18.4$. 
Assuming no intrinsic variability, the resulting $\chi^2_\nu$ values for bands redder than F210M are all significantly greater than unity. 
Such behavior resembles the visible but $<3\sigma$ continuum trend seen in the light curves of A2744-QSO1 \cite{Furtak2025b}, which was interpreted as no continuum variability due to its low statistical significance. 
Compared to that case, \tgta provides a $\sim10$ times longer rest-frame time baseline and exhibits a larger variability amplitude. 
For \tgtb, no significant variability is detected in most bands, with the amplitude constrained to $\Delta m \lesssim 0.3$ mag. 
In several bands, we find tentative hints of variability, with $\Delta m_{\rm max} \sim 0.5\text{--}0.9~\mathrm{mag}$ at $\sim 2.7\,\sigma$ and $\chi^2_{\nu} \gtrsim 2$. 
We caution the faintness of \tgtb\ in the rest-frame UV ($28\sim29$\,mag with lensing magnification) and strong intracluster light background for \tgtb.1/2. Further monitoring with deeper JWST imaging is required to confirm this variability. We also examine the short-term variability of \tgta\ and \tgtb\ by comparing the SLICE F322W2 photometry with the VENUS synthetic F277W$+$F356W measurements over a $\sim$1-year baseline (see Figure \ref{fig:venus_slice}), and find no significant variability.

\medskip

Systematic uncertainties in the lensing magnification are the primary source of error for our variability analysis. We perform tests on 12 other multiply imaged systems with \zph$>3$ in the same field and find no significant flux and color variations (Figure \ref{fig:RX1_vari}c), confirming the reliability of our cluster mass model. Microlensing and ``millilensing'' effects by cluster stars and dark matter subhalos are expected to be negligible (Supplementary Text). 
The lack of variability in \tgta{}’s SW bands and \tgtb{}’s bands further supports that their magnification correction is self-consistent.
We also detected significant bluer-when-brighter rest-frame optical color variations of \tgta\ (Figure \ref{fig:RX1_vari}b), which confirms that the observed variability is intrinsic, since lensing can introduce flux anomalies but not color ones. 
Other possible systematics, such as spatial PSF variation, depth differences, or contamination\cite{Furtak2025b,Stone2025,Zhang2025a}, cannot reproduce the coherent, multi-band behavior we observe. These tests collectively demonstrate that the detected variability arises from the source itself rather than from lensing or instrumental effects.

\subsection*{Signatures of opacity-driven pulsation}
The detection of significant optical variability in \tgta\ provides the first robust evidence that some LRDs undergo continuum luminosity changes, underscoring their unique nature.  
Unlike normal AGNs, which commonly show detectable variability at month-year timescales\cite{MacLeod2010}, \tgta\ (as well as other LRDs) shows no significant short-term variability between the SLICE and \venus\ epochs (rest-frame $\sim 0.2$ yr; see Figure \ref{fig:venus_slice}), but varies on much longer timescales (rest-frame $>20$ years). 
Such behavior is expected if LRDs are surrounded by a dense, optically thick envelope resembling a stellar atmosphere\cite{deGraaff2025_ruby,Inayoshi2025_densegas,Naidu2025,Rusakov2025}, as proposed to explain their strong Balmer breaks and absorption features\cite{Matthee2024,Lin2024,deGraaff2025_ruby,Naidu2025}. 
In this framework, any central rapid accretion fluctuations would be damped by the envelope, while the observed variability in \tgta\ reflects slow changes in the thermal emission (i.e., blackbody) from such an envelope with photospheric radius $R_{\rm ph}\sim 1000$ AU and effective temperature $T_{\rm eff} \sim 5000~{\rm K}$ \cite{Begelman2025,Kido2025,Lin2025_locallrd,LiuH_2025,Inayoshi2025_SpectralUniformity}. The observed rest-frame variability timescale closely matches the dynamical time of the gas envelope \cite{Inayoshi2025_SpectralUniformity}
\begin{equation}
\label{equ:tdyn}
    t_{\mathrm{dyn}} \equiv 2\pi\sqrt\frac{R_{\rm ph}^3}{GM_{\rm BH}} \approx 44.6~{\rm yr}\,\lambda_{\rm Edd}^{3/4}\left(\frac{M_{\rm BH}}{10^{6.28}\,M_\odot}\right)^{1/4}\left(\frac{T_{\rm eff}}{5000\,\rm K}\right)^{-3},
\end{equation}
where $\lambda_{\rm Edd}$ is the Eddington ratio.

\medskip

We thus model the multi-epoch SEDs of \tgta\ with a single-temperature blackbody\cite{Begelman2025,Kido2025,Lin2025_locallrd,Torralba2025}(Materials and Methods), whose total luminosity $L_{\rm BB}\propto R_{\rm ph}^2T_{\rm eff}^4$. 
An epoch-independent power-law component is added to account for the potential host galaxy dominance in rest-frame UV \cite{Ji2025a,Naidu2025,Inayoshi2025_SpectralUniformity}.  
This simple model provides a reasonable fit to the SEDs of \tgta\ (Figure \ref{fig:lc_BB_fit}A), with $T_{\rm eff} \sim 4000\,\rm K$, $R_\mathrm{ph} \sim 2000$\,AU, and $L_{\rm BB} \sim 2\times10^{\rm 44} \rm\,erg\,s^{-1}$. 
The best-fit temperatures increase systematically in the brighter epochs, and the inferred blackbody luminosities roughly follow $L_{\rm BB} \propto T_{\rm eff}^4$ (the best-fit relation is $L_{\rm BB} \propto T_{\rm eff}^{4.9\pm1.0}$) as shown in Figure \ref{fig:lc_BB_fit}(B). The deviation from $L \propto T_{\rm eff}^4$ indicates a modest change in the effective emitting radius of $\sim$200 AU (10\% of the $R_{\rm ph}$) over $\sim$10 years in the rest frame. 

\medskip

Such $L_{\rm BB} \sim T_{\rm eff}^4$ behavior is reminiscent of the classical opacity-driven pulsations ($\kappa$-mechanism) seen in variable super-giant stars such as Cepheids\cite{Cox1980,Li1994}.
In those stars, compression of a partially ionized layer (usually $\rm He^+$ layer) increases its opacity so that radiation is temporarily trapped and energy is stored.
The compression also builds up a pressure gradient, driving the layer to expand further than the original oscillation. 
During expansion, the gas cools and becomes more transparent, the pressure drops, and the layer contracts, completing the cycle.
In effect, the layer acts like a heat engine, converting trapped radiation into mechanical work and sustaining the pulsation cycle\cite{Eddington1917}. 
Similar opacity-driven pulsations have also been predicted for the radiation-dominated envelopes of supermassive stars (SMSs; $\sim10^3$-$10^5\,M_\odot$), which might be the progenitor of high-redshift SMBHs \cite{Osaki1966_Pulsation,Baraffe2001,Inayoshi2013_pulsation} and have even been proposed as a possible explanation for LRDs\cite{Begelman2025,Nandal2025}.
Although the central black hole in an LRD is estimated to be far more massive than typical SMSs, the overall configuration is analogous: both systems are in hydrostatic equilibrium state supported by radiation pressure, and both possess similar surface temperatures and gravity. 
Within the envelope, regions with temperatures of order $\sim 3\times10^4$ K would naturally create a He$^+$ partial ionization zone that acts as an engine region.
Therefore, a similar physical mechanism could give rise to the slow, periodic rest-frame optical variability observed for \tgta, while its rest-frame UV emission, dominated by the host galaxy, remains stable.

\medskip

Previous studies show that SMSs are unstable primarily to radial perturbations\cite{Baraffe2001,Sonoi2012}. 
Motivated by this, we examine whether the current limited multi-epoch data of \tgta\ are compatible with a minimal description of pulsation based on a single fundamental radial mode (f-mode).
In a polytropic envelope, the fundamental eigenfrequency solved by \cite{1941ApJ....94..245S} naturally gives a pulsation period that can be expressed as $T_{0} = f_0t_{\rm dyn}$, where $f_0 = 0.37$ is set by the numerical fundamental eigenvalue from \cite{1941ApJ....94..245S}. 
Substituting the inferred properties of \tgta (i.e., $T_{\rm eff} = 4000\rm\,K$ and $M_{\rm BH} = 10^{6.28}\,M_\odot$) in Equation (1), the resulting pulsation period is $T_{0} = 32$\,yr.
We model the variability in all bands redder than F210M using a sinusoid (i.e., f-mode) model with this fixed period (Materials and Methods). 
The sinusoid phase is tied across all bands, while allowing the amplitude and mean flux to vary independently per band.
The fitting results are shown in Figure \ref{fig:lc_BB_fit}(C). 
All bands are well described by this simple f-mode model, with a total $\chi^2_{\nu} = 0.09$ for 11 degrees of freedom (dof).
The F322W2 band includes eight observational points, augmented by synthetic fluxes from the \venus\ F277W and F356W bands, and is also well reproduced by the model.
While the low $\chi^2_{\nu}$ suggests potential overfitting, comparably good fits occur only within a narrow strip in the $M_{\rm BH}$-$T_{\rm eff}$ space that corresponds to a proper pulsation period range (Figure~\ref{fig:param_lcfit}). 
The independently inferred $M_{\rm BH}$ and $T_{\rm eff}$ of \tgta fall precisely within this optimal region, demonstrating that the fit is physically self-consistent rather than an artifact of flexible modeling.
Overall, the good agreement across all bands indicates that the pulsation scenario dominated by the f-mode offers a reasonable description of the current observed light curve. 
Future modeling with improved sampling will enable stronger constraints on additional pulsation modes and the pulsation period, refining this physical picture.

\medskip

Projecting the illustrative f-mode light curve model into the $T_{\rm eff}$-$L_{\rm BB}$ space with blackbody model visualizes the system's potential temporal sequence in the pulsation scenario (Materials and Methods).
The resulting trajectory in Figure~\ref{fig:lc_BB_fit}(B) shows a counterclockwise loop, resembling that of classical pulsating stars like Cepheids. 
In this pattern, the envelope is hotter (and smaller) during the brightening phase than the dimming phase at a given luminosity, a consequence of the hysteresis between temperature and radius variations \cite{Cox1980,2000A&A...360..245B}(see also Figure \ref{fig:Cep_4313}).
Such a lag arises from finite thermal inertia: when compressed, the gas heats faster than it can radiate, and when expanding, it cools more slowly, mimicking a heat engine in which heating and cooling are out of phase with mechanical motion.
Consistent with this picture, the expansion velocity inferred ($132\,{\rm km\,s^{-1}}$ from \tgta.3 to \tgta.1) is larger than the contraction velocity ($61\,{\rm km\,s^{-1}}$ from \tgta.1 to \tgta.2), as expected for opacity-driven pulsations. 
These velocities are obtained by dividing the change in $R_{\rm ph}$ between the epochs by the corresponding rest-frame time intervals.
While we caution that this trajectory is a model-dependent inference, the data's natural accommodation of such a phase lag lends qualitative support to the stellar-like envelope hypothesis.

\medskip

The pulsation scenario also provides a natural explanation for the observed diversity in the long-term variability of LRDs. 
Analogous to stars that are variable only within a specific instability strip on the Hertzsprung--Russell (HR) diagram \cite{Cox1980}, the opacity-driven instability for LRD envelopes likely also depends on their physical properties like temperature and luminosity (or $M_{\rm BH}$).
\tgta, with a relatively cool effective temperature ($T_{\rm eff}\sim 4000$\,K), exhibits strong variability, whereas the hotter system \tgtb\ ($T_{\rm eff}\sim 5000$\,K) remains stable in the rest-frame optical. 
This potential temperature dependence also offers a context for the absence of detectable variability in local LRDs, which have $T_{\rm eff}\sim 5000$\,K\cite{Lin2025_locallrd}, over the rest-frame $\sim$5-year ZTF baseline\cite{Burke2025}.
These results collectively suggest that the pulsational instability domain for LRDs may be favored in cooler envelopes (e.g., $T_{\rm eff}\lesssim 5000$\,K).

\subsection*{Future monitoring and predictions}
While the $\kappa$-mechanism can plausibly account for the variability of \tgta, the observed $L_{\rm BB} \propto T_{\rm eff}^{4.9\pm1.0}$ relation is somewhat steeper than the $L_{\rm BB}\sim T_{\rm eff}^4$ slope expected for a quasi-static pulsating photosphere, hinting that variations in the accretion state, which naturally present in AGNs through disc instabilities or changes in accretion rate\cite{Ricci2023}, may also play a role. 
Adopting a typical optical depth of $\tau \sim 10$, the photon diffusion time $t_{\rm diff} \sim \tau R_{\rm ph}/c = 0.1\,\rm yr$, implying that the envelope can respond quasi-isobarically to the central luminosity variations.
In this scenario, increases in the central ionizing source luminosity would drive simultaneous rises in $T_{\rm eff}$ and $R_{\rm ph}$ ($PR_{\rm ph}^3\propto T_{\rm eff}$), steepening the $T_{\rm eff}$-$L_{\rm BB}$ slope. 

\medskip

Future JWST multi-cycle spectrophotometric monitoring will be critical to distinguish and confirm these scenarios. 
Such a campaign bridges two distinct time domains: while multi-cycle observations are essential to capture evolution on timescales of days to years, the time delays between lensed images simultaneously provides baseline extending to decades or even a century.
Spectroscopically, pulsation-driven variability should produce detectable velocity shifts ($\sim$100 km s$^{-1}$) in absorption lines (\ha, \hb, He I, etc.), if some of them are produced by the gas envelope. Such shifts arise from envelope expansion/contraction (Figure \ref{fig:cartoon}A), analogous to pulsating stars\cite{2005MNRAS.362.1167P,2020A&A...641A..74H}. Conversely, variability driven by changes in the accretion state may cause the broad lines to vary following the bolometric luminosity change if they are powered by the central BH (Figure \ref{fig:cartoon}B), while narrow forbidden lines originating from the host galaxy remain stable. 
These predictions can be directly tested through future multi-cycle JWST/NIRSpec prism and high-resolution spectroscopy effectively. 
Complementarily, multi-epoch JWST imaging observations offer an independent falsifiable test as shown in Figure \ref{fig:cartoon}: while accretion state changes imply irregular, aperiodic fluctuations, the pulsation model predicts coherent phase shifts across the lensed images detectable variation of $\sim0.1$ mag within 5-10 years. If the real period is significantly shorter, the phase evolution across the images would be correspondingly larger.

\bigskip

The discovery of long-term variability in \tgta\ reveals that the dense gas envelope surrounding some early SMBHs could behave remarkably like an enormous, radiation-pressure-dominated stellar atmosphere, providing strong support for the envelope scenario of LRDs and offering insights into black-hole seeding.
This finding motivates a broader investigation into the parallels between LRDs and stellar physics, particularly whether these objects occupy an instability strip in the HR diagram or adhere to a Period-Luminosity relation analogous to the Leavitt Law for Cepheids. 
While confirming such fundamental relations requires a larger sample, cluster-lensed systems like \tgta\ and \tgtb\ offer a unique opportunity to extend the baseline to the century scale, opening a new window to study the nature of LRDs and the formation of the first SMBHs.

\newpage


\begin{figure*}
    \centering
    \vspace{-12pt}
    \includegraphics[width=\linewidth]{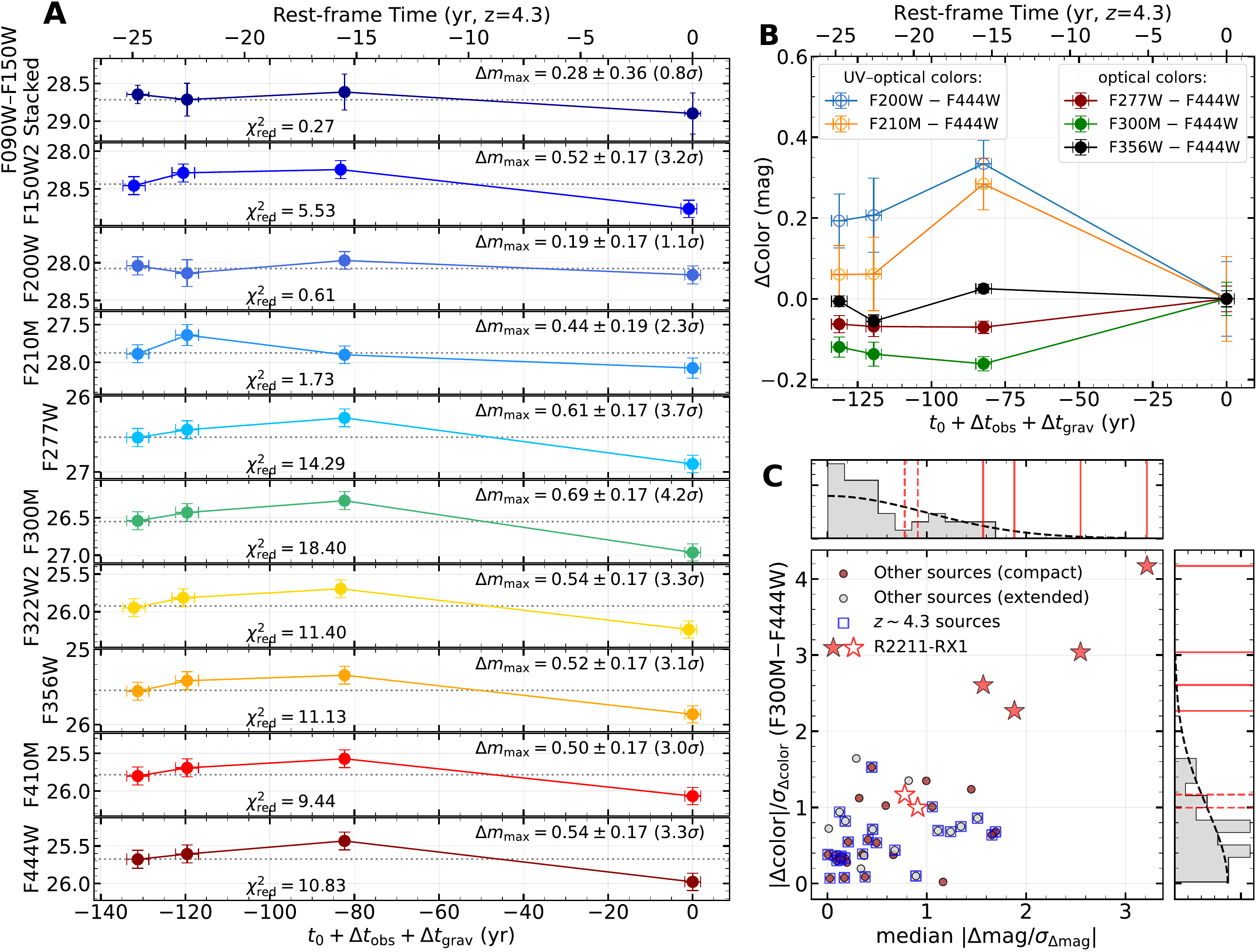}
    \vspace{-36pt}
    \caption{\textbf{Brightness and color variability of \tgta.} \textbf{(A)} De-magnified light curves of \tgta\ in each JWST band, incorporating the gravitational lens time delays $\Delta t_{\rm grav}$ given in Table \ref{tab:lens_param}. $t_0$ is the \venus\ observation date. Dotted lines show the mean magnitude of each band, and the error bars include magnification uncertainties. We show the maximal $\Delta m$ and $\chi_{\nu}^2$ values assuming no variability for each band. \textbf{(B)} Multi-band color evolution of \tgta, with x-axis the same as (A). The y-axis shows the color difference relative to image \tgta.2. \textbf{(C)} S/N distribution of brightness and color (F300M$-$F444W) variability between all epochs pairs for \tgta\ and the other 12  multiply imaged sources with \zph$>3$. Sources with \zph$\sim 4.3$ in Figure~\ref{fig:sourceplane_pos} are marked with blue squares. For each epoch pair, the brightness variability S/N is measured as the median S/N of $\Delta \rm \,mag$ across the F200W and redder bands to reflect the overall variability of the SED. In the top/right panels, red lines indicate \tgta, and the gray histogram represents the other sources. The two \tgta\ pairs with the shortest observed-frame intervals (10 and 36 yr) have S/N\,$\sim1$ and are shown as open stars, while all others have S/N$>2$. The S/N of the other sources are broadly consistent with a one-sided normal distribution and mostly below 1, suggesting that the measured variability of \tgta\ is not biased by systematic errors in magnification.}
    \label{fig:RX1_vari}
\end{figure*}

\begin{figure*}
    \centering
    \includegraphics[width=0.9\linewidth]{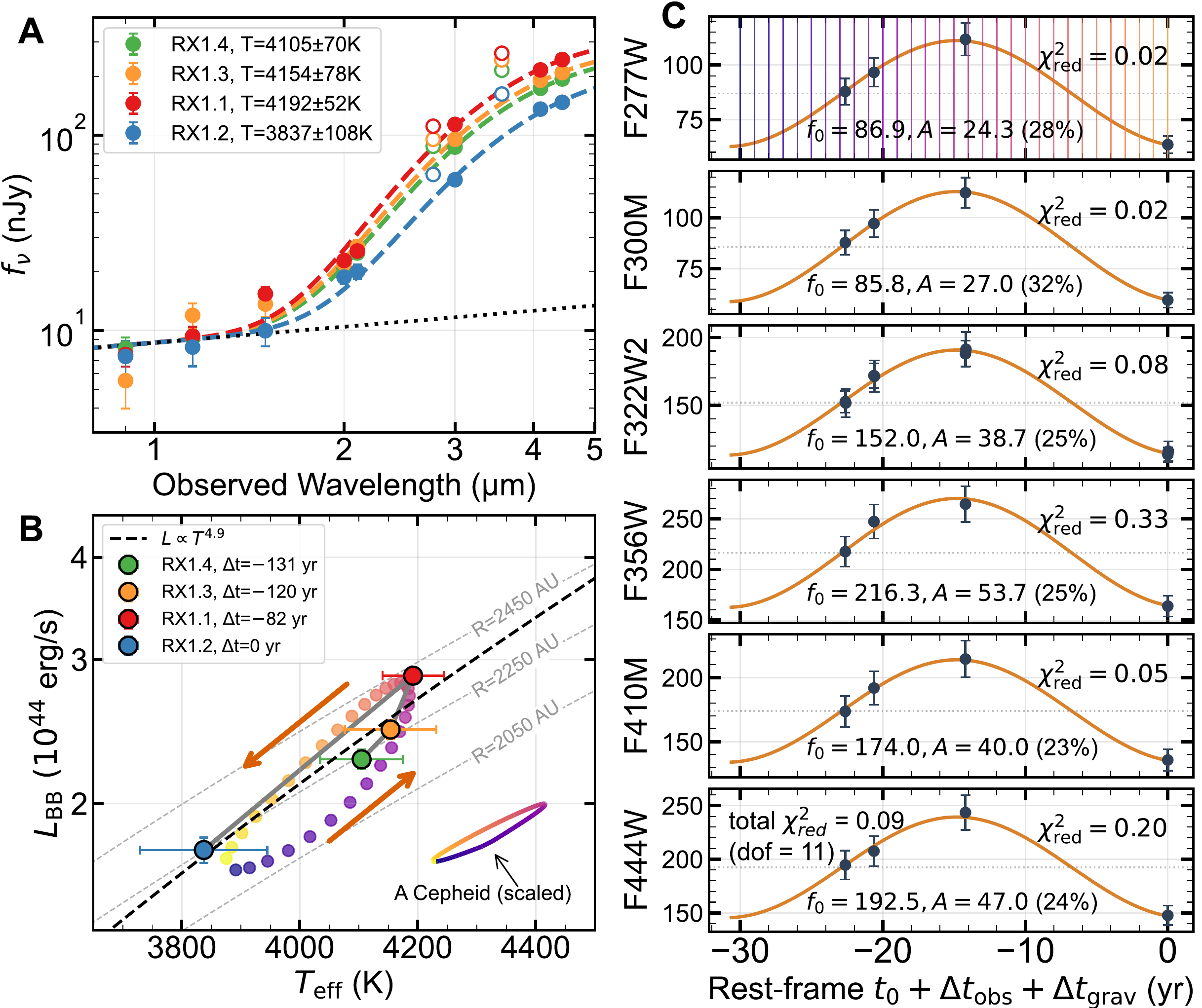}
    \caption{\textbf{SED modeling, light curve modeling, and $T_{\rm eff}$-$L_{\rm BB}$ evolution of \tgta.} \textbf{(A)} Intrinsic SEDs and best-fit models for the four multiple images of \tgta. Filled/open circles mark fitted/excluded (potentially affected by strong emission lines) bands. The dashed colored lines show the best-fit single-temperature blackbody + shared power law (black dotted line). \textbf{(B)} $T_{\rm eff}$ versus $L_{\rm BB}$ for the four images of \tgta. The solid gray line connects the four images, ordered by their relative time delays. The dashed black line indicates the best-fit power-law scaling $L_{\rm BB} \propto T_{\rm eff}^{4.9\pm1.0}$. Colored dots denote the evolutionary sequence derived from the light curve fitting in (C), showing a counterclockwise progression in chronological order. We also show the $T_{\rm eff}$-$L_{\rm BB}$ circle of a Cepheid with mean $T_{\rm eff}$ and $L_{\rm BB}$ scaled and color matched to the phase of the LRD circle in the bottom right. Dashed gray curves denote loci of constant blackbody radius. \textbf{(C)} Light curve (delensed flux in nJy) fitting results of \tgta. Observed light curves (gray points) and best models (orange curves) are shown, with best-fit amplitude, phase, and $\chi^2_{\nu}$ marked in each band. The total $\chi^2_{\nu}$ is indicated in the upper left of the F444W band panel. Vertical lines in the top panel mark the 40 epochs sampled at one-year intervals, with colors corresponding to those in (B).}
    \label{fig:lc_BB_fit}
\end{figure*}

\begin{figure*}
    \centering
    \includegraphics[width=\linewidth]{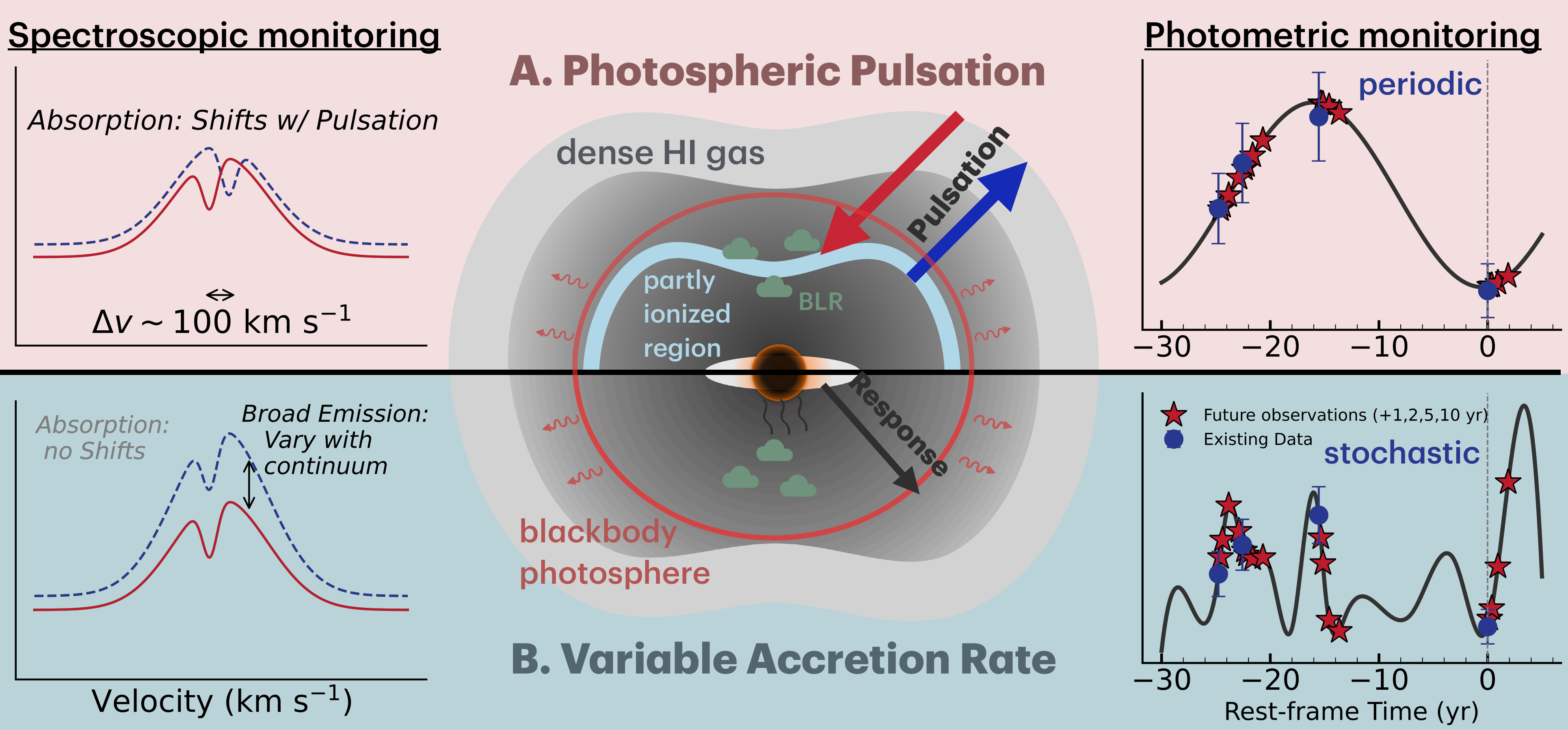}
    \caption{\textbf{Conceptional models of LRD dense gas envelopes under two scenarios.} \textbf{(A)} Under the photospheric pulsation interpretation, we expect the dense \hi\ gas to undergo similar pulsation as the partly ionized region, producing velocity shift of certain absorption lines among multiple images and periodicity in the light curve that can be obtained through long-term spectrophotometric monitoring. \textbf{(B)} Under the variable accretion state interpretation, we expect the variation of broad emission line strength correlated with the underlying continuum while no obvious velocity shift in absorption lines, and the resultant long-term light curve will appear stochastic. Expected outcomes from future spectroscopic (left) and photometric (right) monitoring are indicated.}
    \label{fig:cartoon}
\end{figure*}


\begin{table}[!t]
\centering
\caption{\textbf{Coordinates, lensing magnifications, and gravitational lens time delays of the multiple images of \tgta\ and \tgtb}. Negative lensing magnifications indicate that the image parity is reversed (i.e., the image orientation is flipped).}
\label{tab:lens_param}
\vspace{12pt}
\begin{tabular}{lccccccc}
\hline\hline
ID & R.A. & Decl. & $\mu$ & $\mu_{\rm tan}$ & $\mu_{\rm rad}$ & $-\Delta t_{\rm grav}$ (days) \\\hline
\tgta.1 & 332.95385 & $-3.83207$ & $3.60^{+0.14}_{-0.14}$ & $3.25^{+0.12}_{-0.08}$ & $1.10^{+0.03}_{-0.01}$ & $29990^{+680}_{-660}$ \\
\tgta.2 & 332.92441 & $-3.82768$ & $2.20^{+0.07}_{-0.07}$ & $2.15^{+0.05}_{-0.04}$ & $1.01^{+0.02}_{-0.01}$ & 0 \\
\tgta.3 & 332.94059 & $-3.84067$ & $-2.16^{+0.07}_{-0.09}$ & $-1.70^{+0.05}_{-0.04}$ & $1.31^{+0.01}_{-0.06}$ & $43570^{+970}_{-910}$ \\
\tgta.4 & 332.94437 & $-3.81911$ & $-2.98^{+0.12}_{-0.15}$ & $-1.88^{+0.07}_{-0.06}$ & $1.63^{+0.01}_{-0.07}$ & $47960^{+970}_{-990}$ \\
\tgtb.1 & 332.94046 & $-3.83532$ & $-1.09^{+0.05}_{-0.05}$ & $-0.64^{+0.02}_{-0.01}$ & $1.71^{+0.05}_{-0.03}$ & $58300^{+1100}_{-1100}$ \\
\tgtb.2 & 332.94092 & $-3.83182$ & $1.07^{+0.08}_{-0.10}$ & $-0.38^{+0.02}_{-0.00}$ & $-2.83^{+0.17}_{-0.15}$ & $58800^{+1100}_{-1100}$ \\
\tgtb.3 & 332.95469 & $-3.82245$ & $3.49^{+0.14}_{-0.14}$ & $3.20^{+0.09}_{-0.10}$ & $1.09^{+0.01}_{-0.02}$ & $7860^{+460}_{-380}$ \\
\tgtb.4 & 332.92738 & $-3.81977$ & $3.33^{+0.13}_{-0.11}$ & $3.29^{+0.06}_{-0.08}$ & $1.01^{+0.02}_{-0.02}$ & 0 \\
\tgtb.5 & 332.94432 & $-3.81431$ & $-8.69^{+0.76}_{-0.84}$ & $-6.63^{+0.65}_{-0.52}$ & $1.32^{+0.03}_{-0.02}$ & $14400^{+660}_{-600}$ \\
\hline
\end{tabular}
\end{table}


\clearpage 

%
\bibliography{00_main.bib} 

@ARTICLE{Zhang2025b,
       author = {{Zhang}, Zijian and {Jiang}, Linhua and {Liu}, Weiyang and {Ho}, Luis C. and {Inayoshi}, Kohei},
        title = "{JWST Insights into Narrow-line Little Red Dots}",
      journal = {arXiv e-prints},
         year = 2025,
        month = jun,
          eid = {arXiv:2506.04350},
        pages = {arXiv:2506.04350},
          doi = {10.48550/arXiv.2506.04350},
archivePrefix = {arXiv},
       eprint = {2506.04350},
 primaryClass = {astro-ph.GA},
       adsurl = {https://ui.adsabs.harvard.edu/abs/2025arXiv250604350Z}
}

@ARTICLE{Ricci2023,
       author = {{Ricci}, Claudio and {Trakhtenbrot}, Benny},
        title = "{Changing-look active galactic nuclei}",
      journal = {Nature Astronomy},
         year = 2023,
        month = nov,
       volume = {7},
        pages = {1282-1294},
          doi = {10.1038/s41550-023-02108-4},
archivePrefix = {arXiv},
       eprint = {2211.05132},
 primaryClass = {astro-ph.GA},
       adsurl = {https://ui.adsabs.harvard.edu/abs/2023NatAs...7.1282R}
}

@ARTICLE{2005MNRAS.362.1167P,
       author = {{Petterson}, O.~K.~L. and {Cottrell}, P.~L. and {Albrow}, M.~D. and {Fokin}, A.},
        title = "{A spectroscopic study of bright southern Cepheids - a high-resolution view of Cepheid atmospheres}",
      journal = {\mnras},
         year = 2005,
        month = oct,
       volume = {362},
       number = {4},
        pages = {1167-1182},
          doi = {10.1111/j.1365-2966.2005.09332.x},
       adsurl = {https://ui.adsabs.harvard.edu/abs/2005MNRAS.362.1167P}
}

@ARTICLE{Burke2025,
       author = {{Burke}, Colin J. and {Stone}, Zachary and {Shen}, Yue and {Jiang}, Yan-Fei},
        title = "{Too Quiet for Comfort: Local Little Red Dots Lack Variability over Decades}",
      journal = {arXiv e-prints},
         year = 2025,
        month = nov,
          eid = {arXiv:2511.16082},
        pages = {arXiv:2511.16082},
          doi = {10.48550/arXiv.2511.16082},
archivePrefix = {arXiv},
       eprint = {2511.16082},
 primaryClass = {astro-ph.GA},
       adsurl = {https://ui.adsabs.harvard.edu/abs/2025arXiv251116082B}
}

@ARTICLE{2020A&A...633A.107H,
       author = {{Hawkins}, M.~R.~S.},
        title = "{The signature of primordial black holes in the dark matter halos of galaxies}",
      journal = {\aap},
         year = 2020,
        month = jan,
       volume = {633},
          eid = {A107},
        pages = {A107},
          doi = {10.1051/0004-6361/201936462},
archivePrefix = {arXiv},
       eprint = {2001.07633},
 primaryClass = {astro-ph.GA},
       adsurl = {https://ui.adsabs.harvard.edu/abs/2020A&A...633A.107H}
}

@BOOK{Schneider1992g,
       author = {{Schneider}, Peter and {Ehlers}, J{\"u}rgen and {Falco}, Emilio E.},
        title = "{Gravitational Lenses}",
         year = 1992,
          doi = {10.1007/978-3-662-03758-4},
       adsurl = {https://ui.adsabs.harvard.edu/abs/1992grle.book.....S}
}

@ARTICLE{2019A&A...622A.103B,
       author = {{Boquien}, M. and {Burgarella}, D. and {Roehlly}, Y. and {Buat}, V. and {Ciesla}, L. and {Corre}, D. and {Inoue}, A.~K. and {Salas}, H.},
        title = "{CIGALE: a python Code Investigating GALaxy Emission}",
      journal = {\aap},
         year = 2019,
        month = feb,
       volume = {622},
          eid = {A103},
        pages = {A103},
          doi = {10.1051/0004-6361/201834156},
archivePrefix = {arXiv},
       eprint = {1811.03094},
 primaryClass = {astro-ph.GA},
       adsurl = {https://ui.adsabs.harvard.edu/abs/2019A&A...622A.103B}
}

@ARTICLE{2005MNRAS.360.1413B,
       author = {{Burgarella}, D. and {Buat}, V. and {Iglesias-P{\'a}ramo}, J.},
        title = "{Star formation and dust attenuation properties in galaxies from a statistical ultraviolet-to-far-infrared analysis}",
      journal = {\mnras},
         year = 2005,
        month = jul,
       volume = {360},
       number = {4},
        pages = {1413-1425},
          doi = {10.1111/j.1365-2966.2005.09131.x},
archivePrefix = {arXiv},
       eprint = {astro-ph/0504434},
 primaryClass = {astro-ph},
       adsurl = {https://ui.adsabs.harvard.edu/abs/2005MNRAS.360.1413B}
}

@ARTICLE{Inayoshi2025_SpectralUniformity,
       author = {{Inayoshi}, Kohei and {Murase}, Kohta and {Kashiyama}, Kazumi},
        title = "{Spectral Uniformity of Little Red Dots: A Natural Outcome of Coevolving Seed Black Holes and Nascent Starbursts}",
      journal = {arXiv e-prints},
         year = 2025,
        month = sep,
          eid = {arXiv:2509.19422},
        pages = {arXiv:2509.19422},
          doi = {10.48550/arXiv.2509.19422},
archivePrefix = {arXiv},
       eprint = {2509.19422},
 primaryClass = {astro-ph.GA},
       adsurl = {https://ui.adsabs.harvard.edu/abs/2025arXiv250919422I}
}

@ARTICLE{Begelman2025,
       author = {{Begelman}, Mitchell C. and {Dexter}, Jason},
        title = "{Little Red Dots As Late-stage Quasi-stars}",
      journal = {arXiv e-prints},
         year = 2025,
        month = jul,
          eid = {arXiv:2507.09085},
        pages = {arXiv:2507.09085},
          doi = {10.48550/arXiv.2507.09085},
archivePrefix = {arXiv},
       eprint = {2507.09085},
 primaryClass = {astro-ph.GA},
       adsurl = {https://ui.adsabs.harvard.edu/abs/2025arXiv250709085B}
}

@ARTICLE{Nandal2025,
       author = {{Nandal}, Devesh and {Loeb}, Abraham},
        title = "{Supermassive Stars Match the Spectral Signatures of JWST's Little Red Dots}",
      journal = {arXiv e-prints},
         year = 2025,
        month = jul,
          eid = {arXiv:2507.12618},
        pages = {arXiv:2507.12618},
          doi = {10.48550/arXiv.2507.12618},
archivePrefix = {arXiv},
       eprint = {2507.12618},
 primaryClass = {astro-ph.GA},
       adsurl = {https://ui.adsabs.harvard.edu/abs/2025arXiv250712618N}
}

@BOOK{Cox1980,
       author = {{Cox}, John P.},
        title = "{Theory of Stellar Pulsation. (PSA-2), Volume 2}",
         year = 1980,
       volume = {2},
       adsurl = {https://ui.adsabs.harvard.edu/abs/1980tsp..book.....C}
}

@ARTICLE{1981ApJ...243..140G,
       author = {{Gott}, III, J.~R.},
        title = "{Are heavy halos made of low mass stars - A gravitational lens test}",
      journal = {\apj},
         year = 1981,
        month = jan,
       volume = {243},
        pages = {140-146},
          doi = {10.1086/158576},
       adsurl = {https://ui.adsabs.harvard.edu/abs/1981ApJ...243..140G}
}

@ARTICLE{2020A&A...641A..74H,
       author = {{Hocd{\'e}}, V. and {Nardetto}, N. and {Borgniet}, S. and {Lagadec}, E. and {Kervella}, P. and {M{\'e}rand}, A. and {Evans}, N. and {Gillet}, D. and {Mathias}, Ph. and {Chiavassa}, A. and {Gallenne}, A. and {Breuval}, L. and {Javanmardi}, B.},
        title = "{Pulsating chromosphere of classical Cepheids. Calcium infrared triplet and H{\ensuremath{\alpha}} profile variations}",
      journal = {\aap},
         year = 2020,
        month = sep,
       volume = {641},
          eid = {A74},
        pages = {A74},
          doi = {10.1051/0004-6361/202037795},
archivePrefix = {arXiv},
       eprint = {2007.01365},
 primaryClass = {astro-ph.SR},
       adsurl = {https://ui.adsabs.harvard.edu/abs/2020A&A...641A..74H}
}

@ARTICLE{2000A&A...360..245B,
       author = {{Bono}, G. and {Marconi}, M. and {Stellingwerf}, R.~F.},
        title = "{Classical Cepheid pulsation models --- VI. The Hertzsprung progression}",
      journal = {\aap},
         year = 2000,
        month = aug,
       volume = {360},
        pages = {245-262},
          doi = {10.48550/arXiv.astro-ph/0006229},
archivePrefix = {arXiv},
       eprint = {astro-ph/0006229},
 primaryClass = {astro-ph},
       adsurl = {https://ui.adsabs.harvard.edu/abs/2000A&A...360..245B}
}

@ARTICLE{Eddington1917,
       author = {{Eddington}, A.~S.},
        title = "{The pulsation theory of Cepheid variables}",
      journal = {The Observatory},
         year = 1917,
        month = aug,
       volume = {40},
        pages = {290-293},
       adsurl = {https://ui.adsabs.harvard.edu/abs/1917Obs....40..290E}
}

@ARTICLE{Greene2025Lbol,
       author = {{Greene}, Jenny E. and {Setton}, David J. and {Furtak}, Lukas J. and {Naidu}, Rohan P. and {Volonteri}, Marta and {Dayal}, Pratika and {Labbe}, Ivo and {van Dokkum}, Pieter and {Bezanson}, Rachel and {Brammer}, Gabriel and {Cutler}, Sam E. and {Glazebrook}, Karl and {de Graaff}, Anna and {Hirschmann}, Michaela and {Hviding}, Raphael E. and {Kokorev}, Vasily and {Leja}, Joel and {Liu}, Hanpu and {Ma}, Yilun and {Matthee}, Jorryt and {Nanayakkara}, Themiya and {Oesch}, Pascal A. and {Pan}, Richard and {Price}, Sedona H. and {Spilker}, Justin S. and {Wang}, Bingjie and {Weaver}, John R. and {Whitaker}, Katherine E. and {Williams}, Christina C. and {Zitrin}, Adi},
        title = "{What you see is what you get: empirically measured bolometric luminosities of Little Red Dots}",
      journal = {arXiv e-prints},
         year = 2025,
        month = sep,
          eid = {arXiv:2509.05434},
        pages = {arXiv:2509.05434},
          doi = {10.48550/arXiv.2509.05434},
archivePrefix = {arXiv},
       eprint = {2509.05434},
 primaryClass = {astro-ph.GA},
       adsurl = {https://ui.adsabs.harvard.edu/abs/2025arXiv250905434G}
}

@ARTICLE{LiuH_2025,
       author = {{Liu}, Hanpu and {Jiang}, Yan-Fei and {Quataert}, Eliot and {Greene}, Jenny E. and {Ma}, Yilun},
        title = "{The Balmer Break and Optical Continuum of Little Red Dots From Super-Eddington Accretion}",
      journal = {arXiv e-prints},
         year = 2025,
        month = jul,
          eid = {arXiv:2507.07190},
        pages = {arXiv:2507.07190},
          doi = {10.48550/arXiv.2507.07190},
archivePrefix = {arXiv},
       eprint = {2507.07190},
 primaryClass = {astro-ph.GA},
       adsurl = {https://ui.adsabs.harvard.edu/abs/2025arXiv250707190L}
}

@ARTICLE{2024Sci...384..890H,
       author = {{Heintz}, Kasper E. and {Watson}, Darach and {Brammer}, Gabriel and {Vejlgaard}, Simone and {Hutter}, Anne and {Strait}, Victoria B. and {Matthee}, Jorryt and {Oesch}, Pascal A. and {Jakobsson}, P{\'a}ll and {Tanvir}, Nial R. and {Laursen}, Peter and {Naidu}, Rohan P. and {Mason}, Charlotte A. and {Killi}, Meghana and {Jung}, Intae and {Hsiao}, Tiger Yu-Yang and {Abdurro'uf} and {Coe}, Dan and {Arrabal Haro}, Pablo and {Finkelstein}, Steven L. and {Toft}, Sune},
        title = "{Strong damped Lyman-{\ensuremath{\alpha}} absorption in young star-forming galaxies at redshifts 9 to 11}",
      journal = {Science},
         year = 2024,
        month = may,
       volume = {384},
       number = {6698},
        pages = {890-894},
          doi = {10.1126/science.adj0343},
archivePrefix = {arXiv},
       eprint = {2306.00647},
 primaryClass = {astro-ph.GA},
       adsurl = {https://ui.adsabs.harvard.edu/abs/2024Sci...384..890H}
}

@ARTICLE{Kocevski2025,
       author = {{Kocevski}, Dale D. and {Finkelstein}, Steven L. and {Barro}, Guillermo and {Taylor}, Anthony J. and {Calabr{\`o}}, Antonello and {Laloux}, Brivael and {Buchner}, Johannes and {Trump}, Jonathan R. and {Leung}, Gene C.~K. and {Yang}, Guang and {Dickinson}, Mark and {P{\'e}rez-Gonz{\'a}lez}, Pablo G. and {Pacucci}, Fabio and {Inayoshi}, Kohei and {Somerville}, Rachel S. and {McGrath}, Elizabeth J. and {Akins}, Hollis B. and {Bagley}, Micaela B. and {Bowler}, Rebecca A.~A. and {Bisigello}, Laura and {Carnall}, Adam and {Casey}, Caitlin M. and {Cheng}, Yingjie and {Cleri}, Nikko J. and {Costantin}, Luca and {Cullen}, Fergus and {Davis}, Kelcey and {Donnan}, Callum T. and {Dunlop}, James S. and {Ellis}, Richard S. and {Ferguson}, Henry C. and {Fujimoto}, Seiji and {Fontana}, Adriano and {Giavalisco}, Mauro and {Grazian}, Andrea and {Grogin}, Norman A. and {Hathi}, Nimish P. and {Hirschmann}, Michaela and {Huertas-Company}, Marc and {Holwerda}, Benne W. and {Illingworth}, Garth and {Juneau}, St{\'e}phanie and {Kartaltepe}, Jeyhan S. and {Koekemoer}, Anton M. and {Li}, Wenxiu and {Lucas}, Ray A. and {Magee}, Dan and {Mason}, Charlotte and {McLeod}, Derek J. and {McLure}, Ross J. and {Napolitano}, Lorenzo and {Papovich}, Casey and {Pirzkal}, Nor and {Rodighiero}, Giulia and {Santini}, Paola and {Wilkins}, Stephen M. and {Yung}, L.~Y. Aaron},
        title = "{The Rise of Faint, Red Active Galactic Nuclei at z $>$ 4: A Sample of Little Red Dots in the JWST Extragalactic Legacy Fields}",
      journal = {\apj},
         year = 2025,
        month = jun,
       volume = {986},
       number = {2},
          eid = {126},
        pages = {126},
          doi = {10.3847/1538-4357/adbc7d},
archivePrefix = {arXiv},
       eprint = {2404.03576},
 primaryClass = {astro-ph.GA},
       adsurl = {https://ui.adsabs.harvard.edu/abs/2025ApJ...986..126K}
}

@ARTICLE{cerny25,
       author = {{Cerny}, Catherine and {Mahler}, Guillaume and {Sharon}, Keren and {Jauzac}, Mathilde and {Khullar}, Gourav and {Beauchesne}, Benjamin and {Diego}, Jose M. and {Lagattuta}, David J. and {Limousin}, Marceau and {Patel}, Nency R. and {Richard}, Johan and {Cornil-Baiotto}, Carla and {Gladders}, Michael D. and {Werner}, Stephane and {Doppel}, Jessica E. and {Floyd}, Benjamin and {Gonzalez}, Anthony H. and {Massey}, Richard J. and {Montes}, Mireia and {Bayliss}, Matthew B. and {Bleem}, Lindsey E. and {Canning}, Rebecca E.~A. and {Edge}, Alastair C. and {McDonald}, Michael and {Natarjan}, Priyamvada and {Stark}, Anthony A. and {Gassis}, Raven},
        title = "{Strong LensIng and Cluster Evolution (SLICE) with JWST: Early Results, Lens Models, and High-Redshift Detections}",
      journal = {arXiv e-prints},
         year = 2025,
        month = mar,
          eid = {arXiv:2503.17498},
        pages = {arXiv:2503.17498},
          doi = {10.48550/arXiv.2503.17498},
archivePrefix = {arXiv},
       eprint = {2503.17498},
 primaryClass = {astro-ph.CO},
       adsurl = {https://ui.adsabs.harvard.edu/abs/2025arXiv250317498C}
}

@ARTICLE{2013MNRAS.430..330H,
       author = {{H{\"a}u{\ss}ler}, Boris and {Bamford}, Steven P. and {Vika}, Marina and {Rojas}, Alex L. and {Barden}, Marco and {Kelvin}, Lee S. and {Alpaslan}, Mehmet and {Robotham}, Aaron S.~G. and {Driver}, Simon P. and {Baldry}, I.~K. and {Brough}, Sarah and {Hopkins}, Andrew M. and {Liske}, Jochen and {Nichol}, Robert C. and {Popescu}, Cristina C. and {Tuffs}, Richard J.},
        title = "{MegaMorph - multiwavelength measurement of galaxy structure: complete S{\'e}rsic profile information from modern surveys}",
      journal = {\mnras},
         year = 2013,
        month = mar,
       volume = {430},
       number = {1},
        pages = {330-369},
          doi = {10.1093/mnras/sts633},
archivePrefix = {arXiv},
       eprint = {1212.3332},
 primaryClass = {astro-ph.CO},
       adsurl = {https://ui.adsabs.harvard.edu/abs/2013MNRAS.430..330H}
}

@ARTICLE{Yue2024,
       author = {{Yue}, Minghao and {Eilers}, Anna-Christina and {Ananna}, Tonima Tasnim and {Panagiotou}, Christos and {Kara}, Erin and {Miyaji}, Takamitsu},
        title = "{Stacking X-Ray Observations of ``Little Red Dots'': Implications for Their Active Galactic Nucleus Properties}",
      journal = {\apjl},
         year = 2024,
        month = oct,
       volume = {974},
       number = {2},
          eid = {L26},
        pages = {L26},
          doi = {10.3847/2041-8213/ad7eba},
archivePrefix = {arXiv},
       eprint = {2404.13290},
 primaryClass = {astro-ph.GA},
       adsurl = {https://ui.adsabs.harvard.edu/abs/2024ApJ...974L..26Y}
}

@ARTICLE{Perez2024,
       author = {{P{\'e}rez-Gonz{\'a}lez}, Pablo G. and {Barro}, Guillermo and {Rieke}, George H. and {Lyu}, Jianwei and {Rieke}, Marcia and {Alberts}, Stacey and {Williams}, Christina C. and {Hainline}, Kevin and {Sun}, Fengwu and {Pusk{\'a}s}, D{\'a}vid and {Annunziatella}, Marianna and {Baker}, William M. and {Bunker}, Andrew J. and {Egami}, Eiichi and {Ji}, Zhiyuan and {Johnson}, Benjamin D. and {Robertson}, Brant and {Rodr{\'\i}guez Del Pino}, Bruno and {Rujopakarn}, Wiphu and {Shivaei}, Irene and {Tacchella}, Sandro and {Willmer}, Christopher N.~A. and {Willott}, Chris},
        title = "{What Is the Nature of Little Red Dots and what Is Not, MIRI SMILES Edition}",
      journal = {\apj},
         year = 2024,
        month = jun,
       volume = {968},
       number = {1},
          eid = {4},
        pages = {4},
          doi = {10.3847/1538-4357/ad38bb},
archivePrefix = {arXiv},
       eprint = {2401.08782},
 primaryClass = {astro-ph.GA},
       adsurl = {https://ui.adsabs.harvard.edu/abs/2024ApJ...968....4P}
}

@ARTICLE{Williams2024,
       author = {{Williams}, Christina C. and {Alberts}, Stacey and {Ji}, Zhiyuan and {Hainline}, Kevin N. and {Lyu}, Jianwei and {Rieke}, George and {Endsley}, Ryan and {Suess}, Katherine A. and {Sun}, Fengwu and {Johnson}, Benjamin D. and {Florian}, Michael and {Shivaei}, Irene and {Rujopakarn}, Wiphu and {Baker}, William M. and {Bhatawdekar}, Rachana and {Boyett}, Kristan and {Bunker}, Andrew J. and {Cameron}, Alex J. and {Carniani}, Stefano and {Charlot}, Stephane and {Curtis-Lake}, Emma and {DeCoursey}, Christa and {de Graaff}, Anna and {Egami}, Eiichi and {Eisenstein}, Daniel J. and {Gibson}, Justus L. and {Hausen}, Ryan and {Helton}, Jakob M. and {Maiolino}, Roberto and {Maseda}, Michael V. and {Nelson}, Erica J. and {P{\'e}rez-Gonz{\'a}lez}, Pablo G. and {Rieke}, Marcia J. and {Robertson}, Brant E. and {Saxena}, Aayush and {Tacchella}, Sandro and {Willmer}, Christopher N.~A. and {Willott}, Chris J.},
        title = "{The Galaxies Missed by Hubble and ALMA: The Contribution of Extremely Red Galaxies to the Cosmic Census at 3 $<$ z $<$ 8}",
      journal = {\apj},
         year = 2024,
        month = jun,
       volume = {968},
       number = {1},
          eid = {34},
        pages = {34},
          doi = {10.3847/1538-4357/ad3f17},
archivePrefix = {arXiv},
       eprint = {2311.07483},
 primaryClass = {astro-ph.GA},
       adsurl = {https://ui.adsabs.harvard.edu/abs/2024ApJ...968...34W}
}

@ARTICLE{Mazzolari2024,
       author = {{Mazzolari}, G. and {Gilli}, R. and {Maiolino}, R. and {Prandoni}, I. and {Delvecchio}, I. and {Norman}, C. and {Jimenez-Andrade}, E.~F. and {Belladitta}, S. and {Vito}, F. and {Momjian}, E. and {Chiaberge}, M. and {Trefoloni}, B. and {Signorini}, M. and {Ji}, X. and {D'Amato}, Q. and {Risaliti}, G. and {Baldi}, R.~D. and {Fabian}, A. and {{\"U}bler}, H. and {D'Eugenio}, F. and {Scholtz}, J. and {Juod{\v{z}}balis}, I. and {Mignoli}, M. and {Brusa}, M. and {Murphy}, E. and {Muxlow}, T.~W.~B.},
        title = "{The radio properties of the JWST-discovered AGN}",
      journal = {arXiv e-prints},
         year = 2024,
        month = dec,
          eid = {arXiv:2412.04224},
        pages = {arXiv:2412.04224},
          doi = {10.48550/arXiv.2412.04224},
archivePrefix = {arXiv},
       eprint = {2412.04224},
 primaryClass = {astro-ph.GA},
       adsurl = {https://ui.adsabs.harvard.edu/abs/2024arXiv241204224M}
}

@ARTICLE{Maiolino2025,
       author = {{Maiolino}, Roberto and {Risaliti}, Guido and {Signorini}, Matilde and {Trefoloni}, Bartolomeo and {Juod{\v{z}}balis}, Ignas and {Scholtz}, Jan and {{\"U}bler}, Hannah and {D'Eugenio}, Francesco and {Carniani}, Stefano and {Fabian}, Andy and {Ji}, Xihan and {Mazzolari}, Giovanni and {Bertola}, Elena and {Brusa}, Marcella and {Bunker}, Andrew J. and {Charlot}, Stephane and {Comastri}, Andrea and {Cresci}, Giovanni and {DeCoursey}, Christa Noel and {Egami}, Eiichi and {Fiore}, Fabrizio and {Gilli}, Roberto and {Perna}, Michele and {Tacchella}, Sandro and {Venturi}, Giacomo},
        title = "{JWST meets Chandra: a large population of Compton thick, feedback-free, and intrinsically X-ray weak AGN, with a sprinkle of SNe}",
      journal = {\mnras},
         year = 2025,
        month = apr,
       volume = {538},
       number = {3},
        pages = {1921-1943},
          doi = {10.1093/mnras/staf359},
archivePrefix = {arXiv},
       eprint = {2405.00504},
 primaryClass = {astro-ph.GA},
       adsurl = {https://ui.adsabs.harvard.edu/abs/2025MNRAS.538.1921M}
}

@ARTICLE{Matthee2024,
       author = {{Matthee}, Jorryt and {Naidu}, Rohan P. and {Brammer}, Gabriel and {Chisholm}, John and {Eilers}, Anna-Christina and {Goulding}, Andy and {Greene}, Jenny and {Kashino}, Daichi and {Labbe}, Ivo and {Lilly}, Simon J. and {Mackenzie}, Ruari and {Oesch}, Pascal A. and {Weibel}, Andrea and {Wuyts}, Stijn and {Xiao}, Mengyuan and {Bordoloi}, Rongmon and {Bouwens}, Rychard and {van Dokkum}, Pieter and {Illingworth}, Garth and {Kramarenko}, Ivan and {Maseda}, Michael V. and {Mason}, Charlotte and {Meyer}, Romain A. and {Nelson}, Erica J. and {Reddy}, Naveen A. and {Shivaei}, Irene and {Simcoe}, Robert A. and {Yue}, Minghao},
        title = "{Little Red Dots: An Abundant Population of Faint Active Galactic Nuclei at z {\ensuremath{\sim}} 5 Revealed by the EIGER and FRESCO JWST Surveys}",
      journal = {\apj},
         year = 2024,
        month = mar,
       volume = {963},
       number = {2},
          eid = {129},
        pages = {129},
          doi = {10.3847/1538-4357/ad2345},
archivePrefix = {arXiv},
       eprint = {2306.05448},
 primaryClass = {astro-ph.GA},
       adsurl = {https://ui.adsabs.harvard.edu/abs/2024ApJ...963..129M}
}

@ARTICLE{Harikane2023,
       author = {{Harikane}, Yuichi and {Zhang}, Yechi and {Nakajima}, Kimihiko and {Ouchi}, Masami and {Isobe}, Yuki and {Ono}, Yoshiaki and {Hatano}, Shun and {Xu}, Yi and {Umeda}, Hiroya},
        title = "{A JWST/NIRSpec First Census of Broad-line AGNs at z = 4-7: Detection of 10 Faint AGNs with M $_{BH}$ {}10$^{6}$-{}10$^{8}$ M $_{{\ensuremath{\odot}}}$ and Their Host Galaxy Properties}",
      journal = {\apj},
         year = 2023,
        month = dec,
       volume = {959},
       number = {1},
          eid = {39},
        pages = {39},
          doi = {10.3847/1538-4357/ad029e},
archivePrefix = {arXiv},
       eprint = {2303.11946},
 primaryClass = {astro-ph.GA},
       adsurl = {https://ui.adsabs.harvard.edu/abs/2023ApJ...959...39H}
}

@ARTICLE{Greene2024,
       author = {{Greene}, Jenny E. and {Labbe}, Ivo and {Goulding}, Andy D. and {Furtak}, Lukas J. and {Chemerynska}, Iryna and {Kokorev}, Vasily and {Dayal}, Pratika and {Volonteri}, Marta and {Williams}, Christina C. and {Wang}, Bingjie and {Setton}, David J. and {Burgasser}, Adam J. and {Bezanson}, Rachel and {Atek}, Hakim and {Brammer}, Gabriel and {Cutler}, Sam E. and {Feldmann}, Robert and {Fujimoto}, Seiji and {Glazebrook}, Karl and {de Graaff}, Anna and {Khullar}, Gourav and {Leja}, Joel and {Marchesini}, Danilo and {Maseda}, Michael V. and {Matthee}, Jorryt and {Miller}, Tim B. and {Naidu}, Rohan P. and {Nanayakkara}, Themiya and {Oesch}, Pascal A. and {Pan}, Richard and {Papovich}, Casey and {Price}, Sedona H. and {van Dokkum}, Pieter and {Weaver}, John R. and {Whitaker}, Katherine E. and {Zitrin}, Adi},
        title = "{UNCOVER Spectroscopy Confirms the Surprising Ubiquity of Active Galactic Nuclei in Red Sources at z $>$ 5}",
      journal = {\apj},
         year = 2024,
        month = mar,
       volume = {964},
       number = {1},
          eid = {39},
        pages = {39},
          doi = {10.3847/1538-4357/ad1e5f},
archivePrefix = {arXiv},
       eprint = {2309.05714},
 primaryClass = {astro-ph.GA},
       adsurl = {https://ui.adsabs.harvard.edu/abs/2024ApJ...964...39G}
}

@ARTICLE{2013MNRAS.435..623V,
       author = {{Vika}, Marina and {Bamford}, Steven P. and {H{\"a}u{\ss}ler}, Boris and {Rojas}, Alex L. and {Borch}, Andrea and {Nichol}, Robert C.},
        title = "{MegaMorph - multiwavelength measurement of galaxy structure. S{\'e}rsic profile fits to galaxies near and far}",
      journal = {\mnras},
         year = 2013,
        month = oct,
       volume = {435},
       number = {1},
        pages = {623-649},
          doi = {10.1093/mnras/stt1320},
archivePrefix = {arXiv},
       eprint = {1307.4996},
 primaryClass = {astro-ph.CO},
       adsurl = {https://ui.adsabs.harvard.edu/abs/2013MNRAS.435..623V}
}

@ARTICLE{rieke23b,
       author = {{Rieke}, Marcia J. and {Robertson}, Brant and {Tacchella}, Sandro and {Hainline}, Kevin and {Johnson}, Benjamin D. and {Hausen}, Ryan and {Ji}, Zhiyuan and {Willmer}, Christopher N.~A. and {Eisenstein}, Daniel J. and {Pusk{\'a}s}, D{\'a}vid and {Alberts}, Stacey and {Arribas}, Santiago and {Baker}, William M. and {Baum}, Stefi and {Bhatawdekar}, Rachana and {Bonaventura}, Nina and {Boyett}, Kristan and {Bunker}, Andrew J. and {Cameron}, Alex J. and {Carniani}, Stefano and {Charlot}, Stephane and {Chevallard}, Jacopo and {Chen}, Zuyi and {Curti}, Mirko and {Curtis-Lake}, Emma and {Danhaive}, A. Lola and {DeCoursey}, Christa and {Dressler}, Alan and {Egami}, Eiichi and {Endsley}, Ryan and {Helton}, Jakob M. and {Hviding}, Raphael E. and {Kumari}, Nimisha and {Looser}, Tobias J. and {Lyu}, Jianwei and {Maiolino}, Roberto and {Maseda}, Michael V. and {Nelson}, Erica J. and {Rieke}, George and {Rix}, Hans-Walter and {Sandles}, Lester and {Saxena}, Aayush and {Sharpe}, Katherine and {Shivaei}, Irene and {Skarbinski}, Maya and {Smit}, Renske and {Stark}, Daniel P. and {Stone}, Meredith and {Suess}, Katherine A. and {Sun}, Fengwu and {Topping}, Michael and {{\"U}bler}, Hannah and {Villanueva}, Natalia C. and {Wallace}, Imaan E.~B. and {Williams}, Christina C. and {Willott}, Chris and {Whitler}, Lily and {Witstok}, Joris and {Woodrum}, Charity},
        title = "{JADES Initial Data Release for the Hubble Ultra Deep Field: Revealing the Faint Infrared Sky with Deep JWST NIRCam Imaging}",
      journal = {\apjs},
         year = 2023,
        month = nov,
       volume = {269},
       number = {1},
          eid = {16},
        pages = {16},
          doi = {10.3847/1538-4365/acf44d},
archivePrefix = {arXiv},
       eprint = {2306.02466},
 primaryClass = {astro-ph.GA},
       adsurl = {https://ui.adsabs.harvard.edu/abs/2023ApJS..269...16R}
}

@ARTICLE{brammer08,
       author = {{Brammer}, Gabriel B. and {van Dokkum}, Pieter G. and {Coppi}, Paolo},
        title = "{EAZY: A Fast, Public Photometric Redshift Code}",
      journal = {\apj},
         year = 2008,
        month = oct,
       volume = {686},
       number = {2},
        pages = {1503-1513},
          doi = {10.1086/591786},
archivePrefix = {arXiv},
       eprint = {0807.1533},
 primaryClass = {astro-ph},
       adsurl = {https://ui.adsabs.harvard.edu/abs/2008ApJ...686.1503B}
}

@ARTICLE{hainline24,
       author = {{Hainline}, Kevin N. and {Johnson}, Benjamin D. and {Robertson}, Brant and {Tacchella}, Sandro and {Helton}, Jakob M. and {Sun}, Fengwu and {Eisenstein}, Daniel J. and {Simmonds}, Charlotte and {Topping}, Michael W. and {Whitler}, Lily and {Willmer}, Christopher N.~A. and {Rieke}, Marcia and {Suess}, Katherine A. and {Hviding}, Raphael E. and {Cameron}, Alex J. and {Alberts}, Stacey and {Baker}, William M. and {Baum}, Stefi and {Bhatawdekar}, Rachana and {Bonaventura}, Nina and {Boyett}, Kristan and {Bunker}, Andrew J. and {Carniani}, Stefano and {Charlot}, Stephane and {Chevallard}, Jacopo and {Chen}, Zuyi and {Curti}, Mirko and {Curtis-Lake}, Emma and {D'Eugenio}, Francesco and {Egami}, Eiichi and {Endsley}, Ryan and {Hausen}, Ryan and {Ji}, Zhiyuan and {Looser}, Tobias J. and {Lyu}, Jianwei and {Maiolino}, Roberto and {Nelson}, Erica and {Pusk{\'a}s}, D{\'a}vid and {Rawle}, Tim and {Sandles}, Lester and {Saxena}, Aayush and {Smit}, Renske and {Stark}, Daniel P. and {Williams}, Christina C. and {Willott}, Chris and {Witstok}, Joris},
        title = "{The Cosmos in Its Infancy: JADES Galaxy Candidates at z $>$ 8 in GOODS-S and GOODS-N}",
      journal = {\apj},
         year = 2024,
        month = mar,
       volume = {964},
       number = {1},
          eid = {71},
        pages = {71},
          doi = {10.3847/1538-4357/ad1ee4},
archivePrefix = {arXiv},
       eprint = {2306.02468},
 primaryClass = {astro-ph.GA},
       adsurl = {https://ui.adsabs.harvard.edu/abs/2024ApJ...964...71H}
}

@ARTICLE{1996A&AS..117..393B,
       author = {{Bertin}, E. and {Arnouts}, S.},
        title = "{SExtractor: Software for source extraction.}",
      journal = {\aaps},
         year = "1996",
        month = "Jun",
       volume = {117},
        pages = {393-404},
          doi = {10.1051/aas:1996164},
       adsurl = {https://ui.adsabs.harvard.edu/abs/1996A&AS..117..393B}
}

@ARTICLE{coe19,
       author = {{Coe}, Dan and {Salmon}, Brett and {Brada{\v{c}}}, Maru{\v{s}}a and {Bradley}, Larry D. and {Sharon}, Keren and {Zitrin}, Adi and {Acebron}, Ana and {Cerny}, Catherine and {Cibirka}, Nath{\'a}lia and {Strait}, Victoria and {Paterno-Mahler}, Rachel and {Mahler}, Guillaume and {Avila}, Roberto J. and {Ogaz}, Sara and {Huang}, Kuang-Han and {Pelliccia}, Debora and {Stark}, Daniel P. and {Mainali}, Ramesh and {Oesch}, Pascal A. and {Trenti}, Michele and {Carrasco}, Daniela and {Dawson}, William A. and {Rodney}, Steven A. and {Strolger}, Louis-Gregory and {Riess}, Adam G. and {Jones}, Christine and {Frye}, Brenda L. and {Czakon}, Nicole G. and {Umetsu}, Keiichi and {Vulcani}, Benedetta and {Graur}, Or and {Jha}, Saurabh W. and {Graham}, Melissa L. and {Molino}, Alberto and {Nonino}, Mario and {Hjorth}, Jens and {Selsing}, Jonatan and {Christensen}, Lise and {Kikuchihara}, Shotaro and {Ouchi}, Masami and {Oguri}, Masamune and {Welch}, Brian and {Lemaux}, Brian C. and {Andrade-Santos}, Felipe and {Hoag}, Austin T. and {Johnson}, Traci L. and {Peterson}, Avery and {Past}, Matthew and {Fox}, Carter and {Agulli}, Irene and {Livermore}, Rachael and {Ryan}, Russell E. and {Lam}, Daniel and {Sendra-Server}, Irene and {Toft}, Sune and {Lovisari}, Lorenzo and {Su}, Yuanyuan},
        title = "{RELICS: Reionization Lensing Cluster Survey}",
      journal = {\apj},
         year = 2019,
        month = oct,
       volume = {884},
       number = {1},
          eid = {85},
        pages = {85},
          doi = {10.3847/1538-4357/ab412b},
archivePrefix = {arXiv},
       eprint = {1903.02002},
 primaryClass = {astro-ph.GA},
       adsurl = {https://ui.adsabs.harvard.edu/abs/2019ApJ...884...85C}
}

@ARTICLE{dey19,
       author = {{Dey}, Arjun and {Schlegel}, David J. and {Lang}, Dustin and {Blum}, Robert and {Burleigh}, Kaylan and {Fan}, Xiaohui and {Findlay}, Joseph R. and {Finkbeiner}, Doug and {Herrera}, David and {Juneau}, St{\'e}phanie and {Landriau}, Martin and {Levi}, Michael and {McGreer}, Ian and {Meisner}, Aaron and {Myers}, Adam D. and {Moustakas}, John and {Nugent}, Peter and {Patej}, Anna and {Schlafly}, Edward F. and {Walker}, Alistair R. and {Valdes}, Francisco and {Weaver}, Benjamin A. and {Y{\`e}che}, Christophe and {Zou}, Hu and {Zhou}, Xu and {Abareshi}, Behzad and {Abbott}, T.~M.~C. and {Abolfathi}, Bela and {Aguilera}, C. and {Alam}, Shadab and {Allen}, Lori and {Alvarez}, A. and {Annis}, James and {Ansarinejad}, Behzad and {Aubert}, Marie and {Beechert}, Jacqueline and {Bell}, Eric F. and {BenZvi}, Segev Y. and {Beutler}, Florian and {Bielby}, Richard M. and {Bolton}, Adam S. and {Brice{\~n}o}, C{\'e}sar and {Buckley-Geer}, Elizabeth J. and {Butler}, Karen and {Calamida}, Annalisa and {Carlberg}, Raymond G. and {Carter}, Paul and {Casas}, Ricard and {Castander}, Francisco J. and {Choi}, Yumi and {Comparat}, Johan and {Cukanovaite}, Elena and {Delubac}, Timoth{\'e}e and {DeVries}, Kaitlin and {Dey}, Sharmila and {Dhungana}, Govinda and {Dickinson}, Mark and {Ding}, Zhejie and {Donaldson}, John B. and {Duan}, Yutong and {Duckworth}, Christopher J. and {Eftekharzadeh}, Sarah and {Eisenstein}, Daniel J. and {Etourneau}, Thomas and {Fagrelius}, Parker A. and {Farihi}, Jay and {Fitzpatrick}, Mike and {Font-Ribera}, Andreu and {Fulmer}, Leah and {G{\"a}nsicke}, Boris T. and {Gaztanaga}, Enrique and {George}, Koshy and {Gerdes}, David W. and {Gontcho}, Satya Gontcho A. and {Gorgoni}, Claudio and {Green}, Gregory and {Guy}, Julien and {Harmer}, Diane and {Hernandez}, M. and {Honscheid}, Klaus and {Huang}, Lijuan Wendy and {James}, David J. and {Jannuzi}, Buell T. and {Jiang}, Linhua and {Joyce}, Richard and {Karcher}, Armin and {Karkar}, Sonia and {Kehoe}, Robert and {Kneib}, Jean-Paul and {Kueter-Young}, Andrea and {Lan}, Ting-Wen and {Lauer}, Tod R. and {Le Guillou}, Laurent and {Le Van Suu}, Auguste and {Lee}, Jae Hyeon and {Lesser}, Michael and {Perreault Levasseur}, Laurence and {Li}, Ting S. and {Mann}, Justin L. and {Marshall}, Robert and {Mart{\'\i}nez-V{\'a}zquez}, C.~E. and {Martini}, Paul and {du Mas des Bourboux}, H{\'e}lion and {McManus}, Sean and {Meier}, Tobias Gabriel and {M{\'e}nard}, Brice and {Metcalfe}, Nigel and {Mu{\~n}oz-Guti{\'e}rrez}, Andrea and {Najita}, Joan and {Napier}, Kevin and {Narayan}, Gautham and {Newman}, Jeffrey A. and {Nie}, Jundan and {Nord}, Brian and {Norman}, Dara J. and {Olsen}, Knut A.~G. and {Paat}, Anthony and {Palanque-Delabrouille}, Nathalie and {Peng}, Xiyan and {Poppett}, Claire L. and {Poremba}, Megan R. and {Prakash}, Abhishek and {Rabinowitz}, David and {Raichoor}, Anand and {Rezaie}, Mehdi and {Robertson}, A.~N. and {Roe}, Natalie A. and {Ross}, Ashley J. and {Ross}, Nicholas P. and {Rudnick}, Gregory and {Safonova}, Sasha and {Saha}, Abhijit and {S{\'a}nchez}, F. Javier and {Savary}, Elodie and {Schweiker}, Heidi and {Scott}, Adam and {Seo}, Hee-Jong and {Shan}, Huanyuan and {Silva}, David R. and {Slepian}, Zachary and {Soto}, Christian and {Sprayberry}, David and {Staten}, Ryan and {Stillman}, Coley M. and {Stupak}, Robert J. and {Summers}, David L. and {Sien Tie}, Suk and {Tirado}, H. and {Vargas-Maga{\~n}a}, Mariana and {Vivas}, A. Katherina and {Wechsler}, Risa H. and {Williams}, Doug and {Yang}, Jinyi and {Yang}, Qian and {Yapici}, Tolga and {Zaritsky}, Dennis and {Zenteno}, A. and {Zhang}, Kai and {Zhang}, Tianmeng and {Zhou}, Rongpu and {Zhou}, Zhimin},
        title = "{Overview of the DESI Legacy Imaging Surveys}",
      journal = {\aj},
         year = 2019,
        month = may,
       volume = {157},
       number = {5},
          eid = {168},
        pages = {168},
          doi = {10.3847/1538-3881/ab089d},
archivePrefix = {arXiv},
       eprint = {1804.08657},
 primaryClass = {astro-ph.IM},
       adsurl = {https://ui.adsabs.harvard.edu/abs/2019AJ....157..168D}
}

@ARTICLE{gaiadr2,
       author = {{Gaia Collaboration} and {Brown}, A.~G.~A. and {Vallenari}, A. and {Prusti}, T. and {de Bruijne}, J.~H.~J. and {Babusiaux}, C. and {Bailer-Jones}, C.~A.~L. and {Biermann}, M. and {Evans}, D.~W. and {Eyer}, L. and {Jansen}, F. and {Jordi}, C. and {Klioner}, S.~A. and {Lammers}, U. and {Lindegren}, L. and {Luri}, X. and {Mignard}, F. and {Panem}, C. and {Pourbaix}, D. and {Randich}, S. and {Sartoretti}, P. and {Siddiqui}, H.~I. and {Soubiran}, C. and {van Leeuwen}, F. and {Walton}, N.~A. and {Arenou}, F. and {Bastian}, U. and {Cropper}, M. and {Drimmel}, R. and {Katz}, D. and {Lattanzi}, M.~G. and {Bakker}, J. and {Cacciari}, C. and {Casta{\~n}eda}, J. and {Chaoul}, L. and {Cheek}, N. and {De Angeli}, F. and {Fabricius}, C. and {Guerra}, R. and {Holl}, B. and {Masana}, E. and {Messineo}, R. and {Mowlavi}, N. and {Nienartowicz}, K. and {Panuzzo}, P. and {Portell}, J. and {Riello}, M. and {Seabroke}, G.~M. and {Tanga}, P. and {Th{\'e}venin}, F. and {Gracia-Abril}, G. and {Comoretto}, G. and {Garcia-Reinaldos}, M. and {Teyssier}, D. and {Altmann}, M. and {Andrae}, R. and {Audard}, M. and {Bellas-Velidis}, I. and {Benson}, K. and {Berthier}, J. and {Blomme}, R. and {Burgess}, P. and {Busso}, G. and {Carry}, B. and {Cellino}, A. and {Clementini}, G. and {Clotet}, M. and {Creevey}, O. and {Davidson}, M. and {De Ridder}, J. and {Delchambre}, L. and {Dell'Oro}, A. and {Ducourant}, C. and {Fern{\'a}ndez-Hern{\'a}ndez}, J. and {Fouesneau}, M. and {Fr{\'e}mat}, Y. and {Galluccio}, L. and {Garc{\'\i}a-Torres}, M. and {Gonz{\'a}lez-N{\'u}{\~n}ez}, J. and {Gonz{\'a}lez-Vidal}, J.~J. and {Gosset}, E. and {Guy}, L.~P. and {Halbwachs}, J. -L. and {Hambly}, N.~C. and {Harrison}, D.~L. and {Hern{\'a}ndez}, J. and {Hestroffer}, D. and {Hodgkin}, S.~T. and {Hutton}, A. and {Jasniewicz}, G. and {Jean-Antoine-Piccolo}, A. and {Jordan}, S. and {Korn}, A.~J. and {Krone-Martins}, A. and {Lanzafame}, A.~C. and {Lebzelter}, T. and {L{\"o}ffler}, W. and {Manteiga}, M. and {Marrese}, P.~M. and {Mart{\'\i}n-Fleitas}, J.~M. and {Moitinho}, A. and {Mora}, A. and {Muinonen}, K. and {Osinde}, J. and {Pancino}, E. and {Pauwels}, T. and {Petit}, J. -M. and {Recio-Blanco}, A. and {Richards}, P.~J. and {Rimoldini}, L. and {Robin}, A.~C. and {Sarro}, L.~M. and {Siopis}, C. and {Smith}, M. and {Sozzetti}, A. and {S{\"u}veges}, M. and {Torra}, J. and {van Reeven}, W. and {Abbas}, U. and {Abreu Aramburu}, A. and {Accart}, S. and {Aerts}, C. and {Altavilla}, G. and {{\'A}lvarez}, M.~A. and {Alvarez}, R. and {Alves}, J. and {Anderson}, R.~I. and {Andrei}, A.~H. and {Anglada Varela}, E. and {Antiche}, E. and {Antoja}, T. and {Arcay}, B. and {Astraatmadja}, T.~L. and {Bach}, N. and {Baker}, S.~G. and {Balaguer-N{\'u}{\~n}ez}, L. and {Balm}, P. and {Barache}, C. and {Barata}, C. and {Barbato}, D. and {Barblan}, F. and {Barklem}, P.~S. and {Barrado}, D. and {Barros}, M. and {Barstow}, M.~A. and {Bartholom{\'e} Mu{\~n}oz}, S. and {Bassilana}, J. -L. and {Becciani}, U. and {Bellazzini}, M. and {Berihuete}, A. and {Bertone}, S. and {Bianchi}, L. and {Bienaym{\'e}}, O. and {Blanco-Cuaresma}, S. and {Boch}, T. and {Boeche}, C. and {Bombrun}, A. and {Borrachero}, R. and {Bossini}, D. and {Bouquillon}, S. and {Bourda}, G. and {Bragaglia}, A. and {Bramante}, L. and {Breddels}, M.~A. and {Bressan}, A. and {Brouillet}, N. and {Br{\"u}semeister}, T. and {Brugaletta}, E. and {Bucciarelli}, B. and {Burlacu}, A. and {Busonero}, D. and {Butkevich}, A.~G. and {Buzzi}, R. and {Caffau}, E. and {Cancelliere}, R. and {Cannizzaro}, G. and {Cantat-Gaudin}, T. and {Carballo}, R. and {Carlucci}, T. and {Carrasco}, J.~M. and {Casamiquela}, L. and {Castellani}, M. and {Castro-Ginard}, A. and {Charlot}, P. and {Chemin}, L. and {Chiavassa}, A. and {Cocozza}, G. and {Costigan}, G. and {Cowell}, S. and {Crifo}, F. and {Crosta}, M. and {Crowley}, C. and {Cuypers}, J. and {Dafonte}, C. and {Damerdji}, Y. and {Dapergolas}, A. and {David}, P. and {David}, M. and {de Laverny}, P. and {De Luise}, F.},
        title = "{Gaia Data Release 2. Summary of the contents and survey properties}",
      journal = {\aap},
         year = 2018,
        month = aug,
       volume = {616},
          eid = {A1},
        pages = {A1},
          doi = {10.1051/0004-6361/201833051},
archivePrefix = {arXiv},
       eprint = {1804.09365},
 primaryClass = {astro-ph.GA},
       adsurl = {https://ui.adsabs.harvard.edu/abs/2018A&A...616A...1G}
}

@software{bushouse24,
author = {Bushouse, Howard and Eisenhamer, Jonathan and Dencheva, Nadia and Davies, James and Greenfield, Perry and Morrison, Jane and Hodge, Phil and Simon, Bernie and Grumm, David and Droettboom, Michael and Slavich, Edward and Sosey, Megan and Pauly, Tyler and Miller, Todd and Jedrzejewski, Robert and Hack, Warren and Davis, David and Crawford, Steven and Law, David and Gordon, Karl and Regan, Michael and Cara, Mihai and MacDonald, Ken and Bradley, Larry and Shanahan, Clare and Jamieson, William and Teodoro, Mairan and Williams, Thomas and Pena-Guerrero, Maria and Graham, Brett and Molter, Edward and Brandt, Timothy and Hayes, Christian and Cooper, Rachel and Clarke, Melanie},
doi = {10.5281/zenodo.15178003},
month = apr,
title = {{JWST Calibration Pipeline}},
url = {https://github.com/spacetelescope/jwst},
version = {1.18.0},
year = {2025}
}

@ARTICLE{sun25b,
       author = {{Sun}, Fengwu and {Fudamoto}, Yoshinobu and {Lin}, Xiaojing and {Helton}, Jakob M. and {Hsiao}, Tiger Yu-Yang and {Egami}, Eiichi and {Akhtarkavan}, Arshia and {Bunker}, Andrew J. and {Cai}, Zheng and {DeCoursey}, Christa and {Eisenstein}, Daniel J. and {Fan}, Xiaohui and {Harikane}, Yuichi and {Ji}, Zhiyuan and {Jin}, Xiangyu and {Liu}, Weizhe and {Liu}, Yichen and {Ma}, Zheng and {Maiolino}, Roberto and {Ouchi}, Masami and {Tee}, Wei Leong and {Wang}, Feige and {Willmer}, Christopher N.~A. and {Wu}, Yunjing and {Xu}, Yi and {Yang}, Jinyi and {Zhang}, Junyu and {Zhu}, Yongda},
        title = "{Slitless Areal Pure-Parallel HIgh-Redshift Emission Survey (SAPPHIRES): Early Data Release of Deep JWST/NIRCam Images and Spectra in MACS J0416 Parallel Field}",
      journal = {arXiv e-prints},
         year = 2025,
        month = mar,
          eid = {arXiv:2503.15587},
        pages = {arXiv:2503.15587},
          doi = {10.48550/arXiv.2503.15587},
archivePrefix = {arXiv},
       eprint = {2503.15587},
 primaryClass = {astro-ph.GA},
       adsurl = {https://ui.adsabs.harvard.edu/abs/2025arXiv250315587S}
}

@software{photutils,
  author       = {Larry Bradley and
                  Brigitta Sip{\H o}cz and
                  Thomas Robitaille and
                  Erik Tollerud and
                  Z\`e Vin{\'{\i}}cius and
                  Christoph Deil and
                  Kyle Barbary and
                  Tom J Wilson and
                  Ivo Busko and
                  Axel Donath and
                  Hans Moritz G{\"u}nther and
                  Mihai Cara and
                  P. L. Lim and
                  Sebastian Me{\ss}linger and
                  Zach Burnett and
                  Simon Conseil and
                  Michael Droettboom and
                  Azalee Bostroem and
                  E. M. Bray and
                  Lars Andersen Bratholm and
                  William Jamieson and
                  Adam Ginsburg and
                  Geert Barentsen and
                  Matt Craig and
                  Sergio Pascual and
                  Shivangee Rathi and
                  Marshall Perrin and
                  Brett M. Morris and
                  Gabriel Perren},
  title        = {astropy/photutils: 1.12.0},
  month        = apr,
  year         = 2024,
  publisher    = {Zenodo},
  version      = {1.12.0},
  doi          = {10.5281/zenodo.10967176},
  url          = {https://doi.org/10.5281/zenodo.10967176}
}

@ARTICLE{Lin2024,
       author = {{Lin}, Xiaojing and {Wang}, Feige and {Fan}, Xiaohui and {Cai}, Zheng and {Champagne}, Jaclyn B. and {Sun}, Fengwu and {Volonteri}, Marta and {Yang}, Jinyi and {Hennawi}, Joseph F. and {Ba{\~n}ados}, Eduardo and {Barth}, Aaron and {Eilers}, Anna-Christina and {Farina}, Emanuele Paolo and {Liu}, Weizhe and {Jin}, Xiangyu and {Jun}, Hyunsung D. and {Lupi}, Alessandro and {Kakiichi}, Koki and {Mazzucchelli}, Chiara and {Onoue}, Masafusa and {Pan}, Zhiwei and {Pizzati}, Elia and {Rojas-Ruiz}, Sof{\'\i}a and {Schindler}, Jan-Torge and {Trakhtenbrot}, Benny and {Shen}, Yue and {Trebitsch}, Maxime and {Zhuang}, Ming-Yang and {Endsley}, Ryan and {Meyer}, Romain A. and {Li}, Zihao and {Li}, Mingyu and {Pudoka}, Maria and {Tee}, Wei Leong and {Wu}, Yunjing and {Zhang}, Haowen},
        title = "{A SPectroscopic survey of biased halos In the Reionization Era (ASPIRE): Broad-line AGN at $z=4-5$ revealed by JWST/NIRCam WFSS}",
      journal = {arXiv e-prints},
         year = 2024,
        month = jul,
          eid = {arXiv:2407.17570},
        pages = {arXiv:2407.17570},
          doi = {10.48550/arXiv.2407.17570},
archivePrefix = {arXiv},
       eprint = {2407.17570},
 primaryClass = {astro-ph.GA},
       adsurl = {https://ui.adsabs.harvard.edu/abs/2024arXiv240717570L}
}

@ARTICLE{Juodzbalis2025,
       author = {{Juod{\v{z}}balis}, Ignas and {Maiolino}, Roberto and {Baker}, William M. and {Lake}, Emma Curtis and {Scholtz}, Jan and {D'Eugenio}, Francesco and {Trefoloni}, Bartolomeo and {Isobe}, Yuki and {Tacchella}, Sandro and {Bunker}, Andrew J. and {Carniani}, Stefano and {Charlot}, St{\'e}phane and {Jones}, Gareth C. and {Parlanti}, Eleonora and {Perna}, Michele and {Rinaldi}, Pierluigi and {Robertson}, Brant and {{\"U}bler}, Hannah and {Venturi}, Giacomo and {Willott}, Chris},
        title = "{JADES: comprehensive census of broad-line AGN from Reionization to Cosmic Noon revealed by JWST}",
      journal = {arXiv e-prints},
         year = 2025,
        month = apr,
          eid = {arXiv:2504.03551},
        pages = {arXiv:2504.03551},
archivePrefix = {arXiv},
       eprint = {2504.03551},
 primaryClass = {astro-ph.GA},
       adsurl = {https://ui.adsabs.harvard.edu/abs/2025arXiv250403551J}
}

@ARTICLE{Naidu2025,
       author = {{Naidu}, Rohan P. and {Matthee}, Jorryt and {Katz}, Harley and {de Graaff}, Anna and {Oesch}, Pascal and {Smith}, Aaron and {Greene}, Jenny E. and {Brammer}, Gabriel and {Weibel}, Andrea and {Hviding}, Raphael and {Chisholm}, John and {Labb\textbackslash'e}, Ivo and {Simcoe}, Robert A. and {Witten}, Callum and {Atek}, Hakim and {Baggen}, Josephine F.~W. and {Belli}, Sirio and {Bezanson}, Rachel and {Boogaard}, Leindert A. and {Bose}, Sownak and {Covelo-Paz}, Alba and {Dayal}, Pratika and {Fudamoto}, Yoshinobu and {Furtak}, Lukas J. and {Giovinazzo}, Emma and {Goulding}, Andy and {Gronke}, Max and {Heintz}, Kasper E. and {Hirschmann}, Michaela and {Illingworth}, Garth and {Inoue}, Akio K. and {Johnson}, Benjamin D. and {Leja}, Joel and {Leonova}, Ecaterina and {McConachie}, Ian and {Maseda}, Michael V. and {Natarajan}, Priyamvada and {Nelson}, Erica and {Setton}, David J. and {Shivaei}, Irene and {Sobral}, David and {Stefanon}, Mauro and {Tacchella}, Sandro and {Toft}, Sune and {Torralba}, Alberto and {van Dokkum}, Pieter and {van der Wel}, Arjen and {Volonteri}, Marta and {Walter}, Fabian and {Wang}, Bingjie and {Watson}, Darach},
        title = "{A ``Black Hole Star'' Reveals the Remarkable Gas-Enshrouded Hearts of the Little Red Dots}",
      journal = {arXiv e-prints},
         year = 2025,
        month = mar,
          eid = {arXiv:2503.16596},
        pages = {arXiv:2503.16596},
          doi = {10.48550/arXiv.2503.16596},
archivePrefix = {arXiv},
       eprint = {2503.16596},
 primaryClass = {astro-ph.GA},
       adsurl = {https://ui.adsabs.harvard.edu/abs/2025arXiv250316596N}
}

@ARTICLE{Wang2024b,
       author = {{Wang}, Bingjie and {Leja}, Joel and {de Graaff}, Anna and {Brammer}, Gabriel B. and {Weibel}, Andrea and {van Dokkum}, Pieter and {Baggen}, Josephine F.~W. and {Suess}, Katherine A. and {Greene}, Jenny E. and {Bezanson}, Rachel and {Cleri}, Nikko J. and {Hirschmann}, Michaela and {Labb{\'e}}, Ivo and {Matthee}, Jorryt and {McConachie}, Ian and {Naidu}, Rohan P. and {Nelson}, Erica and {Oesch}, Pascal A. and {Setton}, David J. and {Williams}, Christina C.},
        title = "{RUBIES: Evolved Stellar Populations with Extended Formation Histories at z {\ensuremath{\sim}} 7{\textendash}8 in Candidate Massive Galaxies Identified with JWST/NIRSpec}",
      journal = {\apjl},
         year = 2024,
        month = jul,
       volume = {969},
       number = {1},
          eid = {L13},
        pages = {L13},
          doi = {10.3847/2041-8213/ad55f7},
archivePrefix = {arXiv},
       eprint = {2405.01473},
 primaryClass = {astro-ph.GA},
       adsurl = {https://ui.adsabs.harvard.edu/abs/2024ApJ...969L..13W}
}

@ARTICLE{deGraaff2025_ruby,
       author = {{de Graaff}, Anna and {Rix}, Hans-Walter and {Naidu}, Rohan P. and {Labb{\'e}}, Ivo and {Wang}, Bingjie and {Leja}, Joel and {Matthee}, Jorryt and {Katz}, Harley and {Greene}, Jenny E. and {Hviding}, Raphael E. and {Baggen}, Josephine and {Bezanson}, Rachel and {Boogaard}, Leindert A. and {Brammer}, Gabriel and {Dayal}, Pratika and {van Dokkum}, Pieter and {Goulding}, Andy D. and {Hirschmann}, Michaela and {Maseda}, Michael V. and {McConachie}, Ian and {Miller}, Tim B. and {Nelson}, Erica and {Oesch}, Pascal A. and {Setton}, David J. and {Shivaei}, Irene and {Weibel}, Andrea and {Whitaker}, Katherine E. and {Williams}, Christina C.},
        title = "{A remarkable ruby: Absorption in dense gas, rather than evolved stars, drives the extreme Balmer break of a little red dot at z = 3.5}",
      journal = {\aap},
         year = 2025,
        month = sep,
       volume = {701},
          eid = {A168},
        pages = {A168},
          doi = {10.1051/0004-6361/202554681},
archivePrefix = {arXiv},
       eprint = {2503.16600},
 primaryClass = {astro-ph.GA},
       adsurl = {https://ui.adsabs.harvard.edu/abs/2025A&A...701A.168D}
}

@ARTICLE{OGLE2017,
       author = {{Soszy{\'n}ski}, I. and {Udalski}, A. and {Szyma{\'n}ski}, M.~K. and {Wyrzykowski}, {\L}. and {Ulaczyk}, K. and {Poleski}, R. and {Pietrukowicz}, P. and {Koz{\l}owski}, S. and {Skowron}, D.~M. and {Skowron}, J. and {Mr{\'o}z}, P. and {Pawlak}, M.},
        title = "{Concluding Henrietta Leavitt's Work on Classical Cepheids in the Magellanic System and Other Updates of the OGLE Collection of Variable Stars}",
      journal = {Acta Astronomica},
         year = 2017,
        month = jun,
       volume = {67},
       number = {2},
        pages = {103-113},
          doi = {10.32023/0001-5237/67.2.1},
archivePrefix = {arXiv},
       eprint = {1706.09452},
 primaryClass = {astro-ph.SR},
       adsurl = {https://ui.adsabs.harvard.edu/abs/2017AcA....67..103S}
}

@ARTICLE{OGLE2015,
       author = {{Soszy{\'n}ski}, I. and {Udalski}, A. and {Szyma{\'n}ski}, M.~K. and {Skowron}, D. and {Pietrzy{\'n}ski}, G. and {Poleski}, R. and {Pietrukowicz}, P. and {Skowron}, J. and {Mr{\'o}z}, P. and {Koz{\l}owski}, S. and {Wyrzykowski}, {\L}. and {Ulaczyk}, K. and {Pawlak}, M.},
        title = "{The OGLE Collection of Variable Stars. Classical Cepheids in the Magellanic System}",
      journal = {Acta Astronomica},
         year = 2015,
        month = dec,
       volume = {65},
       number = {4},
        pages = {297-312},
          doi = {10.48550/arXiv.1601.01318},
archivePrefix = {arXiv},
       eprint = {1601.01318},
 primaryClass = {astro-ph.SR},
       adsurl = {https://ui.adsabs.harvard.edu/abs/2015AcA....65..297S}
}

@ARTICLE{YBC2019,
       author = {{Chen}, Yang and {Girardi}, L{\'e}o and {Fu}, Xiaoting and {Bressan}, Alessandro and {Aringer}, Bernhard and {Dal Tio}, Piero and {Pastorelli}, Giada and {Marigo}, Paola and {Costa}, Guglielmo and {Zhang}, Xing},
        title = "{YBC: a stellar bolometric corrections database with variable extinction coefficients. Application to PARSEC isochrones}",
      journal = {\aap},
         year = 2019,
        month = dec,
       volume = {632},
          eid = {A105},
        pages = {A105},
          doi = {10.1051/0004-6361/201936612},
archivePrefix = {arXiv},
       eprint = {1910.09037},
 primaryClass = {astro-ph.SR},
       adsurl = {https://ui.adsabs.harvard.edu/abs/2019A&A...632A.105C}
}

@ARTICLE{Kokubo2024,
       author = {{Kokubo}, Mitsuru and {Harikane}, Yuichi},
        title = "{Challenging the AGN scenario for JWST/NIRSpec broad H$\alpha$ emitters/Little Red Dots in light of non-detection of NIRCam photometric variability and X-ray}",
      journal = {arXiv e-prints},
         year = 2024,
        month = jul,
          eid = {arXiv:2407.04777},
        pages = {arXiv:2407.04777},
          doi = {10.48550/arXiv.2407.04777},
archivePrefix = {arXiv},
       eprint = {2407.04777},
 primaryClass = {astro-ph.GA},
       adsurl = {https://ui.adsabs.harvard.edu/abs/2024arXiv240704777K}
}

@ARTICLE{Tee2025,
       author = {{Tee}, Wei Leong and {Fan}, Xiaohui and {Wang}, Feige and {Yang}, Jinyi},
        title = "{Lack of Rest-frame Ultraviolet Variability in Little Red Dots Based on HST and JWST Observations}",
      journal = {\apjl},
         year = 2025,
        month = apr,
       volume = {983},
       number = {1},
          eid = {L26},
        pages = {L26},
          doi = {10.3847/2041-8213/adc5e3},
archivePrefix = {arXiv},
       eprint = {2412.05242},
 primaryClass = {astro-ph.GA},
       adsurl = {https://ui.adsabs.harvard.edu/abs/2025ApJ...983L..26T}
}

@ARTICLE{Zhang2025a,
       author = {{Zhang}, Zijian and {Jiang}, Linhua and {Liu}, Weiyang and {Ho}, Luis C.},
        title = "{Analysis of Multi-epoch JWST Images of {\ensuremath{\sim}}300 Little Red Dots: Tentative Detection of Variability in a Minority of Sources}",
      journal = {\apj},
         year = 2025,
        month = may,
       volume = {985},
       number = {1},
          eid = {119},
        pages = {119},
          doi = {10.3847/1538-4357/adcb3e},
archivePrefix = {arXiv},
       eprint = {2411.02729},
 primaryClass = {astro-ph.GA},
       adsurl = {https://ui.adsabs.harvard.edu/abs/2025ApJ...985..119Z}
}

@ARTICLE{Ji2025a,
       author = {{Ji}, Xihan and {Maiolino}, Roberto and {{\"U}bler}, Hannah and {Scholtz}, Jan and {D'Eugenio}, Francesco and {Sun}, Fengwu and {Perna}, Michele and {Turner}, Hannah and {Arribas}, Santiago and {Bennett}, Jake S. and {Bunker}, Andrew and {Carniani}, Stefano and {Charlot}, St{\'e}phane and {Cresci}, Giovanni and {Curti}, Mirko and {Egami}, Eiichi and {Fabian}, Andy and {Inayoshi}, Kohei and {Isobe}, Yuki and {Jones}, Gareth and {Juod{\v{z}}balis}, Ignas and {Kumari}, Nimisha and {Lyu}, Jianwei and {Mazzolari}, Giovanni and {Parlanti}, Eleonora and {Robertson}, Brant and {Rodr{\'\i}guez Del Pino}, Bruno and {Schneider}, Raffaella and {Sijacki}, Debora and {Tacchella}, Sandro and {Trinca}, Alessandro and {Valiante}, Rosa and {Venturi}, Giacomo and {Volonteri}, Marta and {Willott}, Chris and {Witten}, Callum and {Witstok}, Joris},
        title = "{BlackTHUNDER -- A non-stellar Balmer break in a black hole-dominated little red dot at $z=7.04$}",
      journal = {arXiv e-prints},
         year = 2025,
        month = jan,
          eid = {arXiv:2501.13082},
        pages = {arXiv:2501.13082},
          doi = {10.48550/arXiv.2501.13082},
archivePrefix = {arXiv},
       eprint = {2501.13082},
 primaryClass = {astro-ph.GA},
       adsurl = {https://ui.adsabs.harvard.edu/abs/2025arXiv250113082J}
}

@ARTICLE{Rusakov2025,
       author = {{Rusakov}, V. and {Watson}, D. and {Nikopoulos}, G.~P. and {Brammer}, G. and {Gottumukkala}, R. and {Harvey}, T. and {Heintz}, K.~E. and {Nielsen}, R.~D. and {Sim}, S.~A. and {Sneppen}, A. and {Vijayan}, A.~P. and {Adams}, N. and {Austin}, D. and {Conselice}, C.~J. and {Goolsby}, C.~M. and {Toft}, S. and {Witstok}, J.},
        title = "{JWST's little red dots: an emerging population of young, low-mass AGN cocooned in dense ionized gas}",
      journal = {arXiv e-prints},
         year = 2025,
        month = mar,
          eid = {arXiv:2503.16595},
        pages = {arXiv:2503.16595},
          doi = {10.48550/arXiv.2503.16595},
archivePrefix = {arXiv},
       eprint = {2503.16595},
 primaryClass = {astro-ph.GA},
       adsurl = {https://ui.adsabs.harvard.edu/abs/2025arXiv250316595R}
}

@ARTICLE{Furtak2025b,
       author = {{Furtak}, Lukas J. and {Secunda}, Amy R. and {Greene}, Jenny E. and {Zitrin}, Adi and {Labb{\'e}}, Ivo and {Golubchik}, Miriam and {Bezanson}, Rachel and {Kokorev}, Vasily and {Atek}, Hakim and {Brammer}, Gabriel B. and {Chemerynska}, Iryna and {Cutler}, Sam E. and {Dayal}, Pratika and {Feldmann}, Robert and {Fujimoto}, Seiji and {Glazebrook}, Karl and {Leja}, Joel and {Ma}, Yilun and {Matthee}, Jorryt and {Naidu}, Rohan P. and {Nelson}, Erica J. and {Oesch}, Pascal A. and {Pan}, Richard and {Price}, Sedona H. and {Suess}, Katherine A. and {Wang}, Bingjie and {Weaver}, John R. and {Whitaker}, Katherine E.},
        title = "{Investigating photometric and spectroscopic variability in the multiply imaged little red dot A2744-QSO1}",
      journal = {\aap},
         year = 2025,
        month = jun,
       volume = {698},
          eid = {A227},
        pages = {A227},
          doi = {10.1051/0004-6361/202554110},
archivePrefix = {arXiv},
       eprint = {2502.07875},
 primaryClass = {astro-ph.GA},
       adsurl = {https://ui.adsabs.harvard.edu/abs/2025A&A...698A.227F}
}

@ARTICLE{Inayoshi2025_densegas,
       author = {{Inayoshi}, Kohei and {Maiolino}, Roberto},
        title = "{Extremely Dense Gas around Little Red Dots and High-redshift Active Galactic Nuclei: A Nonstellar Origin of the Balmer Break and Absorption Features}",
      journal = {\apjl},
         year = 2025,
        month = feb,
       volume = {980},
       number = {2},
          eid = {L27},
        pages = {L27},
          doi = {10.3847/2041-8213/adaebd},
archivePrefix = {arXiv},
       eprint = {2409.07805},
 primaryClass = {astro-ph.GA},
       adsurl = {https://ui.adsabs.harvard.edu/abs/2025ApJ...980L..27I}
}

@ARTICLE{Kido2025,
       author = {{Kido}, Daisaburo and {Ioka}, Kunihito and {Hotokezaka}, Kenta and {Inayoshi}, Kohei and {Irwin}, Christopher M.},
        title = "{Black Hole Envelopes in Little Red Dots}",
      journal = {arXiv e-prints},
         year = 2025,
        month = may,
          eid = {arXiv:2505.06965},
        pages = {arXiv:2505.06965},
          doi = {10.48550/arXiv.2505.06965},
archivePrefix = {arXiv},
       eprint = {2505.06965},
 primaryClass = {astro-ph.HE},
       adsurl = {https://ui.adsabs.harvard.edu/abs/2025arXiv250506965K}
}

@ARTICLE{Stone2025,
       author = {{Stone}, Zachary and {Shen}, Yue and {Zhuang}, Ming-Yang and {Hu}, Lei and {Pierel}, Justin and {Li}, Junyao and {Burgasser}, Adam J. and {Greene}, Jenny E. and {Pan}, Zhiwei and {Shapley}, Alice E. and {Sun}, Fengwu and {Venkatraman}, Padmavathi and {Wang}, Feige},
        title = "{NEXUS: A Search for Nuclear Variability with the First Two JWST NIRCam Epochs}",
      journal = {arXiv e-prints},
         year = 2025,
        month = sep,
          eid = {arXiv:2509.19585},
        pages = {arXiv:2509.19585},
          doi = {10.48550/arXiv.2509.19585},
archivePrefix = {arXiv},
       eprint = {2509.19585},
 primaryClass = {astro-ph.GA},
       adsurl = {https://ui.adsabs.harvard.edu/abs/2025arXiv250919585S}
}

@ARTICLE{Furtak2023,
       author = {{Furtak}, Lukas J. and {Zitrin}, Adi and {Plat}, Ad{\`e}le and {Fujimoto}, Seiji and {Wang}, Bingjie and {Nelson}, Erica J. and {Labb{\'e}}, Ivo and {Bezanson}, Rachel and {Brammer}, Gabriel B. and {van Dokkum}, Pieter and {Endsley}, Ryan and {Glazebrook}, Karl and {Greene}, Jenny E. and {Leja}, Joel and {Price}, Sedona H. and {Smit}, Renske and {Stark}, Daniel P. and {Weaver}, John R. and {Whitaker}, Katherine E. and {Atek}, Hakim and {Chevallard}, Jacopo and {Curtis-Lake}, Emma and {Dayal}, Pratika and {Feltre}, Anna and {Franx}, Marijn and {Fudamoto}, Yoshinobu and {Marchesini}, Danilo and {Mowla}, Lamiya A. and {Pan}, Richard and {Suess}, Katherine A. and {Vidal-Garc{\'\i}a}, Alba and {Williams}, Christina C.},
        title = "{JWST UNCOVER: Extremely Red and Compact Object at z $_{phot}\simeq$ 7.6 Triply Imaged by A2744}",
      journal = {\apj},
         year = 2023,
        month = aug,
       volume = {952},
       number = {2},
          eid = {142},
        pages = {142},
          doi = {10.3847/1538-4357/acdc9d},
archivePrefix = {arXiv},
       eprint = {2212.10531},
 primaryClass = {astro-ph.GA},
       adsurl = {https://ui.adsabs.harvard.edu/abs/2023ApJ...952..142F}
}

@ARTICLE{Lin2025_locallrd,
       author = {{Lin}, Xiaojing and {Fan}, Xiaohui and {Cai}, Zheng and {Bian}, Fuyan and {Liu}, Hanpu and {Sun}, Fengwu and {Ma}, Yilun and {Greene}, Jenny E. and {Strauss}, Michael A. and {Green}, Richard and {Lyu}, Jianwei and {Champagne}, Jaclyn B. and {Goulding}, Andy D. and {Inayoshi}, Kohei and {Jin}, Xiangyu and {Leung}, Gene C.~K. and {Li}, Mingyu and {Liu}, Yichen and {Mao}, Junjie and {Pudoka}, Maria Anne and {Tee}, Wei Leong and {Wang}, Ben and {Wang}, Feige and {Wu}, Yunjing and {Yang}, Jinyi and {Zhang}, Haowen and {Zhu}, Yongda},
        title = "{The Discovery of Little Red Dots in the Local Universe: Signatures of Cool Gas Envelopes}",
      journal = {arXiv e-prints},
         year = 2025,
        month = jul,
          eid = {arXiv:2507.10659},
        pages = {arXiv:2507.10659},
          doi = {10.48550/arXiv.2507.10659},
archivePrefix = {arXiv},
       eprint = {2507.10659},
 primaryClass = {astro-ph.GA},
       adsurl = {https://ui.adsabs.harvard.edu/abs/2025arXiv250710659L}
}

@ARTICLE{Oguri2010,
       author = {{Oguri}, Masamune},
        title = "{The Mass Distribution of SDSS J1004+4112 Revisited}",
      journal = {\pasj},
         year = 2010,
        month = aug,
       volume = {62},
        pages = {1017},
          doi = {10.1093/pasj/62.4.1017},
archivePrefix = {arXiv},
       eprint = {1005.3103},
 primaryClass = {astro-ph.CO},
       adsurl = {https://ui.adsabs.harvard.edu/abs/2010PASJ...62.1017O}
}

@ARTICLE{Oguri2021,
       author = {{Oguri}, Masamune},
        title = "{Fast Calculation of Gravitational Lensing Properties of Elliptical Navarro-Frenk-White and Hernquist Density Profiles}",
      journal = {\pasp},
         year = 2021,
        month = jul,
       volume = {133},
       number = {1025},
          eid = {074504},
        pages = {074504},
          doi = {10.1088/1538-3873/ac12db},
archivePrefix = {arXiv},
       eprint = {2106.11464},
 primaryClass = {astro-ph.IM},
       adsurl = {https://ui.adsabs.harvard.edu/abs/2021PASP..133g4504O}
}

@ARTICLE{2023ApJ...946L..12B,
       author = {{Bagley}, Micaela B. and {Finkelstein}, Steven L. and {Koekemoer}, Anton M. and {Ferguson}, Henry C. and {Arrabal Haro}, Pablo and {Dickinson}, Mark and {Kartaltepe}, Jeyhan S. and {Papovich}, Casey and {P{\'e}rez-Gonz{\'a}lez}, Pablo G. and {Pirzkal}, Nor and {Somerville}, Rachel S. and {Willmer}, Christopher N.~A. and {Yang}, Guang and {Yung}, L.~Y. Aaron and {Fontana}, Adriano and {Grazian}, Andrea and {Grogin}, Norman A. and {Hirschmann}, Michaela and {Kewley}, Lisa J. and {Kirkpatrick}, Allison and {Kocevski}, Dale D. and {Lotz}, Jennifer M. and {Medrano}, Aubrey and {Morales}, Alexa M. and {Pentericci}, Laura and {Ravindranath}, Swara and {Trump}, Jonathan R. and {Wilkins}, Stephen M. and {Calabr{\`o}}, Antonello and {Cooper}, M.~C. and {Costantin}, Luca and {de la Vega}, Alexander and {Hilbert}, Bryan and {Hutchison}, Taylor A. and {Larson}, Rebecca L. and {Lucas}, Ray A. and {McGrath}, Elizabeth J. and {Ryan}, Russell and {Wang}, Xin and {Wuyts}, Stijn},
        title = "{CEERS Epoch 1 NIRCam Imaging: Reduction Methods and Simulations Enabling Early JWST Science Results}",
      journal = {\apjl},
         year = 2023,
        month = mar,
       volume = {946},
       number = {1},
          eid = {L12},
        pages = {L12},
          doi = {10.3847/2041-8213/acbb08},
archivePrefix = {arXiv},
       eprint = {2211.02495},
 primaryClass = {astro-ph.IM},
       adsurl = {https://ui.adsabs.harvard.edu/abs/2023ApJ...946L..12B}
}

@ARTICLE{Cerny2018,
       author = {{Cerny}, Catherine and {Sharon}, Keren and {Andrade-Santos}, Felipe and {Avila}, Roberto J. and {Brada{\v{c}}}, Maru{\v{s}}a and {Bradley}, Larry D. and {Carrasco}, Daniela and {Coe}, Dan and {Czakon}, Nicole G. and {Dawson}, William A. and {Frye}, Brenda L. and {Hoag}, Austin and {Huang}, Kuang-Han and {Johnson}, Traci L. and {Jones}, Christine and {Lam}, Daniel and {Lovisari}, Lorenzo and {Mainali}, Ramesh and {Oesch}, Pascal A. and {Ogaz}, Sara and {Past}, Matthew and {Paterno-Mahler}, Rachel and {Peterson}, Avery and {Riess}, Adam G. and {Rodney}, Steven A. and {Ryan}, Russell E. and {Salmon}, Brett and {Sendra-Server}, Irene and {Stark}, Daniel P. and {Strolger}, Louis-Gregory and {Trenti}, Michele and {Umetsu}, Keiichi and {Vulcani}, Benedetta and {Zitrin}, Adi},
        title = "{RELICS: Strong Lens Models for Five Galaxy Clusters from the Reionization Lensing Cluster Survey}",
      journal = {\apj},
         year = 2018,
        month = jun,
       volume = {859},
       number = {2},
          eid = {159},
        pages = {159},
          doi = {10.3847/1538-4357/aabe7b},
archivePrefix = {arXiv},
       eprint = {1710.09329},
 primaryClass = {astro-ph.GA},
       adsurl = {https://ui.adsabs.harvard.edu/abs/2018ApJ...859..159C}
}

@ARTICLE{Liu2023,
       author = {{Liu}, Yuting and {Oguri}, Masamune and {Cao}, Shuo},
        title = "{Hubble constant from the cluster-lensed quasar system SDSS J 1004 +4112 : Investigation of the lens model dependence}",
      journal = {Physical Review D},
         year = 2023,
        month = oct,
       volume = {108},
       number = {8},
          eid = {083532},
        pages = {083532},
          doi = {10.1103/PhysRevD.108.083532},
archivePrefix = {arXiv},
       eprint = {2307.14833},
 primaryClass = {astro-ph.CO},
       adsurl = {https://ui.adsabs.harvard.edu/abs/2023PhRvD.108h3532L}
}

@ARTICLE{Setton2024,
       author = {{Setton}, David J. and {Greene}, Jenny E. and {de Graaff}, Anna and {Ma}, Yilun and {Leja}, Joel and {Matthee}, Jorryt and {Bezanson}, Rachel and {Boogaard}, Leindert A. and {Cleri}, Nikko J. and {Katz}, Harley and {Labbe}, Ivo and {Maseda}, Michael V. and {McConachie}, Ian and {Miller}, Tim B. and {Price}, Sedona H. and {Suess}, Katherine A. and {van Dokkum}, Pieter and {Wang}, Bingjie and {Weibel}, Andrea and {Whitaker}, Katherine E. and {Williams}, Christina C.},
        title = "{Little Red Dots at an Inflection Point: Ubiquitous ``V-Shaped'' Turnover Consistently Occurs at the Balmer Limit}",
      journal = {arXiv e-prints},
         year = 2024,
        month = nov,
          eid = {arXiv:2411.03424},
        pages = {arXiv:2411.03424},
          doi = {10.48550/arXiv.2411.03424},
archivePrefix = {arXiv},
       eprint = {2411.03424},
 primaryClass = {astro-ph.GA},
       adsurl = {https://ui.adsabs.harvard.edu/abs/2024arXiv241103424S}
}

@ARTICLE{Hviding2025,
       author = {{Hviding}, Raphael E. and {de Graaff}, Anna and {Miller}, Tim B. and {Setton}, David J. and {Greene}, Jenny E. and {Labb{\'e}}, Ivo and {Brammer}, Gabriel and {Bezanson}, Rachel and {Boogaard}, Leindert A. and {Cleri}, Nikko J. and {Leja}, Joel and {Maseda}, Michael V. and {McConachie}, Ian and {Matthee}, Jorryt and {Naidu}, Rohan P. and {Oesch}, Pascal A. and {Wang}, Bingjie and {Whitaker}, Katherine E. and {Williams}, Christina},
        title = "{RUBIES: A Spectroscopic Census of Little Red Dots; All V-Shaped Point Sources Have Broad Lines}",
      journal = {arXiv e-prints},
         year = 2025,
        month = jun,
          eid = {arXiv:2506.05459},
        pages = {arXiv:2506.05459},
          doi = {10.48550/arXiv.2506.05459},
archivePrefix = {arXiv},
       eprint = {2506.05459},
 primaryClass = {astro-ph.GA},
       adsurl = {https://ui.adsabs.harvard.edu/abs/2025arXiv250605459H}
}

@ARTICLE{deGraaff_2025_RUBIES,
       author = {{de Graaff}, Anna and {Brammer}, Gabriel and {Weibel}, Andrea and {Lewis}, Zach and {Maseda}, Michael V. and {Oesch}, Pascal A. and {Bezanson}, Rachel and {Boogaard}, Leindert A. and {Cleri}, Nikko J. and {Cooper}, Olivia R. and {Gottumukkala}, Rashmi and {Greene}, Jenny E. and {Hirschmann}, Michaela and {Hviding}, Raphael E. and {Katz}, Harley and {Labb{\'e}}, Ivo and {Leja}, Joel and {Matthee}, Jorryt and {McConachie}, Ian and {Miller}, Tim B. and {Naidu}, Rohan P. and {Price}, Sedona H. and {Rix}, Hans-Walter and {Setton}, David J. and {Suess}, Katherine A. and {Wang}, Bingjie and {Whitaker}, Katherine E. and {Williams}, Christina C.},
        title = "{RUBIES: A complete census of the bright and red distant Universe with JWST/NIRSpec}",
      journal = {\aap},
         year = 2025,
        month = may,
       volume = {697},
          eid = {A189},
        pages = {A189},
          doi = {10.1051/0004-6361/202452186},
archivePrefix = {arXiv},
       eprint = {2409.05948},
 primaryClass = {astro-ph.GA},
       adsurl = {https://ui.adsabs.harvard.edu/abs/2025A&A...697A.189D}
}

@ARTICLE{Bezanson2024_UNCOVER,
       author = {{Bezanson}, Rachel and {Labbe}, Ivo and {Whitaker}, Katherine E. and {Leja}, Joel and {Price}, Sedona H. and {Franx}, Marijn and {Brammer}, Gabriel and {Marchesini}, Danilo and {Zitrin}, Adi and {Wang}, Bingjie and {Weaver}, John R. and {Furtak}, Lukas J. and {Atek}, Hakim and {Coe}, Dan and {Cutler}, Sam E. and {Dayal}, Pratika and {van Dokkum}, Pieter and {Feldmann}, Robert and {F{\"o}rster Schreiber}, Natascha M. and {Fujimoto}, Seiji and {Geha}, Marla and {Glazebrook}, Karl and {de Graaff}, Anna and {Greene}, Jenny E. and {Juneau}, St{\'e}phanie and {Kassin}, Susan and {Kriek}, Mariska and {Khullar}, Gourav and {Maseda}, Michael and {Mowla}, Lamiya A. and {Muzzin}, Adam and {Nanayakkara}, Themiya and {Nelson}, Erica J. and {Oesch}, Pascal A. and {Pacifici}, Camilla and {Pan}, Richard and {Papovich}, Casey and {Setton}, David J. and {Shapley}, Alice E. and {Smit}, Renske and {Stefanon}, Mauro and {Taylor}, Edward N. and {Williams}, Christina C.},
        title = "{The JWST UNCOVER Treasury Survey: Ultradeep NIRSpec and NIRCam Observations before the Epoch of Reionization}",
      journal = {\apj},
         year = 2024,
        month = oct,
       volume = {974},
       number = {1},
          eid = {92},
        pages = {92},
          doi = {10.3847/1538-4357/ad66cf},
archivePrefix = {arXiv},
       eprint = {2212.04026},
 primaryClass = {astro-ph.GA},
       adsurl = {https://ui.adsabs.harvard.edu/abs/2024ApJ...974...92B}
}

@ARTICLE{2000ApJ...533..682C,
       author = {{Calzetti}, Daniela and {Armus}, Lee and {Bohlin}, Ralph C. and {Kinney}, Anne L. and {Koornneef}, Jan and {Storchi-Bergmann}, Thaisa},
        title = "{The Dust Content and Opacity of Actively Star-forming Galaxies}",
      journal = {\apj},
         year = 2000,
        month = apr,
       volume = {533},
       number = {2},
        pages = {682-695},
          doi = {10.1086/308692},
archivePrefix = {arXiv},
       eprint = {astro-ph/9911459},
 primaryClass = {astro-ph},
       adsurl = {https://ui.adsabs.harvard.edu/abs/2000ApJ...533..682C}
}

@ARTICLE{Eisenstein2023_JADES,
       author = {{Eisenstein}, Daniel J. and {Willott}, Chris and {Alberts}, Stacey and {Arribas}, Santiago and {Bonaventura}, Nina and {Bunker}, Andrew J. and {Cameron}, Alex J. and {Carniani}, Stefano and {Charlot}, Stephane and {Curtis-Lake}, Emma and {D'Eugenio}, Francesco and {Endsley}, Ryan and {Ferruit}, Pierre and {Giardino}, Giovanna and {Hainline}, Kevin and {Hausen}, Ryan and {Jakobsen}, Peter and {Johnson}, Benjamin D. and {Maiolino}, Roberto and {Rieke}, Marcia and {Rieke}, George and {Rix}, Hans-Walter and {Robertson}, Brant and {Stark}, Daniel P. and {Tacchella}, Sandro and {Williams}, Christina C. and {Willmer}, Christopher N.~A. and {Baker}, William M. and {Baum}, Stefi and {Bhatawdekar}, Rachana and {Boyett}, Kristan and {Chen}, Zuyi and {Chevallard}, Jacopo and {Circosta}, Chiara and {Curti}, Mirko and {Danhaive}, A. Lola and {DeCoursey}, Christa and {de Graaff}, Anna and {Dressler}, Alan and {Egami}, Eiichi and {Helton}, Jakob M. and {Hviding}, Raphael E. and {Ji}, Zhiyuan and {Jones}, Gareth C. and {Kumari}, Nimisha and {L{\"u}tzgendorf}, Nora and {Laseter}, Isaac and {Looser}, Tobias J. and {Lyu}, Jianwei and {Maseda}, Michael V. and {Nelson}, Erica and {Parlanti}, Eleonora and {Perna}, Michele and {Pusk{\'a}s}, D{\'a}vid and {Rawle}, Tim and {Rodr{\'\i}guez Del Pino}, Bruno and {Sandles}, Lester and {Saxena}, Aayush and {Scholtz}, Jan and {Sharpe}, Katherine and {Shivaei}, Irene and {Silcock}, Maddie S. and {Simmonds}, Charlotte and {Skarbinski}, Maya and {Smit}, Renske and {Stone}, Meredith and {Suess}, Katherine A. and {Sun}, Fengwu and {Tang}, Mengtao and {Topping}, Michael W. and {{\"U}bler}, Hannah and {Villanueva}, Natalia C. and {Wallace}, Imaan E.~B. and {Whitler}, Lily and {Witstok}, Joris and {Woodrum}, Charity},
        title = "{Overview of the JWST Advanced Deep Extragalactic Survey (JADES)}",
      journal = {arXiv e-prints},
         year = 2023,
        month = jun,
          eid = {arXiv:2306.02465},
        pages = {arXiv:2306.02465},
          doi = {10.48550/arXiv.2306.02465},
archivePrefix = {arXiv},
       eprint = {2306.02465},
 primaryClass = {astro-ph.GA},
       adsurl = {https://ui.adsabs.harvard.edu/abs/2023arXiv230602465E}
}

@ARTICLE{Inayoshi2013_pulsation,
       author = {{Inayoshi}, Kohei and {Hosokawa}, Takashi and {Omukai}, Kazuyuki},
        title = "{Pulsational instability of supergiant protostars: do they grow supermassive by accretion?}",
      journal = {\mnras},
         year = 2013,
        month = jun,
       volume = {431},
       number = {4},
        pages = {3036-3044},
          doi = {10.1093/mnras/stt362},
archivePrefix = {arXiv},
       eprint = {1302.6065},
 primaryClass = {astro-ph.SR},
       adsurl = {https://ui.adsabs.harvard.edu/abs/2013MNRAS.431.3036I}
}

@ARTICLE{Osaki1966_Pulsation,
       author = {{Osaki}, Yoji},
        title = "{The Pulsation and Evolution of Super-Massive Stars}",
      journal = {\pasj},
         year = 1966,
        month = dec,
       volume = {18},
       number = {4},
        pages = {384-420},
          doi = {10.1093/pasj/18.4.384},
       adsurl = {https://ui.adsabs.harvard.edu/abs/1966PASJ...18..384O}
}

@ARTICLE{Baraffe2001,
       author = {{Baraffe}, I. and {Heger}, A. and {Woosley}, S.~E.},
        title = "{On the Stability of Very Massive Primordial Stars}",
      journal = {\apj},
         year = 2001,
        month = apr,
       volume = {550},
       number = {2},
        pages = {890-896},
          doi = {10.1086/319808},
archivePrefix = {arXiv},
       eprint = {astro-ph/0009410},
 primaryClass = {astro-ph},
       adsurl = {https://ui.adsabs.harvard.edu/abs/2001ApJ...550..890B}
}

@ARTICLE{Li1994,
       author = {{Li}, Y. and {Gong}, Z.~G.},
        title = "{Red supergiant variables in the Large Magellanic Cloud : their evolution and pulsations.}",
      journal = {\aap},
         year = 1994,
        month = sep,
       volume = {289},
        pages = {449-457},
       adsurl = {https://ui.adsabs.harvard.edu/abs/1994A&A...289..449L}
}

@ARTICLE{Sonoi2012,
       author = {{Sonoi}, Takafumi and {Umeda}, Hideyuki},
        title = "{Vibrational instability of Population III very massive main-sequence stars due to the $\epsilon$-mechanism}",
      journal = {\mnras},
         year = 2012,
        month = mar,
       volume = {421},
       number = {1},
        pages = {L34-L38},
          doi = {10.1111/j.1745-3933.2011.01201.x},
archivePrefix = {arXiv},
       eprint = {1201.0884},
 primaryClass = {astro-ph.SR},
       adsurl = {https://ui.adsabs.harvard.edu/abs/2012MNRAS.421L..34S}
}

@ARTICLE{1941ApJ....94..245S,
       author = {{Schwarzschild}, Martin},
        title = "{Overtone Pulsations for the Standard Model.}",
      journal = {\apj},
         year = 1941,
        month = sep,
       volume = {94},
        pages = {245},
          doi = {10.1086/144329},
       adsurl = {https://ui.adsabs.harvard.edu/abs/1941ApJ....94..245S}
}

@ARTICLE{inayoshi2025_lrd1stagn,
       author = {{Inayoshi}, Kohei},
        title = "{Little Red Dots as the Very First Activity of Black Hole Growth}",
      journal = {\apjl},
         year = 2025,
        month = jul,
       volume = {988},
       number = {1},
          eid = {L22},
        pages = {L22},
          doi = {10.3847/2041-8213/adea66},
archivePrefix = {arXiv},
       eprint = {2503.05537},
 primaryClass = {astro-ph.GA},
       adsurl = {https://ui.adsabs.harvard.edu/abs/2025ApJ...988L..22I}
}

@ARTICLE{Torralba2025,
       author = {{Torralba}, Alberto and {Matthee}, Jorryt and {Pezzulli}, Gabriele and {Naidu}, Rohan P. and {Ishikawa}, Yuzo and {Brammer}, Gabriel B. and {Chang}, Seok-Jun and {Chisholm}, John and {de Graaff}, Anna and {D'Eugenio}, Francesco and {Di Cesare}, Claudia and {Eilers}, Anna-Christina and {Greene}, Jenny E. and {Gronke}, Max and {Iani}, Edoardo and {Kokorev}, Vasily and {Kotiwale}, Gauri and {Kramarenko}, Ivan and {Ma}, Yilun and {Mascia}, Sara and {Navarrete}, Benjam{\'\i}n and {Nelson}, Erica and {Oesch}, Pascal and {Simcoe}, Robert A. and {Wuyts}, Stijn},
        title = "{The warm outer layer of a Little Red Dot as the source of [Fe II] and collisional Balmer lines with scattering wings}",
      journal = {arXiv e-prints},
         year = 2025,
        month = sep,
          eid = {arXiv:2510.00103},
        pages = {arXiv:2510.00103},
          doi = {10.48550/arXiv.2510.00103},
archivePrefix = {arXiv},
       eprint = {2510.00103},
 primaryClass = {astro-ph.GA},
       adsurl = {https://ui.adsabs.harvard.edu/abs/2025arXiv251000103T}
}

@ARTICLE{Chen2025c,
       author = {{Chen}, Kejian and {Li}, Zhengrong and {Inayoshi}, Kohei and {Ho}, Luis C.},
        title = "{Dust Budget Crisis in Little Red Dots}",
      journal = {arXiv e-prints},
         year = 2025,
        month = may,
          eid = {arXiv:2505.22600},
        pages = {arXiv:2505.22600},
          doi = {10.48550/arXiv.2505.22600},
archivePrefix = {arXiv},
       eprint = {2505.22600},
 primaryClass = {astro-ph.GA},
       adsurl = {https://ui.adsabs.harvard.edu/abs/2025arXiv250522600C}
}

@ARTICLE{Setton2025b,
       author = {{Setton}, David J. and {Greene}, Jenny E. and {Spilker}, Justin S. and {Williams}, Christina C. and {Labb{\'e}}, Ivo and {Ma}, Yilun and {Wang}, Bingjie and {Whitaker}, Katherine E. and {Leja}, Joel and {de Graaff}, Anna and {Alberts}, Stacey and {Bezanson}, Rachel and {Boogaard}, Leindert A. and {Brammer}, Gabriel and {Cutler}, Sam E. and {Cleri}, Nikko J. and {Cooper}, Olivia R. and {Dayal}, Pratika and {Fujimoto}, Seiji and {Furtak}, Lukas J. and {Goulding}, Andy D. and {Hirschmann}, Michaela and {Kokorev}, Vasily and {Maseda}, Michael V. and {McConachie}, Ian and {Matthee}, Jorryt and {Miller}, Tim B. and {Naidu}, Rohan P. and {Oesch}, Pascal A. and {Pan}, Richard and {Price}, Sedona H. and {Suess}, Katherine A. and {Weaver}, John R. and {Xiao}, Mengyuan and {Zhang}, Yunchong and {Zitrin}, Adi},
        title = "{A Confirmed Deficit of Hot and Cold Dust Emission in the Most Luminous Little Red Dots}",
      journal = {\apjl},
         year = 2025,
        month = sep,
       volume = {991},
       number = {1},
          eid = {L10},
        pages = {L10},
          doi = {10.3847/2041-8213/ade78b},
archivePrefix = {arXiv},
       eprint = {2503.02059},
 primaryClass = {astro-ph.GA},
       adsurl = {https://ui.adsabs.harvard.edu/abs/2025ApJ...991L..10S}
}

@ARTICLE{2003MNRAS.344.1000B,
       author = {{Bruzual}, G. and {Charlot}, S.},
        title = "{Stellar population synthesis at the resolution of 2003}",
      journal = {\mnras},
         year = 2003,
        month = oct,
       volume = {344},
       number = {4},
        pages = {1000-1028},
          doi = {10.1046/j.1365-8711.2003.06897.x},
archivePrefix = {arXiv},
       eprint = {astro-ph/0309134},
 primaryClass = {astro-ph},
       adsurl = {https://ui.adsabs.harvard.edu/abs/2003MNRAS.344.1000B}
}

@ARTICLE{2003PASP..115..763C,
       author = {{Chabrier}, Gilles},
        title = "{Galactic Stellar and Substellar Initial Mass Function}",
      journal = {\pasp},
         year = 2003,
        month = jul,
       volume = {115},
       number = {809},
        pages = {763-795},
          doi = {10.1086/376392},
archivePrefix = {arXiv},
       eprint = {astro-ph/0304382},
 primaryClass = {astro-ph},
       adsurl = {https://ui.adsabs.harvard.edu/abs/2003PASP..115..763C}
}

@ARTICLE{1992ApJ...395..130P,
       author = {{Pei}, Yichuan C.},
        title = "{Interstellar Dust from the Milky Way to the Magellanic Clouds}",
      journal = {\apj},
         year = 1992,
        month = aug,
       volume = {395},
        pages = {130},
          doi = {10.1086/171637},
       adsurl = {https://ui.adsabs.harvard.edu/abs/1992ApJ...395..130P}
}

@ARTICLE{2023ApJ...944L..58W,
       author = {{Wang}, Bingjie and {Leja}, Joel and {Bezanson}, Rachel and {Johnson}, Benjamin D. and {Khullar}, Gourav and {Labb{\'e}}, Ivo and {Price}, Sedona H. and {Weaver}, John R. and {Whitaker}, Katherine E.},
        title = "{Inferring More from Less: Prospector as a Photometric Redshift Engine in the Era of JWST}",
      journal = {\apjl},
         year = 2023,
        month = feb,
       volume = {944},
       number = {2},
          eid = {L58},
        pages = {L58},
          doi = {10.3847/2041-8213/acba99},
archivePrefix = {arXiv},
       eprint = {2302.08486},
 primaryClass = {astro-ph.GA},
       adsurl = {https://ui.adsabs.harvard.edu/abs/2023ApJ...944L..58W}
}

@ARTICLE{MacLeod2010,
       author = {{MacLeod}, C.~L. and {Ivezi{\'c}}, {\v{Z}}. and {Kochanek}, C.~S. and {Koz{\l}owski}, S. and {Kelly}, B. and {Bullock}, E. and {Kimball}, A. and {Sesar}, B. and {Westman}, D. and {Brooks}, K. and {Gibson}, R. and {Becker}, A.~C. and {de Vries}, W.~H.},
        title = "{Modeling the Time Variability of SDSS Stripe 82 Quasars as a Damped Random Walk}",
      journal = {\apj},
         year = 2010,
        month = oct,
       volume = {721},
       number = {2},
        pages = {1014-1033},
          doi = {10.1088/0004-637X/721/2/1014},
archivePrefix = {arXiv},
       eprint = {1004.0276},
 primaryClass = {astro-ph.CO},
       adsurl = {https://ui.adsabs.harvard.edu/abs/2010ApJ...721.1014M}
}

@ARTICLE{NFW1997,
       author = {{Navarro}, Julio F. and {Frenk}, Carlos S. and {White}, Simon D.~M.},
        title = "{A Universal Density Profile from Hierarchical Clustering}",
      journal = {\apj},
         year = 1997,
        month = dec,
       volume = {490},
       number = {2},
        pages = {493-508},
          doi = {10.1086/304888},
archivePrefix = {arXiv},
       eprint = {astro-ph/9611107},
 primaryClass = {astro-ph},
       adsurl = {https://ui.adsabs.harvard.edu/abs/1997ApJ...490..493N}
}

@ARTICLE{wambsganss91,
       author = {{Wambsganss}, J. and {Paczynski}, B.},
        title = "{Expected Color Variations of the Gravitationally Microlensed QSO 2237+0305}",
      journal = {\aj},
         year = 1991,
        month = sep,
       volume = {102},
        pages = {864},
          doi = {10.1086/115916},
       adsurl = {https://ui.adsabs.harvard.edu/abs/1991AJ....102..864W}
}

@ARTICLE{diego24,
       author = {{Diego}, Jose M. and {Li}, Sung Kei and {Amruth}, Alfred and {Meena}, Ashish K. and {Broadhurst}, Tom J. and {Kelly}, Patrick L. and {Filippenko}, Alexei V. and {Williams}, Liliya L.~R. and {Zitrin}, Adi and {Harris}, William E. and {Reina-Campos}, Marta and {Giocoli}, Carlo and {Dai}, Liang and {Struble}, Mitchell F. and {Treu}, Tommaso and {Fudamoto}, Yoshinobu and {Gilman}, Daniel and {Koekemoer}, Anton M. and {Lim}, Jeremy and {Palencia}, Jose Mar{\'\i}a and {Sun}, Fengwu and {Windhorst}, Rogier A.},
        title = "{Imaging dark matter at the smallest scales with z {\ensuremath{\approx}} 1 lensed stars}",
      journal = {\aap},
         year = 2024,
        month = sep,
       volume = {689},
          eid = {A167},
        pages = {A167},
          doi = {10.1051/0004-6361/202450474},
archivePrefix = {arXiv},
       eprint = {2404.08033},
 primaryClass = {astro-ph.CO},
       adsurl = {https://ui.adsabs.harvard.edu/abs/2024A&A...689A.167D}
}

@ARTICLE{dai18,
       author = {{Dai}, Liang and {Venumadhav}, Tejaswi and {Kaurov}, Alexander A. and {Miralda-Escud}, Jordi},
        title = "{Probing Dark Matter Subhalos in Galaxy Clusters Using Highly Magnified Stars}",
      journal = {\apj},
         year = 2018,
        month = nov,
       volume = {867},
       number = {1},
          eid = {24},
        pages = {24},
          doi = {10.3847/1538-4357/aae478},
archivePrefix = {arXiv},
       eprint = {1804.03149},
 primaryClass = {astro-ph.CO},
       adsurl = {https://ui.adsabs.harvard.edu/abs/2018ApJ...867...24D}
}

@ARTICLE{Vegetti14,
       author = {{Vegetti}, S. and {Koopmans}, L.~V.~E. and {Auger}, M.~W. and {Treu}, T. and {Bolton}, A.~S.},
        title = "{Inference of the cold dark matter substructure mass function at z = 0.2 using strong gravitational lenses}",
      journal = {\mnras},
         year = 2014,
        month = aug,
       volume = {442},
       number = {3},
        pages = {2017-2035},
          doi = {10.1093/mnras/stu943},
archivePrefix = {arXiv},
       eprint = {1405.3666},
 primaryClass = {astro-ph.GA},
       adsurl = {https://ui.adsabs.harvard.edu/abs/2014MNRAS.442.2017V}
}

@ARTICLE{Hsueh20,
       author = {{Hsueh}, J.-W. and {Enzi}, W. and {Vegetti}, S. and {Auger}, M.~W. and {Fassnacht}, C.~D. and {Despali}, G. and {Koopmans}, L.~V.~E. and {McKean}, J.~P.},
        title = "{SHARP - VII. New constraints on the dark matter free-streaming properties and substructure abundance from gravitationally lensed quasars}",
      journal = {\mnras},
         year = 2020,
        month = feb,
       volume = {492},
       number = {2},
        pages = {3047-3059},
          doi = {10.1093/mnras/stz3177},
archivePrefix = {arXiv},
       eprint = {1905.04182},
 primaryClass = {astro-ph.CO},
       adsurl = {https://ui.adsabs.harvard.edu/abs/2020MNRAS.492.3047H}
}

@ARTICLE{Perera25,
       author = {{Perera}, Derek and {Gilman}, Daniel and {Williams}, Liliya L.~R. and {Dai}, Liang and {Du}, Xiaolong and {Rihtarsic}, Gregor and {Becerra-Espinoza}, Joaquin and {Keen}, Allison},
        title = "{Dents in the Mirror: A Novel Probe of Dark Matter Substructure in Galaxy Clusters from the Astrometric Asymmetry of Lensed Arcs}",
      journal = {arXiv e-prints},
         year = 2025,
        month = nov,
          eid = {arXiv:2511.04748},
        pages = {arXiv:2511.04748},
          doi = {10.48550/arXiv.2511.04748},
archivePrefix = {arXiv},
       eprint = {2511.04748},
 primaryClass = {astro-ph.CO},
       adsurl = {https://ui.adsabs.harvard.edu/abs/2025arXiv251104748P}
}

@ARTICLE{Golubchik25,
       author = {{Golubchik}, Miriam and {Furtak}, Lukas J. and {Allingham}, Joseph F.~V. and {Zitrin}, Adi and {Akins}, Hollis B. and {Kokorev}, Vasily and {Fujimoto}, Seiji and {Abdurro'uf} and {Amor{\'\i}n}, Ricardo O. and {Bauer}, Franz E. and {Bezanson}, Rachel and {Brada{\v{c}}}, Marusa and {Bradley}, Larry D. and {Brammer}, Gabriel B. and {Chisholm}, John and {Coe}, Dan and {Conselice}, Christopher J. and {Dayal}, Pratika and {Dessauges-Zavadsky}, Miroslava and {Diego}, Jose M. and {Faisst}, Andreas L. and {Fei}, Qinyue and {Ferguson}, Henry C. and {Finkelstein}, Steven L. and {Frye}, Brenda L. and {Gonz{\'a}lez-Otero}, Mauro and {Greene}, Jenny E. and {Harikane}, Yuichi and {Hsiao}, Tiger Yu-Yang and {Inayoshi}, Kohei and {Jim{\'e}nez-Teja}, Yolanda and {Knudsen}, Kirsten and {Koekemoer}, Anton M. and {Labb{\'e}}, Ivo and {Lucas}, Ray A. and {Magdis}, Georgios E. and {Matthee}, Jorryt and {Messa}, Matteo and {Naidu}, Rohan P. and {Nakane}, Minami and {Noirot}, Ga{\"e}l and {Pan}, Richard and {Papovich}, Casey and {Richard}, Johan and {Ricotti}, Massimo and {Robbins}, Luke and {Stark}, Daniel P. and {Sun}, Fengwu and {Treu}, Tommaso and {Tripodi}, Roberta and {Vanzella}, Eros and {Willott}, Chris and {Windhorst}, Rogier A.},
        title = "{VENUS: When Red meets Blue -- A multiply imaged Little Red Dot with an apparent blue companion behind the galaxy cluster Abell 383}",
      journal = {arXiv e-prints},
         year = 2025,
        month = dec,
          eid = {arXiv:2512.02117},
        pages = {arXiv:2512.02117},
archivePrefix = {arXiv},
       eprint = {2512.02117},
 primaryClass = {astro-ph.GA},
       adsurl = {https://ui.adsabs.harvard.edu/abs/2025arXiv251202117G}
}
\bibliographystyle{sciencemag}

%
%
%
%
%
%


\section*{Acknowledgments}

The authors acknowledge the use of the Canadian Advanced Network for Astronomy Research (CANFAR) Science Platform operated by the Canadian Astronomy Data Center (CADC) and the Digital Research Alliance of Canada (DRAC), with support from the National Research Council of Canada (NRC), the Canadian Space Agency (CSA), CANARIE, and the Canadian Foundation for Innovation (CFI).
This research is based on observations made with the NASA/ESA Hubble Space Telescope obtained from the Space Telescope Science Institute, which is operated by the Association of Universities for Research in Astronomy, Inc., under NASA contract NAS 5-26555. These observations are associated with program(s) GO-14096 (RELICS).
This work is based on observations made with the NASA/ESA/CSA James Webb Space Telescope. The data were obtained from the Mikulski Archive for Space Telescopes at the Space Telescope Science Institute, which is operated by the Association of Universities for Research in Astronomy, Inc., under NASA contract NAS 5-03127 for JWST. These observations are associated with program \#5594 (SLICE) and 6882 (VENUS). Some of the data presented herein were retrieved from the Dawn JWST Archive (DJA). DJA is an initiative of the Cosmic Dawn Center (DAWN), which is funded by the Danish National Research Foundation under grant DNRF140.
F.S.\ specially thank Dr.\ Roberto Decarli for hosting the Bologna meeting in 2025 summer that conceptualizes this project.

\paragraph*{Funding:}
Z.Z.\ and L.J.\ acknowledge support from the National Science Foundation of China (12225301).
M.O.\ acknowledges support from JSPS KAKENHI Grant Numbers JP25H00662, JP22K21349.
K.I.\ acknowledges support from the National Natural Science Foundation of China (12573015, 1251101148, 12233001), the Beijing Natural Science Foundation (IS25003), and the China Manned Space Program (CMS-CSST-2025-A09).
J.M.D.\ acknowledges the support of projects PID2022-138896NB-C51 (MCIU/AEI/MINECO/FEDER, UE) Ministerio de Ciencia, Investigaci\'on y Universidades and SA101P24. 
R.A.\ acknowledges support of Grant PID2023-147386NB-I00 funded by MICIU/AEI/10.13039/501100011033 and by ERDF/EU, and the Severo Ochoa award to the IAA-CSIC CEX2021-001131-S.
H.A.\ acknowledges support from CNES, focused on the JWST mission, and the French National Research Agency (ANR) under grant ANR-21-CE31-0838.  
P.D.\ warmly acknowledges support from an NSERC discovery grant (RGPIN-2025-06182).
D.J.E.\ and J.A.A.T.\ acknowledge support from the Simons Foundation and JWST program 3215.
D.J.E.\ was further supported by the JWST/NIRCam contract to the University of Arizona, NAS5-02015.
G.E.M.\ acknowledges the Villum Fonden research grants 37440 and 13160.
The Cosmic Dawn Center (DAWN) is funded by the Danish National Research Foundation under grant DNRF140.
M.M.\ and E.V.\ acknowledge financial support through grants INAF GO Grant 2022 ``The revolution is around the corner: JWST will probe globular cluster precursors and Population III stellar clusters at cosmic dawn'',  INAF GO Grant 2024 ``Mapping Star Cluster Feedback in a Galaxy 450 Myr after the Big Bang'' and by the European Union - NextGenerationEU within PRIN 2022 project n.20229YBSAN - Globular clusters in cosmological simulations and lensed fields: from their birth to the present epoch.
R.A.W.\ acknowledges support from NASA JWST Interdisciplinary Scientist grants NAG5-12460, NNX14AN10G and 80NSSC18K0200 from GSFC.
A.Z.\ acknowledges support by the Israel Science Foundation Grant No. 864/23.
F.S.\ acknowledge support from JWST program 4924 and 6434. 
Support for JWST program \#3215, 4924, 5594, 6434 and 6882 was provided by NASA through a grant from the Space Telescope Science Institute, which is operated by the Association of Universities for Research in Astronomy, Inc., under NASA contract NAS 5-03127.

\paragraph*{Author contributions:}
Z.Z.\ led the data analyses, physical interpretation, figure production and the paper writing.
M.L.\ discovered the LRDs, contributed to the Keck data and figure production.
M.O.\ led the multiple image identification and constructed the cluster mass model.
X.L., K.I., F.S.\ contributed to the physical interpretation.
Z.Z., M.L., X.L.\ and F.S.\ contributed to the JWST data processing.
C.C, G.M.\ and K.S.\ led the JWST SLICE program design, data acquisition and contributed to the multiple image identification and cluster mass model.
S.F.\ and D.C.\ led the JWST VENUS program design and data acquisition.
S.C., M.L., Y.L., N.R.\ and C.C.S.\ contributed to the Keck data.
F.S.\ led the project conceptualization and administration. 
All authors contributed to the results, discussions and manuscript preparation.

\paragraph*{Competing interests:}
The authors declare that they have no competing interests.

\paragraph*{Data and materials availability:}
All JWST data and HST data are publicly available at MAST: \url{https://archive.stsci.edu/}. 
Image reduction and analysis used publicly available pipelines: CEERS NIRCam \cite{2023ApJ...946L..12B}, CIGALE\cite{2005MNRAS.360.1413B,2019A&A...622A.103B}, 
GalfitM \cite{2013MNRAS.430..330H,2013MNRAS.435..623V}, 
Source Extractor \cite{1996A&AS..117..393B}.  

\subsection*{Supplementary materials}
Materials and Methods\\
Supplementary Text\\
Figs. S1 to S9\\
Tables S1 to S2\\
References \textit{(51-\arabic{enumiv})}\\ 


\newpage


\renewcommand{\thefigure}{S\arabic{figure}}
\renewcommand{\thetable}{S\arabic{table}}
\renewcommand{\theequation}{S\arabic{equation}}
\renewcommand{\thepage}{S\arabic{page}}
\setcounter{figure}{0}
\setcounter{table}{0}
\setcounter{equation}{0}
\setcounter{page}{1} 


\begin{center}
\section*{Supplementary Materials for\\ \scititle}

Zijian~Zhang$^{1,2}$,
    Mingyu~Li$^{3}$,
    Masamune~Oguri$^{4,5}$,
    Xiaojing~Lin$^{3}$,
    Kohei~Inayoshi$^{1}$,  \\
    Catherine~Cerny$^{6}$,
    Dan~Coe$^{7,8,9}$,
    Jose~M.~Diego$^{10}$,
    Seiji~Fujimoto$^{11,12}$,
    Linhua~Jiang$^{1,2,\ast}$,  \\
    Guillaume~Mahler$^{13}$,
    Jorryt~Matthee$^{14}$,
    Rohan~P.~Naidu$^{15}$,
    Keren~Sharon$^{6}$,
    Yue~Shen$^{16,17}$,  \\
    Adi~Zitrin$^{18}$,
    Abdurro'uf$^{19}$,
    Hollis~Akins$^{20}$,
    Joseph~F.~V.~Allingham$^{18}$,
    Ricardo~Amor\'in$^{21}$, \\
    Yoshihisa~Asada$^{11,12}$, 
    Hakim~Atek$^{22}$,
    Franz~E.~Bauer$^{23}$,
    Maru\v{s}a~Brada\v{c}$^{24}$, \\
    Larry~D.~Bradley$^{7}$,
    Zheng~Cai$^{3}$, 
    Sebastiano~Cantalupo$^{25}$,
    Christopher~Conselice$^{26}$,\\
    Liang~Dai$^{27}$,  
    Pratika~Dayal$^{28,11,29}$,
    Eiichi~Egami$^{30}$,
    Daniel~J.~Eisenstein$^{31}$,\\
    Andreas~L.~Faisst$^{32}$,  
    Xiaohui~Fan$^{30}$, 
    Qinyue~Fei$^{11}$,
    Brenda~L.~Frye$^{30}$,\\
    Yoshinobu~Fudamoto$^{4}$, 
    Lukas~J.~Furtak$^{20,33}$,
    Miriam~Golubchik$^{18}$,\\
    Mauro~Gonz\'{a}lez-Otero$^{21}$,
    Yuichi~Harikane$^{34}$,
    \pending{Tiger~Yu-Yang~Hsiao$^{20,33}$}, \\
    Yolanda~Jim\'enez-Teja$^{21,35}$, 
    Jeyhan~S.~Kartaltepe$^{36}$,
    Tomokazu~Kiyota$^{37,38}$, \\
    Anton~M.~Koekemoer$^{7}$, 
    \pending{Kotaro~Kohno$^{39,40}$}, 
    Vasily~Kokorev$^{20,33}$,
    \pending{Nimisha~Kumari$^{9}$}, \\
    Ivo~Labbe$^{41}$,
    Claudia~D.~P.~Lagos$^{42,43}$,
    Conor~Larison$^{7}$,
    Yongming~Liang$^{34,38}$, \\
    Ray~A.~Lucas$^{7}$,
    Jianwei~Lyu$^{30}$,
    Nicholas~S.~Martis$^{24}$,
    Georgios~E.~Magdis$^{43,44}$,\\
    Matteo~Messa$^{45}$,
    Minami~Nakane$^{34,46}$,
    Ga\"el~Noirot$^{7}$,
    Rafael~Ortiz~III$^{47}$, \\
    Masami Ouchi$^{38,34,XX,48}$,
    Justin~D.~R.~Pierel$^{7}$,
    Marc~Postman$^{7}$,
    Naveen~Reddy$^{49}$,\\
    Massimo~Ricotti$^{50}$,
    Daniel~Schaerer$^{51,52}$,
    Raffaella~Schneider$^{53}$,
    Charles~C.~Steidel$^{54}$, \\
    Wei~Leong~Tee$^{55}$,
    Roberta~Tripodi$^{56,24,57}$,
    James~A.~A.~Trussler$^{31}$,
    Hiroya~Umeda$^{34,46}$,\\
    Francesco~Valentino$^{43,44}$,
    Eros~Vanzella$^{45}$,
    Feige~Wang$^{6}$,
    Rogier~Windhorst$^{47}$,\\
    Yunjing~Wu$^{3}$,
    Zihao~Wu$^{31}$,
    Hiroto~Yanagisawa$^{34,46}$,
    Jinyi~Yang$^{6}$,
    Fengwu~Sun$^{31,\ast}$ \\ 
	\small$^\ast$Corresponding authors. Email: \href{mailto:fengwu.sun@cfa.harvard.edu}{fengwu.sun@cfa.harvard.edu} (Fengwu Sun) \& \href{mailto:jiangKIAA@pku.edu.cn}{jiangKIAA@pku.edu.cn} (Linhua Jiang)
\end{center}

\subsubsection*{This PDF file includes:}
Materials and Methods\\
Supplementary Text\\
Figures S1 to S9\\
Tables S1 to S2\\


\newpage


\subsection*{Materials and Methods}

\subsubsection*{Imaging Data and Photometry} 

In this work, we use the JWST/NIRCam data taken by the Vast Exploration for Nascent, Unexplored Sources (\venus) program (PID: 6882; PI: Fujimoto) and Strong LensIng and Cluster Evolution (SLICE) program\cite{cerny25} (PID: 5594; PI: Mahler). 
The \venus\ program is designed to deliver uniform ten-band NIRCam imaging across 0.8--5.0\,\micron\ for a sample of 60 massive and well-studied galaxy clusters.
As part of this survey, \rxc\ was observed in ten NIRCam filters: F090W, F115W, F150W, F200W, F210M, F277W, F300M, F356W, F410M, and F444W on UT October 16, 2025. 
The SLICE program obtained NIRCam imaging of \rxc\ in the ultra-wide F150W2 and F322W2 bands on UT November 17, 2024.
We also include archival HST/ACS imaging data in the F435W, F606W, and F814W bands from the Reionization Lensing Cluster Survey\cite{coe19} (RELICS) for part of our analyses.

All NIRCam imaging data were uniformly processed using a customized JWST pipeline\footnote{\href{https://github.com/spacetelescope/jwst}{https://github.com/spacetelescope/jwst}}\cite{bushouse24} (\verb|v1.18.0|) and CRDS context map \verb|jwst_1364.pmap|. 
The customized steps include: (\romannumeral1) 1/f noise stripe removal in both row and column directions; 
(\romannumeral2) wisp removal in the NIRCam short-wavelength (SW) detectors using STScI-provided templates; 
(\romannumeral3) bad-pixel masking based on sigma-clipped median images from each detector; 
(\romannumeral4) median sky-background subtraction; and 
(\romannumeral5) world coordinate system (WCS) correction using the DESI Legacy Imaging Survey \cite{dey19} DR10 source catalog, 
which is tied to the Gaia DR2 astrometric reference frame \cite{gaiadr2}.
Calibrated NIRCam images in each band were mosaicked through a standard stage-3 pipeline with an
output pixel size of 0\farcs03, \texttt{pixfrac = 1.0} and a common
north-up-east-left WCS frame for all filters. 
The diffuse intracluster light (ICL) is not subtracted from the images.

Following methods in previous work \cite{sun25b}, we create a stacked NIRCam detection image combining the F200W--F444W bands to optimize the S/N and apply iterative image sharpening to improve source deblending in the crowded cluster field. 
Source detection and photometric measurements are then performed using a customized \texttt{photutils} pipeline \cite{photutils}.
We perform aperture photometry on NIRCam-detected sources using a range of aperture radii from 0\farcs1 to 0\farcs5 as well as Kron apertures, with local background measured through rectangular annuli and subtracted. 
Aperture losses are corrected using empirical PSF models constructed from stacks of $\sim10$ bright, isolated stars within the imaging footprint.
Photometric uncertainties are estimated from both error extensions and random-aperture experiments, adopting the larger of the two as the final error.
The median $5\sigma$ depth for point sources measured with an $r=0\farcs1$ aperture is $28.0 \pm 0.1$\,mag ($28.3 \pm 0.2$\,mag) in the five \venus\ NIRCam band in the SW (LW) channel, respectively.
For the SLICE F150W2 and F322W2 bands, the corresponding depth is 28.8\,mag.

Since all images of \tgta\ are perfect point sources, for this source we use the simple aperture photometry with $r=$\,0\farcs15 and aperture correction to improve S/N. For \tgtb.2, \tgtb.3, and \tgtb.4, we also use the above photometry. \tgtb.1, however, has a nearby companion that is not physically associated, as inferred from the prior of its other images, the \zph, and magnification values. We therefore fit \tgtb.1 and this companion with a central PSF and an offset \sersic\ model simultaneously using \texttt{GALFITM} and adopt the best-fit PSF model as its final photometry. For \tgtb.5, it has a significant off-centered extended component.
Its elongation direction is aligned with the direction of $\mu_{\rm tan}$. The large tangential magnification and consistent photometric redshift also support that the extended component is physically associated with \tgtb.5. We therefore fit it using a PSF+\sersic\ model and adopt the sum of these two components as its final photometry. Both \tgtb.1 and \tgtb.2 suffer from strong intracluster light (ICL), which biases the fitting. Before running \texttt{GALFITM}, we therefore subtract a smooth 2D background using a median-based estimator and Gaussian filtering (via \texttt{Background2D}), ensuring that the PSF+\sersic\ modeling is not affected by large-scale ICL residuals.
Photometry of \tgta\ and \tgtb\ are presented in Table~\ref{tab:phot}.

\subsubsection*{Strong Lens Mass Modeling}

We use \texttt{glafic} \cite{Oguri2010,Oguri2021} to construct a strong lens mass model of \rxc. For this cluster, 3 multiply imaged sources are reported previously \cite{Cerny2018} and 6 more multiply imaged sources are reported in later work \cite{cerny25}. We search for multiple images in the \venus\ data to identify additional 34 multiply imaged sources. 
In practice, multiple images are discovered iteratively based on their similar morphologies, colors, photometric redshifts and consistency with the predicted image-plane positions from the pervious best cluster mass model.
In total, we use 146 multiple images from 43 sources for the mass modeling, which are summarized in Figure~\ref{fig:lensing_model} and Table~\ref{tab:multi}. The spectroscopic redshift of one of the sources is reported in previous work \cite{Cerny2018}, and spectroscopic redshifts for two additional sources are obtained from our recent Keck/KCWI observation (private communication from Sebastiano Cantalupo, Mingyu Li et al.). 
When accurate photometric redshifts are available, we include Gaussian priors on their redshifts, typically assuming the width of $\sigma_z=0.2$ according to the photometric redshift posteriors. For \tgta and \tgtb, we adopt a conservative Gaussian prior to their redshift of $4.3\pm 0.4$ (See Photometric Redshift section). We fit the multiple image positions assuming two halo components modeled by the elliptical density profile\cite{NFW1997}, member galaxies modeled by the pseudo-Jaffe ellipsoids, external shear, and third and fourth order multipole perturbations (see e.g., \cite{Liu2023} for more detailed explanations of these lens components). 

Assuming the positional error of $0\farcs 6$ (but $0\farcs 2$ for a few close pairs of multiple images), our best-fitting mass model has $\chi^2=174.5$ for the degree of freedom  of 180. The root-mean-square between observed and model-predicted multiple image positions is $0\farcs 61$. Figure~\ref{fig:lensing_model} indicates that there are many other multiple images near each multiple image of \tgta\ and \tgtb, implying that their magnifications and time delays are tightly constrained by the mass modeling. Errors on model parameters are obtained by the Markov chain Monte Carlo (MCMC).

We use the cluster mass model to examine the redshift of the source. We systematically test a grid of candidate source redshifts spanning 0 to 10. For each assumed redshift, we use the cluster mass model to inverse-map all observed multiple images of the source back to the source plane. A valid redshift should produce a consistent source plane position for all the images. Using this method, our model strongly suggest that the redshifts of \tgta\ and \tgtb\ are in the range of 4-5.5, which is consistent with our $z_{\rm phot}$ estimation.

\subsubsection*{Construction of Empirical LRD Templates}

As LRDs are a newly identified population not represented in current photometric redshift template libraries, such mismatch can bias the photometric redshift estimates. 
Therefore, we construct empirical LRD templates based on the JWST/NIRSpec prism spectra of 44 confirmed LRDs at $z = 2.3$-$7$. These LRD spectra are mainly collected from previous work\cite{Setton2024,Hviding2025}, primarily originating from the RUBIES \cite{deGraaff_2025_RUBIES}, UNCOVER \cite{Bezanson2024_UNCOVER}, and JADES \cite{Eisenstein2023_JADES} programs. We also include several well studied LRDs \cite{Furtak2023,Wang2024b,deGraaff2025_ruby,Naidu2025}. Their NIRSpec spectra are from the DAWN JWST Archive (DJA\footnote{\url{https://dawn-cph.github.io/dja/}.}). We use the version 4 publicly released NIRSpec datasets. These data are reduced using the latest version of \texttt{msaexp} and the \textit{JWST Calibration Pipeline} \cite{2024Sci...384..890H,deGraaff_2025_RUBIES}. The spectroscopic redshifts are obtained through the template fitting algorithms employed in the \texttt{msaexp} pipeline.

To construct a representative empirical template, we first shifted all LRD spectra to the rest frame and interpolated them onto a common logarithmic wavelength grid between 0.7 and 2.5 $\mu \rm m$. Each spectrum was normalized at 5500\,\AA. To suppress noisy data in the blue end, we identified regions with ${\rm S/N}<3$ shortward of the Lyman limit (912\,\AA) and masked those pixels. We first obtain a full-sample composite spectrum of all objects by taking the sigma-clipped ($3\sigma$) median of all fluxes at each wavelength, followed by a mild boxcar smoothing below 1216\,\AA\ to mitigate residual noise. The resulting composite spectrum cover a wide redshift range thanks to the wide redshift range of the sample.

To capture the observed diversity in LRD continua, we further divided the sample in the $(\beta_{\rm UV}, \beta_{\rm opt})$ plane as shown in Figure \ref{fig:lrd_temp}(A), where $\beta_{\rm UV}$ and $\beta_{\rm opt}$ were measured from line-free photometry\cite{Zhang2025b}. We adopted bin edges of $\beta_{\rm UV} = [-2.0, -1.0]$ and $\beta_{\rm opt} = [-0.65, 0, 0.5, 1.4]$, yielding twelve subsamples with distinct UV-optical slopes. Each subsample typically contained 2-7 LRDs. For each bin, we repeated the same stacking procedure above. Because many individual spectra have limited wavelength coverage or low ${\rm S/N}$ in the rest-frame far-UV, we further patched each subsample’s median stack using the full-sample composite. Specifically, pixels below 2000\,\AA\ with ${\rm S/N}<1$ or missing data were replaced by the full-sample template, scaled to match the median flux around 2500\,\AA. Typically this replacement occurs near/blueward the Lyman-break. This approach preserves each group’s overall continuum slope while ensuring smooth, noise-free behavior around the Lyman-break region. 

These twelve empirical templates, together with the global median composite are shown in Figure \ref{fig:lrd_temp}(B). They provide a flexible basis set for \texttt{EAZY} \zph\ estimation, allowing the continuum slope and Balmer break strength to vary smoothly. The resulting library effectively captures the observed diversity of LRDs across UV-optical colors while maintaining realistic spectral shapes at all wavelengths. The library is available at \url{https://github.com/Zijian-astro/LRD-template-Zhang25}.

\subsubsection*{Photometric Redshift} 

We first use \textsc{eazy} \cite{brammer08} to measure photometric redshifts of all sources in the \rxc\ field, following a similar method to that adopted by the JADES team \cite{rieke23b,hainline24}. We include ten NIRCam bands from the \venus\ survey. We use photometry with aperture of $r=$\,0\farcs15 and impose a minimum uncertainty of 5\% on the flux for \textsc{eazy} SED fitting. The spectral template set from \cite{hainline24} is adopted. We adopt the \textsc{eazy} error template \texttt{TEMPLATE\_ERROR.v2.0.zfourge} to account for wavelength-dependent uncertainties in the templates. We explore a redshift range of $z = 0.01$-30 with a step size of $\Delta z = 0.01$, adopting the redshift corresponding to the global minimum of $\chi^2(z)$ as our best-fit \zph. The output probability distribution $P(z) \propto \exp[-\chi^2(z)/2]$ is used to estimate the \zph\ uncertainties. Although the redshift grid formally extends to $z=30$, the multiply imaged sources considered here are mostly to lie at z $\lesssim$ 7 (Table~\ref{tab:multi}), where Ly$\alpha$-damping and DLA absorption have minimal impact on the NIRCam SEDs. Therefore, the adopted template set from \cite{hainline24} is sufficient for our analysis.

For both \tgta\ and \tgtb, the initial \zph\ estimates of the multiple images show two degenerate redshift peaks around $z \sim 4.3$ and $z \sim 6.3$. To further improve the constraints, we include the HST F435W, F606W, and F814W bands. The HST measurements of the individual images are all upper limits, which are estimated from random-aperture experiments. To enhance the overall constraint, we perform \zph\ fitting on the stacked photometry of the multiple images, since stacking improves the S/N in the relatively faint bands. For \tgta, we stack all four images, whereas for \tgtb\ we stack only the three brightest ones (2.3, 2.4, and 2.5). We obtain photometry with radius of 0\farcs15 and aperture correction. Another source of \zph\ uncertainty arises from the mismatch between the observed SEDs of LRDs and the galaxy templates used in the fitting.
We thus apply the LRD templates constructed above to the stacked photometry using \textsc{eazy}. 
To assess \zph\ uncertainty, we also fit the stacked SED of \tgta\ and \tgtb\ using two additional SED fitting codes, \textsc{Prospector} \cite{2023ApJ...944L..58W} and \textsc{CIGALE} \cite{2005MNRAS.360.1413B,2019A&A...622A.103B}. The \textsc{Prospector} setup follows those in previous work \cite{2023ApJ...944L..58W}, and the \textsc{CIGALE} setup are detailed below.

The SEDs, best-fit \zph\ results, and image cutouts of \tgta\ and \tgtb\ are shown in Figure \ref{fig:photz}. The stacked images of both \tgta\ and \tgtb\ show a $\sim 3\sigma$ detection in the F814W band, while F435W and F606W remain undetected. The \textsc{eazy} fitting using the empirical LRD templates obtain $z_{\mathrm{phot}} = 4.53_{-0.16}^{+0.09}$ for \tgta\ and $z_{\mathrm{phot}} = 4.10_{-0.19}^{+0.59}$ for \tgtb. These values are slightly lower than those derived using galaxy templates with \textsc{eazy}, \textsc{CIGALE}, and \textsc{Prospector}. The primary constraints arise from the combination of significant flux excesses in the F277W and F356W bands (relative to F210M, F300M, and F410M) and the non-detections in the F435W and F606W bands, together indicating a redshift in the range $z \approx 3.9$-$4.7$.
It is insufficient to provide tighter constraints within this redshift range based on the photometric data alone.
Notably, both LRDs are located in a region associated with a potential galaxy overdensity (Figure \ref{fig:sourceplane_pos}). Considering the \zph\ of nearby member galaxies, we estimate the most probable redshift of this structure to be $z \sim 4.3$, and therefore adopt $z = 4.3$ as the fiducial redshift for \tgta\ and \tgtb. We note that adopting any redshift within $z = 3.9$-$4.7$ would not affect the conclusions of this work.

\subsubsection*{CIGALE SED modeling setup}
For \texttt{CIGALE} SED modeling, we assume a delayed-$\tau$ star formation history (SFH), where ${\rm SFR} \propto te^{-t/\tau}$ and $\tau \in [0.01, 2]\rm \,Gyr$. We also allow an optional late starburst in the latest 20$\, \rm Myr$. We use the \cite{2003MNRAS.344.1000B} stellar population synthesis models and adopt the \cite{2003PASP..115..763C} initial mass function (IMF). Nebular continuum and emission lines are included, allowing a metallicity range of [0.2, 1]$\,Z_\odot$ and an ionization parameter $\log U\in[-4, -1]$. We adopt the SMC extinction curve \cite{1992ApJ...395..130P} for nebular emission and the modified \cite{2000ApJ...533..682C} attenuation curve for the stellar continuum, with the power-law slope $n \in [-0.6, 0.2]$. We add a relative error of 5\% in quadrature to the uncertainties of the fluxes.

\subsubsection*{V-shape Continua and Compactness}

Using the estimated photometric redshift, we derive the rest-frame UV and optical slopes of \tgta\ and \tgtb, excluding the F277W and F356W bands where strong emission-line contamination from \oiii+\hb\ and \ha\ is expected. Figure~\ref{fig:Vshape_compact}(A) shows their locations in the $\beta_{\rm UV}$-$\beta_{\rm opt}$ plane. The scatter in positions of the different images of the same source reflects both the large uncertainties in the rest-frame UV measurements and potential variability. According to the commonly adopted V-shape criterion of $\beta_{\rm opt}>0$ and $\beta_{\rm UV}<-0.37$\cite{Kocevski2025}, \tgta\ clearly qualifies as an LRD. Although \tgtb\ has a slightly smaller $\beta_{\rm opt}$ and thus falls slightly outside the canonical boundary, it remains on the red side of the distribution. We overlay in Figure~\ref{fig:Vshape_compact}(A) the empirical relation between the UV/optical slopes and the fraction of broad H$\alpha$ emitters from \cite{Zhang2025b}. The position of \tgtb\ strongly suggests that it has broad lines.

To demonstrate their compact morphologies, we show the relation between $m_{\rm F444W}$ and $R_{\rm F444W}$ in Figure~\ref{fig:Vshape_compact}(B). We derive a relation between $m_{\rm F444W}$ and $R_{\rm F444W}$ for point sources as a green line by fitting the stellar locus. Both \tgta\ and \tgtb\ have radii smaller than 1.5 times this relation, confirming their point-like morphologies in the F444W band. Taken together, their V-shaped continua and compact morphologies qualify both objects as LRDs, awaiting further spectroscopic confirmation.

\subsubsection*{Blackbody SED Fitting} 

To investigate the physical origin of the observed coupling between color and luminosity variability of \tgta, we model the SEDs of \tgta\ at each epoch using a single-temperature blackbody component following recent studies \cite{Begelman2025,Kido2025,Lin2025_locallrd,Torralba2025}.  
To account for the potential host galaxy dominated in the rest-frame UV, we add a power-law component.
Its normalization and slope are shared across all epochs but remain free parameters in the fit.
The F277W and F356W bands are excluded from the fit because they are strongly contaminated by emission lines. The blackbody temperature is degenerate with the dust extinction level, as illustrated in Figure 5 of \cite{Kido2025}. However, only mild dust extinction is expected for LRDs \cite{Chen2025c,Lin2025_locallrd,Setton2025b}. 
Therefore, we do not consider dust attenuation in our fitting.
This simple model provides a reasonable fit to the SEDs of \tgta\ (Figure \ref{fig:lc_BB_fit}A), with $T_{\rm eff} \sim 4000\,\rm K$, $R_\mathrm{ph} \sim 2000$\,AU, and $L_{\rm BB} \sim 2\times10^{\rm 44} \rm\,erg\,s^{-1}$. We also fit the SED of \tgtb\ using the eame model, obtaining $T_{\rm eff} \sim 5000\,\rm K$, $R_\mathrm{ph} \sim 600$\,AU, and $L_{\rm BB} \sim 2\times10^{\rm 43} \rm\,erg\,s^{-1}$.
The final $L_{\rm BB}$ used as the bolometric luminosity of \tgta\ and \tgtb\ are the mean $L_{\rm BB}$ of all epoch, with the error obtained by MC sampling.
We also try to include dust attenuation using the Calzetti dust extinction law\cite{2000ApJ...533..682C} with different $A_{\rm V}$ for the blackbody. 
Altering the $A_{\rm V}$ value in the range of 0-2 would change the fitted blackbody temperature in the range of $3500$-$6000$ K, but does not change the main conclusions of this work.

\subsubsection*{Light Curve Fitting} 

Previous studies suggest that supermassive stars are unstable only under radial perturbations \cite{Baraffe2001,Sonoi2012}. 
By analogy, we consider only the f-mode for LRDs, which is expected to dominate their pulsational instability. 
In this case, the characteristic pulsation period of the LRD envelope is expected to be $T_{0} = f_0t_{\rm dyn}$, where we adopt $f_0 = 0.37$ assuming an $n=3$ polytropic equation of state (fully convective, radiation-pressure dominated envelopes)\cite{1941ApJ....94..245S}. 
Using the inferred properties of \tgta (i.e., $T_{\rm eff} = 4000\rm\,K$ and $M_{\rm BH} = 10^{6.28}\,M_\odot$), we calculate the pulsation period as
\begin{equation}
T_{0} \approx 32~{\rm yr}\,\lambda_{\rm Edd}^{3/4}\left(\frac{f_0}{0.37}\right)\left(\frac{M_{\rm BH}}{10^{6.28}\,M_\odot}\right)^{1/4}\left(\frac{T_{\rm eff}}{4000\,\rm K}\right)^{-3}.
\end{equation}
The variability in each band is then modeled as a single shared-phase sinusoid
\begin{equation}
    f(t) =  \bar{f} + A_0 \sin\left(\frac{t}{T_0} + \phi_0\right),
\end{equation}
where $ \bar{f}$ is the mean flux, $A_0$ is the variability amplitude, and $\phi_0$ is the common oscillation phase across all fitted bands. 
We fit this model to the \tgta\ light curves for bands longer than F210M.
For the F322W2 band from SLICE, we additionally include the synthetic mean fluxes of the \venus\ F277W and F356W bands, since F322W2 is equivalent to the combination of these two bands. 
We fit six bands simultaneously, each band containing four independent epochs, so the total $N_{\rm data}=6\times4=24$ measurements. The model has two free parameters (mean flux $\bar f_i$ and amplitude $A_i$) for each band (12 parameters total) plus a single global phase $\phi$ tied across all bands. 
Hence the total number of fitted parameters is $N_{\rm par}=2\times6+1=13$, giving
${\rm dof} = N_{\rm data}-N_{\rm par} = 24-13 = 11$. All bands are well fitted by this model, with a total $\chi^2_{\nu} = 0.09$.

To map the regions of parameter space that yield reasonable fit to the observed data, we perform a grid search over the black hole mass and effective temperature. We evaluate a two-dimensional grid spanning
$\log (M_{\rm BH}/M_\odot)\in[4,9]$ and $T_{\rm eff}\in[3000,6000]~{\rm K}$, with 500 uniformly spaced samples in each dimension.
For each grid point, the corresponding pulsation period $T_0$ is computed, and this fixed period is adopted in the multi-band sinusoidal model.
At every ($M_{\rm BH}$, $T_{\rm eff}$), we refit the full set of six bands simultaneously using the same procedure described above. The final $\chi^2_\nu$ map is shown in Figure~\ref{fig:param_lcfit}.
Only a narrow strip in the $M_{\rm BH}$-$T_{\rm eff}$ plane yields comparably good fits, forming the valley of low $\chi^2_\nu$ values that identifies the reasonable pulsation periods.
The independently inferred parameters of \tgta\ lie precisely within this region, demonstrating that the fit is physically self-consistent rather than an artifact of flexible modeling.

We note that with the current sparse sampling, smaller-timescale variability could still be present (there are some very narrow low $\chi^2_\nu$ valley in the upper-right corner of Figure~\ref{fig:param_lcfit}). 
Nevertheless, prior observations shows that LRDs do not exhibit strong variability on rest-frame $\sim$1-year timescales\cite{Zhang2025a}, and the VENUS-SLICE observed one-year baseline shows no significant flux changes (Figure \ref{fig:venus_slice}). 
Furthermore, shorter pulsation periods of only a few years would imply implausibly low LRD black hole masses of $< 10^{4}\,M_\odot$, reinforcing the plausibility of the multi-year characteristic timescale implied by our best-fit model.

Using the best-fit light curve models, we sample 40 epochs at one-year intervals and fit each epoch with the same blackbody+power-law model as in Figure~\ref{fig:lc_BB_fit}(A). This yields the temporal sequence of luminosity and temperature variations, which is shown in Figure~\ref{fig:lc_BB_fit}(B).

\subsubsection*{Cepheid Temperature and Luminosity Evolution} 
To illustrate the similarity between the observed variability of \tgta\ and the $\kappa$-mechanism-driven pulsations of Cepheid-like stars, we compare their behaviors in the luminosity--temperature ($T_{\rm eff}$-$L_{\rm BB}$) plane. We retrieve observed light curve of Cepheids from the OGLE Collection of Variable Stars\cite{OGLE2015,OGLE2017}, which is the largest set of variable stars and provides well-sampled $V$- and $I$-band light curves of $\sim5000$ classical Cepheids. For each source, we fold the light curves with the published period and construct smooth phase-dependent $V$ and $I$ magnitudes using a spline fit because the $V$ and $I$ data points are usually not observed simutaneously. The color curve is then converted to $T_{\rm eff}$ through interpolation of the bolometric correction obtained from YBC, a stellar bolometric corrections database\footnote{\url{https://sec.center/YBC/}.}\cite{YBC2019}. The $L_{\rm BB}$ is also derived from the corresponding bolometric corrections from YBC and normalized to its cycle-averaged value.

This procedure yields a closed trajectory in the $T_{\rm eff}$-$L_{\rm BB}$ plane for every Cepheid. As shown in Figure~\ref{fig:Cep_4313}, the track of a representative Cepheid (OGLE-LMC-CEP-4313) exhibits a characteristic counterclockwise loop: the temperature rises rapidly near maximum compression, reaches its peak slightly before luminosity maximum, and then decreases gradually during the expansion phase. 
The phase offset between luminosity and temperature extrema, which results in the counterclockwise trajectory in the $T_{\rm eff}$-$L_{\rm BB}$ plane, is a hallmark of $\kappa$-mechanism pulsation and is also seen in the variability pattern of \tgta\ (Figure \ref{fig:lc_BB_fit}b).







\subsection*{Supplementary Text}

\subsubsection*{Cluster Mass Model and Magnification Uncertainty}

The main source of uncertainty for this lensing variability study arises from systematic errors in the lensing magnification. 
We note that our cluster mass model is constructed using a large number of multiple images, three of which have spectroscopic redshifts.
All images of \tgta\ are considerably far away from the critical curve ($\gtrsim15^{\prime\prime}$), and therefore the magnification uncertainty from the cluster mass model is estimated to be small ($\sigma_\mu / \mu \lesssim 0.05$; see Strong Lens Mass Modeling section).
Given the large distance and compact size of LRDs, we also estimate a negligible uncertainty in color variation attributed to differential lensing magnification ($\sigma_\mu / \mu \lesssim 0.003$), unlike the color variation observed in galaxy-lensed quasars \cite{wambsganss91}.

To further assess potential systematic cluster-lensing uncertainty propagated to the variability signal, we compare \tgta\ with 12 other multiply imaged sources that have $z_{\rm phot} >3$ in the same field. 
Among them, six have similar redshift of \tgta\ and \tgtb, as shown in Figure \ref{fig:sourceplane_pos}.
Several of these sources lie very close to \tgta\ in the source plane and should have similar light path (e.g., ID16 and ID37). 
For each source, we compute the S/N of the de-magnified magnitude difference (i.e., $|\Delta \rm mag/\sigma_{\Delta \rm mag}|$) between all epoch pairs in the F200W and longer-wavelength bands.
$\sigma_{\Delta \rm mag}$ is the quadrature sum of the magnitude uncertainties of the two epochs, including both the measurement errors and the uncertainties in the lensing magnification. 
The median of these band-wise S/N values is adopted as the variability S/N for the epoch pair, which is designed to quantify the overall SED variability.
Similarly, we calculate the color (F300M$-$F444W) variability S/N for each epoch pair. This provides a measure of changes in the spectral shape independent of the overall brightness.
As shown in Figure \ref{fig:RX1_vari}(C), the distribution of the magnitude and color variability S/N for the other sources is predominantly below unity, whereas \tgta\ shows a clear excess and reaches $>3$. The lack of significant variability in the comparison sample confirms that the observed flux variability in \tgta\ should be intrinsic, rather than caused by the systematic uncertainties of cluster magnification model. The non-detection of variability in the SW bands of \tgta\ and LW bands of \tgtb\ further supports the reliability of their magnification corrections.

In addition, we assess whether microlensing could contribute to the observed variability (e.g., \cite{1981ApJ...243..140G}). We estimate the expected microlensing optical depth from the stellar surface mass density of the cluster.
The stellar convergence $\kappa_\ast$ is derived from the observed surface brightness profile of the cluster at the F210M band (corresponds to rest-frame H-band), scaled by a mass-to-light ratio of $M/L_{\rm H} = 0.57 \pm 0.26$ \cite{2020A&A...633A.107H}, and compared to the total convergence $\kappa_{\rm tot}$ from the cluster mass model. The optical depth is expressed as\cite{Schneider1992g}
\begin{equation}
    \tau = \frac{\kappa_\ast}{1 - \kappa_c}~(\tau \lesssim 0.1, \kappa_c<1),
\label{eq:tau}
\end{equation}
where $\kappa_c = \kappa_{\rm tot} - \kappa_\ast$ represents the convergence of the smoothly distributed matter. This value quantifies the probability that light rays from the source are significantly perturbed by compact stellar lenses. For the positions of \tgta.1 and \tgta.2, we find $\kappa_{\rm tot}\sim 0.5$, $\kappa_\ast<5\times10^{-4}$, and $\tau<1\times10^{-3}$, implying that the possibility of a microlensing event is negligible. Notably, \tgta.1 and \tgta.2 correspond to the brightest and dimmest images, respectively. For \tgta.3 and \tgta.4, where $\kappa_c \sim 1.3$, the optical-depth expression is not strictly valid, but given their similarly small $\kappa_\ast$, the probability of microlensing remains extremely low. Moreover, even if rare microlensing events occur, it would be in the finite-source regime. For a typical stellar lens, an Einstein radius of $\Theta_\mathrm{E} \sim$1~$\mu$as corresponds to $\sim$1000 AU at $z\simeq4.3$, while the source size inferred from our blackbody fit is $\sim$2000 AU. Taking a typical transverse velocity of $v\sim1000$~km\,s$^{-1}$ from the cluster velocity dispersion, the corresponding crossing time of $\sim4.7~\rm yr$ would make microlensing-induced variability evolve measurably over the JWST mission lifetime. In fact, the F150W2 and F322W2 photometry obtained in 2024 agrees well with the 2025 measurements at similar wavelengths for all four images, further suggesting that microlensing is unlikely to be responsible for the observed variability.

We also assess whether ``millilenses'', such as globular clusters or dark matter subhalos with masses $M \simeq 10^6 - 10^9$\,\msun\ (corresponding to Einstein radii $\Theta_\mathrm{E}\simeq 0.001 - 0.1$\arcsec), could contribute to the observed brightnesses variation with a characteristic timescale of a few hundred years in the observed frame\cite{dai18,diego24}.
Such perturbers would introduce differential magnification across the source, potentially producing color differences among the multiple images of lensed LRDs.
In particular, if the LRD is more compact in the rest-frame optical than in the UV, millilensing would induce larger brightness changes at optical wavelengths.
At the depth of \venus\ imaging, we are able to rule out the presence of dwarf galaxy or massive globular cluster with luminosities $L\gtrsim2\times10^7$\,\lsun\ along the sightlines of LRD images. 
Below this detection limit, \cite{diego24} shows that globular clusters in the galaxy cluster only contribute to $\sim10^{-2} - 10^{-1}$ of the microlensing cross sections above a given magnification, and thus can be ruled out.
On the other hand, the mass fraction of dark matter subhalo at $M\lesssim10^9$\,\msun\ is typically constrained as $f_\mathrm{sub}\lesssim10^{-2}$ in low-redshift strong gravitational lenses\cite{Vegetti14,Hsueh20,Perera25}. 
The optical depth of subhalo millilens follows similar formalism as that of microlenses (Equation~\ref{eq:tau}) and thus $\tau \sim f_\mathrm{sub} \lesssim10^{-2}$.
Therefore, we conclude that millilenses are unlikely the main cause of the observed brightness and color variation seen in the multiple images of \tgta.

Other uncertainties of the variability measurement arise from variations in depth, spatial PSF, statistical noise, and contamination from nearby sources, etc. These effects lead to scatter among non-variable sources \cite{Zhang2025a}. We carefully inspected the image cutouts for every band and for every image: in all cases the sources are isolated, there is no evidence of flux contamination from nearby objects, and the local background and PSF structure do not show anomalies that could bias the aperture photometry. Crucially, the variability follows the same trend across all bands. Random measurement fluctuations or intermittent contamination (e.g., cosmic rays or hot pixels) in individual images would produce uncorrelated or band-dependent behavior, yet we observe coherent variability even between datasets taken at different epochs and under different configurations from SLICE and \venus. 
We therefore conclude that the variability signals are genuine.





\begin{figure}[!t]
\centering
\includegraphics[width=\linewidth]{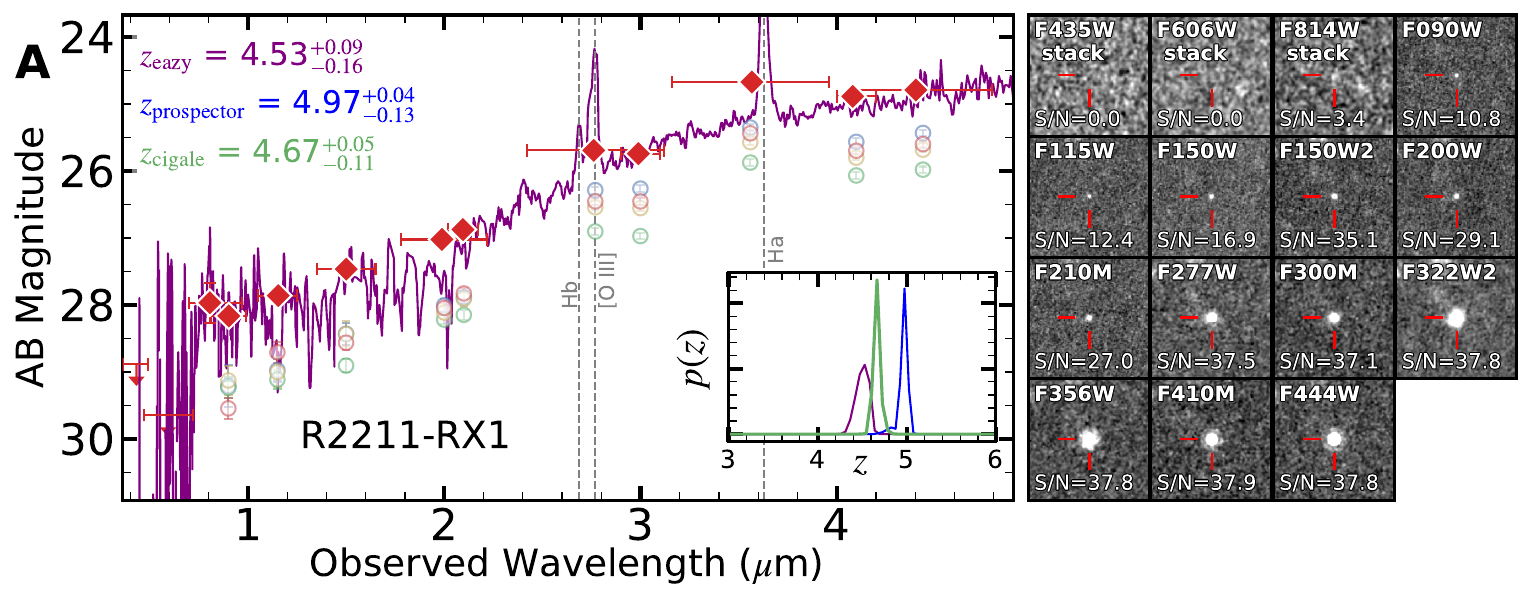}
\includegraphics[width=\linewidth]{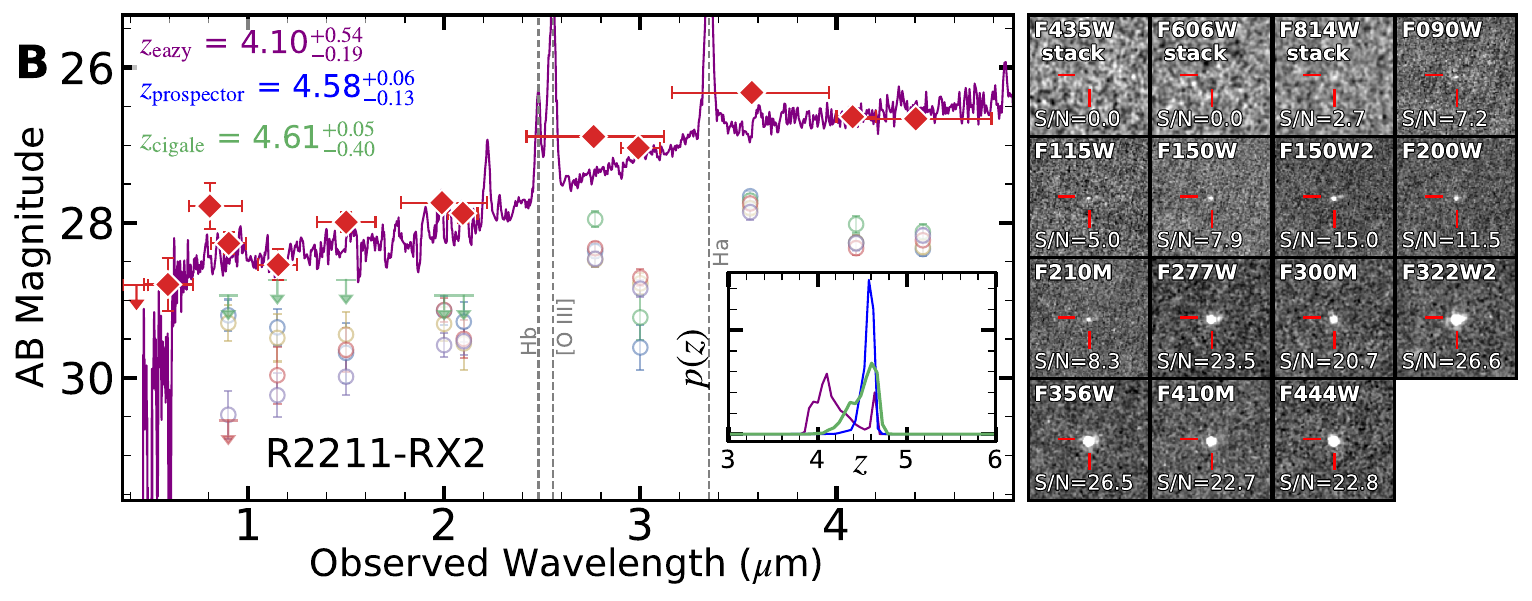}
\caption{\textbf{The SED and images of \tgta\ (A) and \tgtb\ (B)}. The left panels show the stacked observed photometry and $2\sigma$ upper limits from \venus\ JWST bands and RELICS HST F435W/F606W/F814W data (dark red diamonds). The best-match LRD template is overplotted in purple, and the best-fit redshifts ($z_{\rm map}$) with their 5th-95th percentile ranges from EAZY, CIGALE, and Prospector are indicated. 
The photometry from individual images (corrected for magnification $\mu$) is shown as lighter/semi-transparent markers in the background. The inset panels display the full redshift probability distributions $p(z)$ from the same three codes. 
The right panels show 3\arcsec\,$\times$\,3\arcsec\ cutouts in the available JWST bands of \tgta.4 and \tgtb.5, as well as the stacked HST F435W/F606W/F814W images. We show these two images because they have high magnifications and no nearby contamination.
For \tgta, the HST stack combines all multiple lensed images of the same source, while for \tgtb, the stack includes only images \tgtb.3, \tgtb.4, and \tgtb.5. The HST stacked images are smoothed with a Gaussian kernel ($\sigma=1$ pixel).
}
\label{fig:photz}
\end{figure}

\begin{figure}[!t]
\hspace{-0.5cm}
 \includegraphics[width=\textwidth]{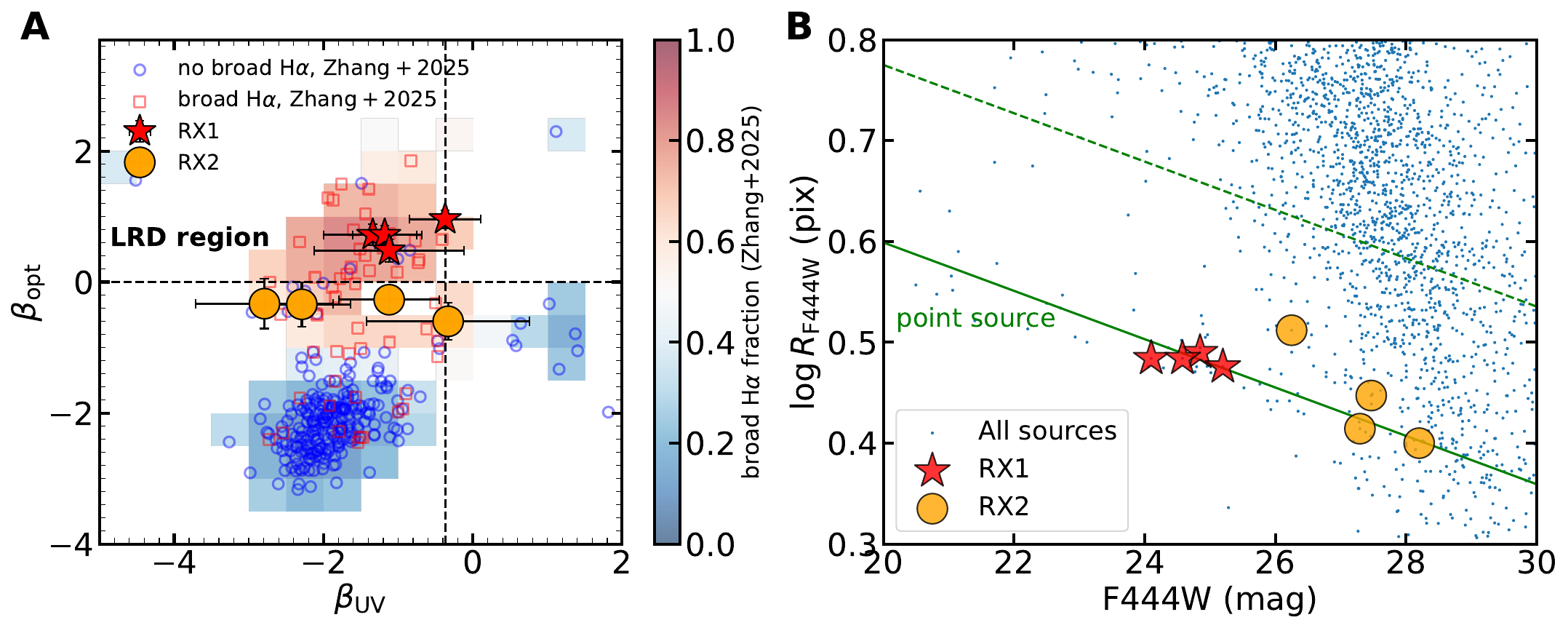}
 \centering
 \caption{\textbf{The colors, sizes and magnitudes of \tgta\ and \tgtb\ as LRDs.} \textbf{(A)} Distribution of the best-fit UV and optical slope using the line-free photometry for \tgta\ and \tgtb. Dashed lines represent the LRD selection criteria (upper left region)\cite{Kocevski2025}. The color of the background squares represents the fraction of broad-line sources adopted from \cite{Zhang2025b}. The red squares and blue circles represent sources with and without broad \ha\ lines in \cite{Zhang2025b}. \textbf{(B)} F444W magnitude versus half-light radius diagram. Green solid and dashed lines denote the best-fit relation to the stellar locus and 1.5 times this relation. Blue dots in the background show all sources in \rxc.}
 \label{fig:Vshape_compact}
\end{figure}

\begin{figure}[!t]
\centering
\includegraphics[width=\linewidth]{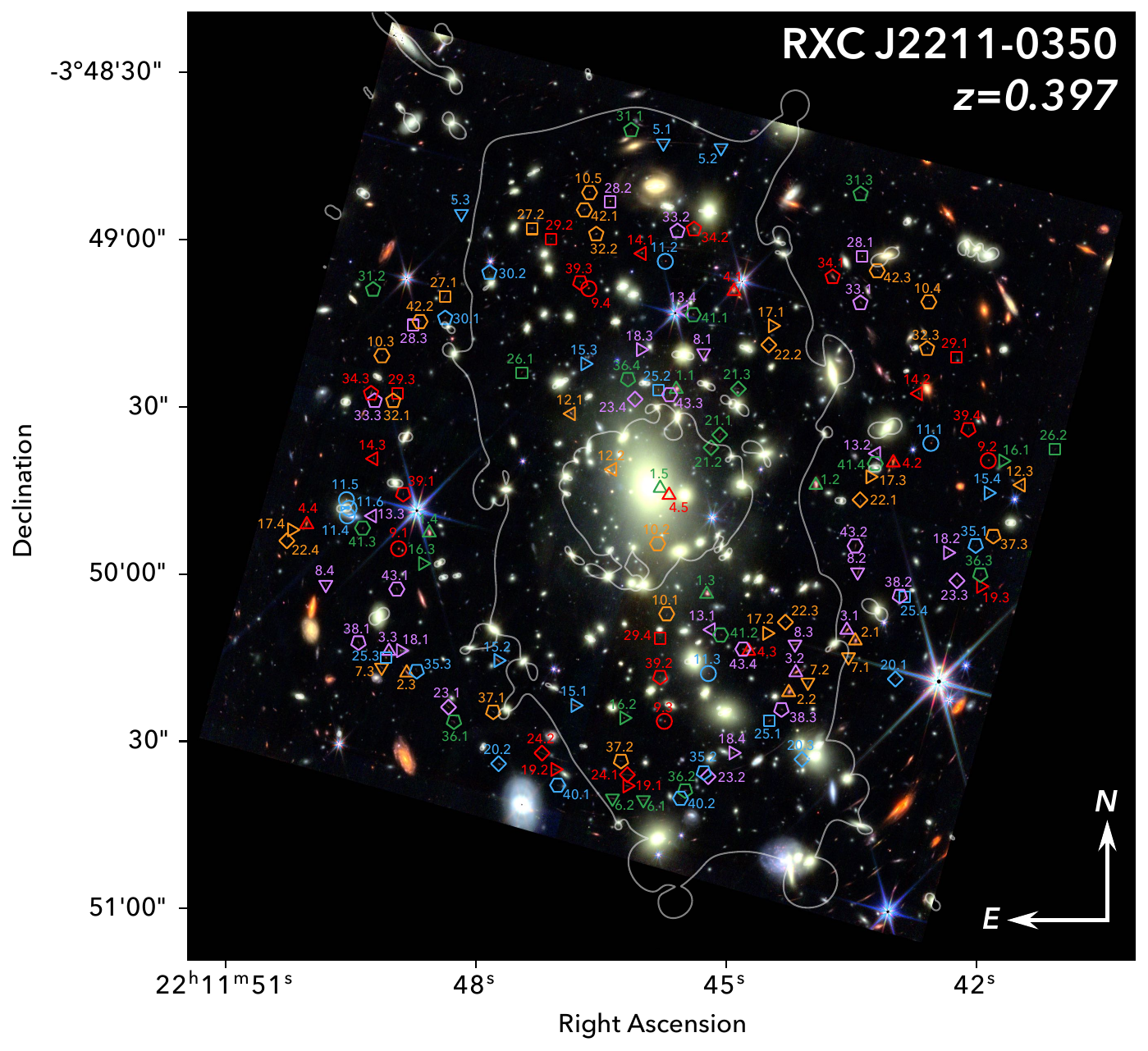}
\caption{\textbf{Multiple images used for the construction of the cluster mass models.} Positions of 146 multiple images from 43 sources in \rxc\ are shown by symbols. In this Figure, ID 9 and 10 correspond to \tgta and \tgtb, respectively. White solid lines indicate critical curves of the best-fitting model for the source redshift of $4.3$.}
\label{fig:lensing_model}
\end{figure}

\begin{figure}[!t]
\centering
\includegraphics[width=0.48 \textwidth]{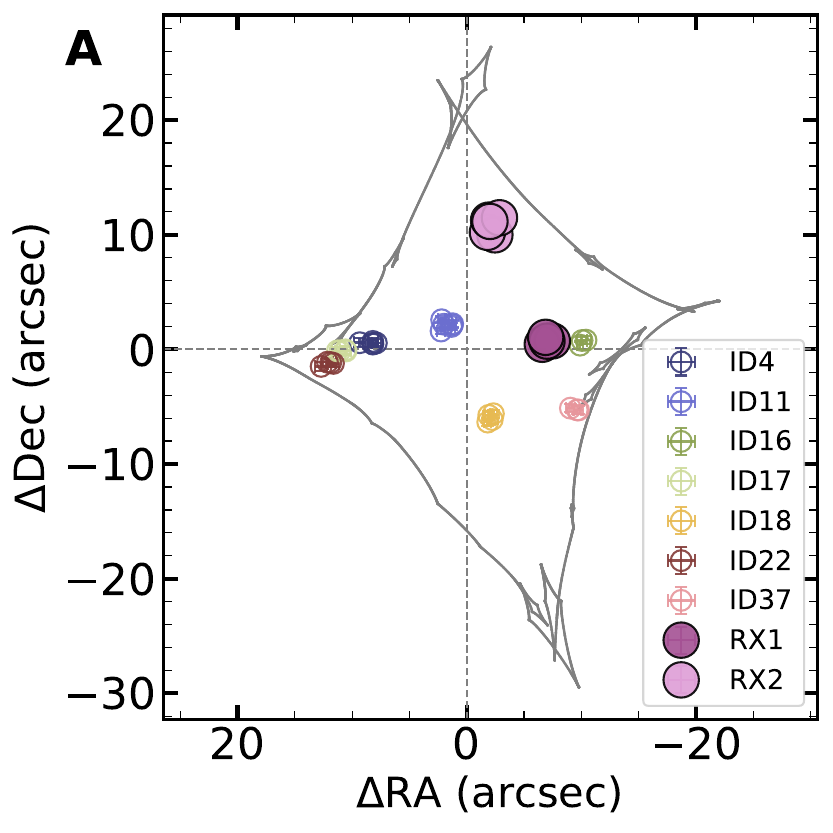}
 \centering
\includegraphics[width=0.5\textwidth]{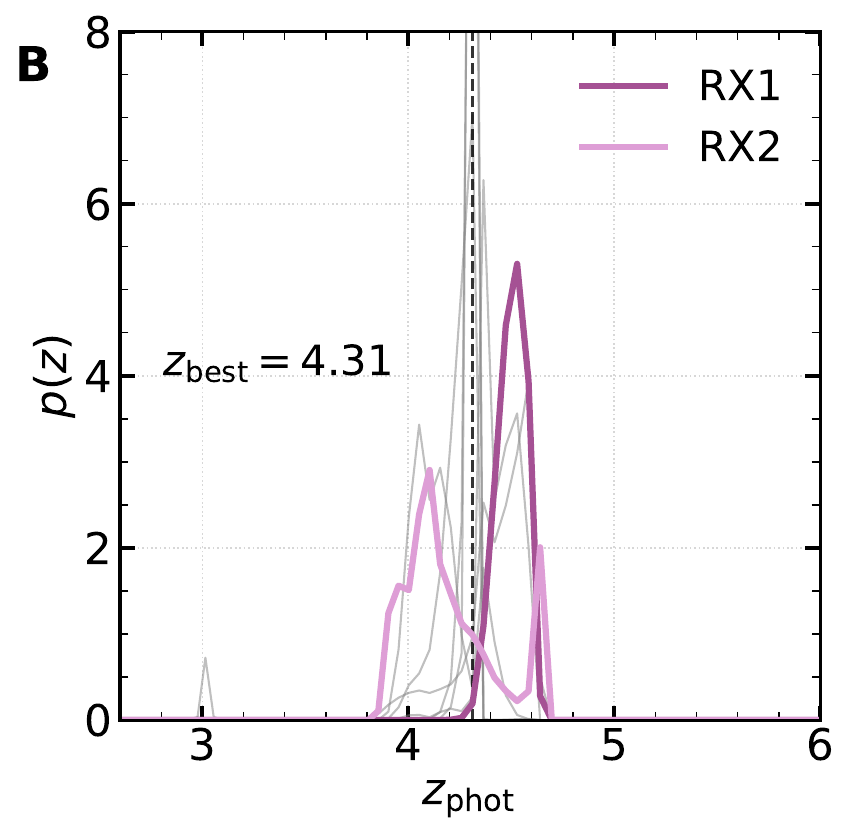}
\caption{\textbf{A potential overdensity of $z\sim4.3$ galaxies including two LRDs.} \textbf{(A)} Source-plane positions of  multiply imaged sources with \zph$\sim4.3$ in \rxc. The gray line represents the caustic curve for $z_{\rm s} = 4.3$. \textbf{(B)} Photometric redshift distribution of sources in (A). The distribution of \tgta\ and \tgtb\ are shown as purples, while other sources are shown as gray curves. The vertical dashed line $z_{\rm best}=4.31$ is the best \zph\ jointly constrained by the \zph\ distribution of other sources.}
\label{fig:sourceplane_pos}
\end{figure}

\begin{figure}[!t]
\centering
\includegraphics[width=\linewidth]{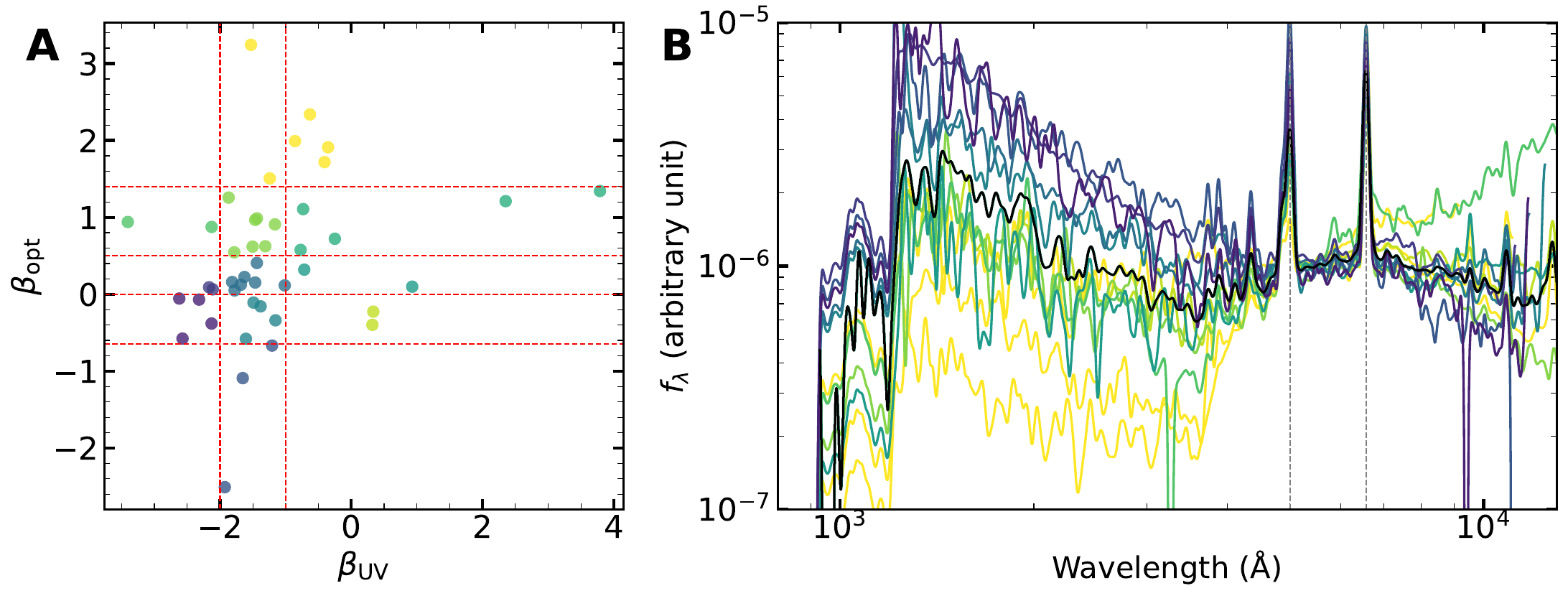}
\caption{\textbf{The construction of LRD spectral templates.} \textbf{(A)} Distribution of LRDs used to construct \zph\ fitting templates in the $(\beta_{\rm UV}, \beta_{\rm opt})$ plane, color-coded to be consistent with the stacked spectra in (B). The dashed red lines denote the adopted binning boundaries ($\beta_{\rm UV} = -2.0, -1.0$; $\beta_{\rm opt} = -0.65, 0.0, 0.5, 1.4$), which divide the sample into twelve subsamples with distinct UV-optical slopes. \textbf{(B)} Empirical rest-frame templates derived for each $(\beta_{\rm UV}, \beta_{\rm opt})$ bin (colored curves), ordered by their relative flux at 2000 $\rm \AA$. The black curve shows the global median stack of all 44 LRDs. Vertical dashed lines mark the wavelengths of \oiii\ and \ha\ emission line. All spectra are normalized at 5500 $\rm \AA$ and smoothed with a Gaussian kernel with $\sigma = 5$.}
\label{fig:lrd_temp}
\end{figure}

\begin{figure}[!t]
\hspace{-0.5cm}
 \includegraphics[width=0.8\textwidth]{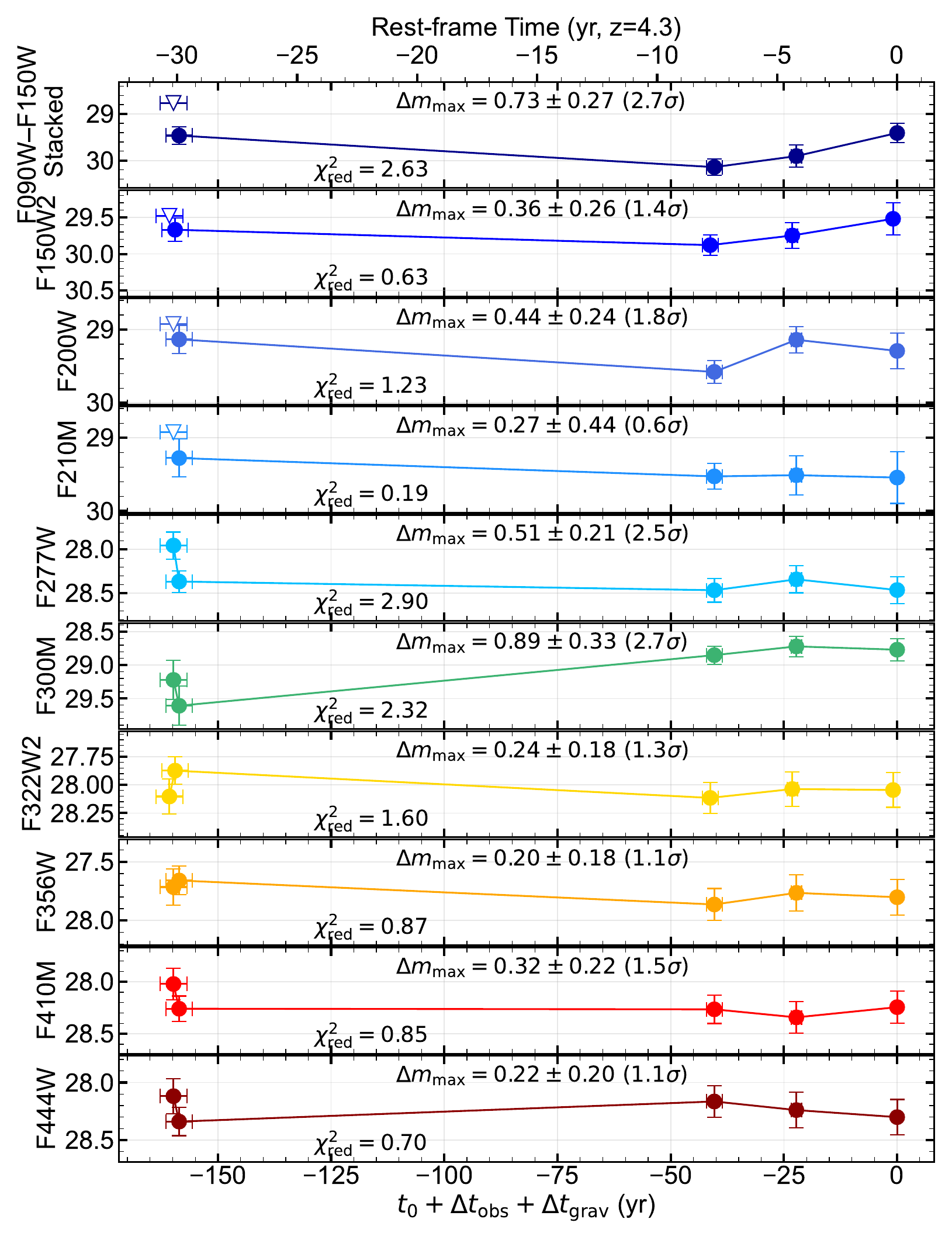}
 \centering
 \caption{\textbf{The same as Figure \ref{fig:RX1_vari}(A), but for \tgtb}. Filled circles indicate detections, and open downward triangles represent $2\sigma$ upper limits.}
 \label{fig:RX2_lc}
\end{figure}

\begin{figure}[!t]
\centering
\includegraphics[width=0.49\textwidth]{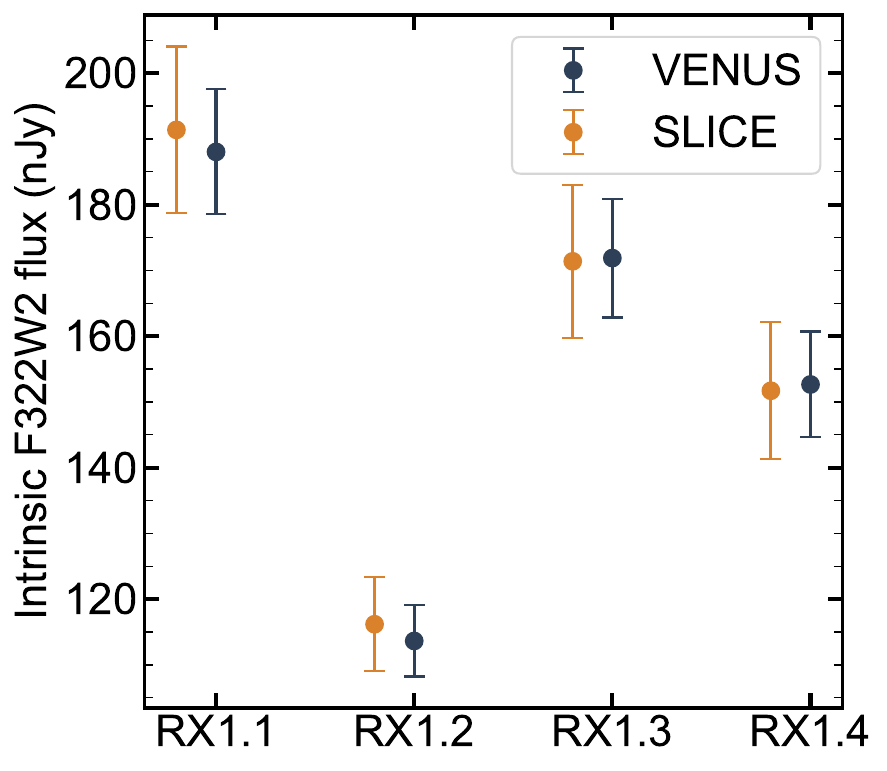}
 \centering
\includegraphics[width=0.48\textwidth]{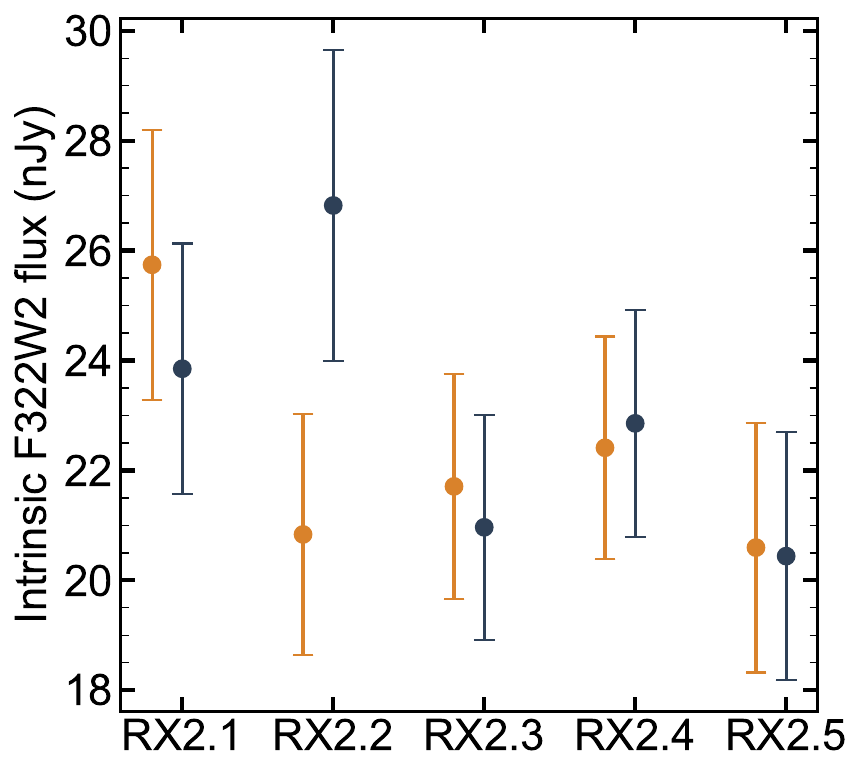}
\caption{\textbf{Comparison of intrinsic F322W2 fluxes from SLICE measurements and synthetic estimates derived from the mean of VENUS F277W and F356W measurements for \tgta\ and \tgtb.} The error bars shown include both photometric uncertainties and magnification uncertainties. The rest-frame observation interval between SLICE and VENUS is $\sim 0.2$\,yr. Overall, \tgta\ and \tgtb do not show signficant variability between the SLICE and VENUS observations (the difference for \tgtb.2 is $<2\sigma$).}
\label{fig:venus_slice}
\end{figure}

\begin{figure}[!t]
\hspace{-0.5cm}
 \includegraphics[width=0.8\textwidth]{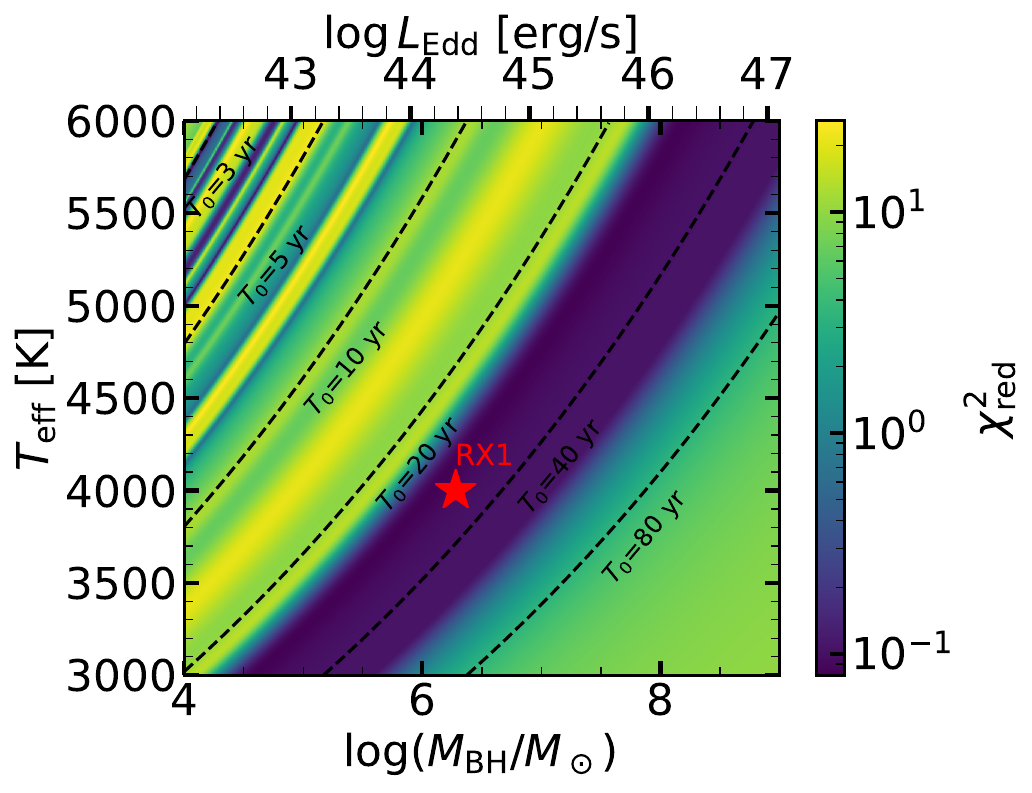}
 \centering
 \caption{\textbf{Total $\chi_{\nu}^2$ values of the light curves modeling over a grid of $M_{\rm BH}$ and $T_{\rm eff}$.} The color scale uses a logarithmic normalization to highlight the structure of the $\chi_{\nu}^2$ landscape. The red star marks the inferred properties of RX1 at $M_{\rm BH} = 10^{6.28}\,M_\odot$ and $T_{\rm eff} = 4000$ K. Black dashed lines show the loci of different characteristic pulsation periods $T_0 = 3$, 5, 10, 20, 40, and 80 yr. The top x-axis indicates the corresponding $\log L_{\rm Edd}$ for each $M_{\rm BH}$. }
 \label{fig:param_lcfit}
\end{figure}

\begin{figure}[!t]
\hspace{-0.5cm}
 \includegraphics[width=0.99\textwidth]{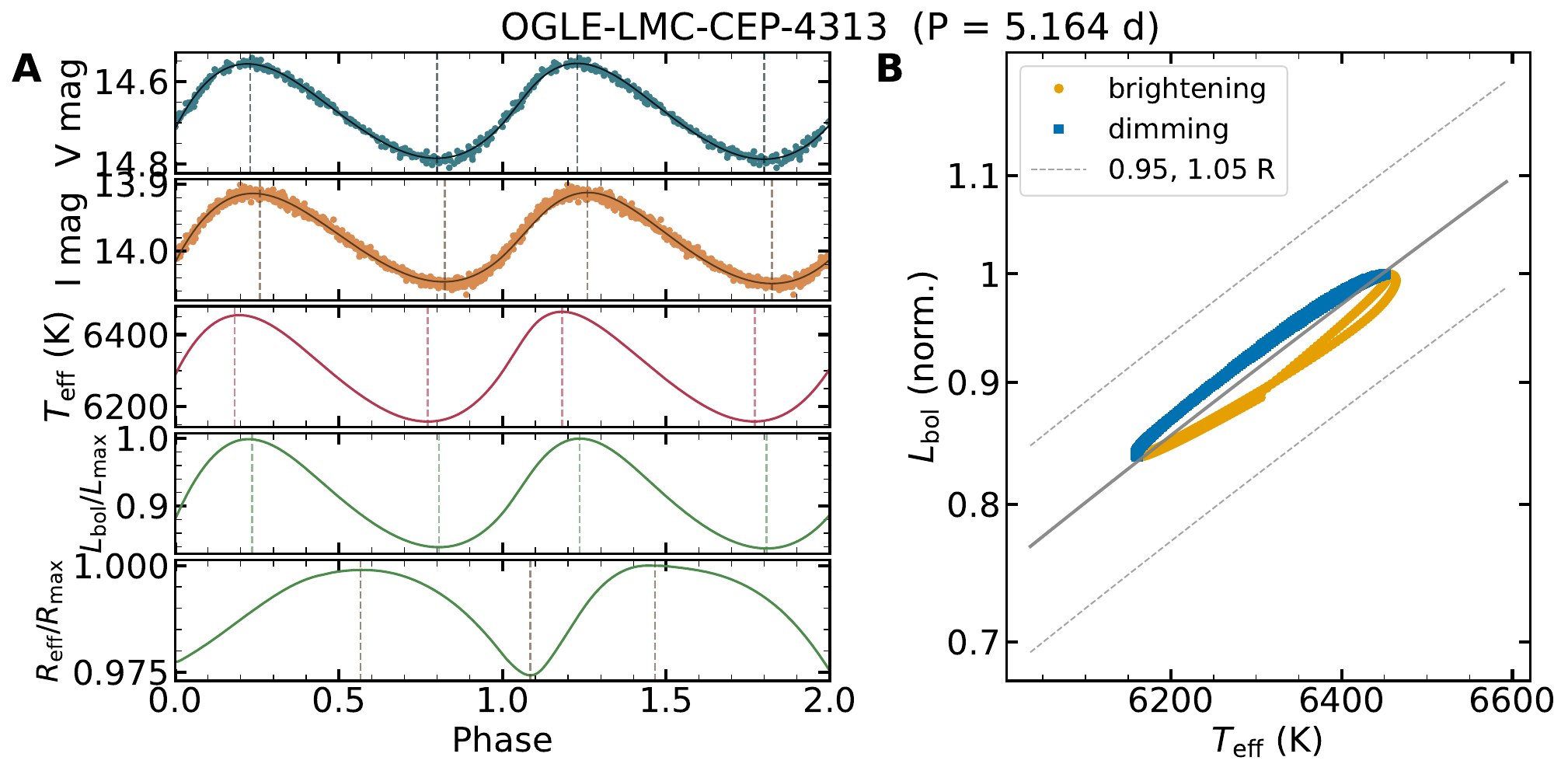}
 \centering
 \caption{\textbf{Multi-band light curves, temperature and luminosity evolution, and the corresponding HR loop for a Cepheid (OGLE-LMC-CEP-4313).} \textbf{(A)} Phased V-band (top) and I-band (second panel) light curves, together with the best-fit spline models. Dashed vertical lines mark the phases of maximum and minimum brightness in each band. The third, fourth, and fifth panels show the inferred $T_{\rm eff}$, $L_{\rm BB}$, and $R_{\rm eff}$ as functions of phase, derived from the spline-interpolated V--I color using the YBC color-temperature and bolometric correction relations; their extrema are similarly indicated. \textbf{(B)} The resulting $T_{\rm eff}$--$L_{\rm BB}$ loop, separated into brightening (orange) and dimming (blue) branches. The loop illustrates the hysteresis between temperature and luminosity through the pulsation cycle. Gray curves denote constant mean radii and for radii offset by $\pm5$\% around the mean.}
 \label{fig:Cep_4313}
\end{figure}



\begin{table}[!t]
\centering
\caption{\label{tab:phot} \textbf{Photometry of the multiple images of \tgta\ and \tgtb.} Fluxes in the units of nJy are measured from the HST (RELICS program) and JWST/NIRCam (\venus\ and SLICES programs) imaging. Photometry is not corrected for lensing magnification. For \tgta and \tgtb.3-5, we use the simple aperture photometry with $r=$\,0\farcs15 and apply the aperture correction. The flux of \tgtb.1 and \tgtb.2 are measured using \texttt{GALFITM} with a PSF+\sersic\ model. For bands in which the images are not detected, we list the $2\sigma$ upper limits.}
\vspace{12pt}
\footnotesize
\begin{tabular}{lcccccccc}
\hline\hline
ID & F435W & F606W & F814W & F090W & F115W & F150W & F150W2 & F200W \\
\hline
\tgta.1 & $<33$ & $<17$ & $<25$ & $28\pm4$ & $34\pm4$ & $57\pm5$ & $66\pm3$ & $84\pm4$ \\
\tgta.2 & $<31$ & $<15$ & $<25$ & $16\pm3$ & $18\pm4$ & $22\pm4$ & $25\pm2$ & $42\pm3$ \\
\tgta.3 & $<32$ & $<14$ & $<24$ & $12\pm4$ & $26\pm4$ & $29\pm5$ & $35\pm2$ & $47\pm4$ \\
\tgta.4 & $<52$ & $<20$ & $<40$ & $24\pm4$ & $27\pm4$ & $46\pm4$ & $45\pm2$ & $61\pm4$ \\
\tgta\ (stack) & $<15$ & $<8$ & $24\pm7$ & $20\pm2$ & $26\pm2$ & $37\pm2$ & $37\pm2$ & $56\pm3$ \\
\tgtb.1 & $<56$ & $<21$ & $<40$ & $9\pm3$ & $8\pm3$ & $6\pm2$ & $6\pm2$ & $9\pm2$ \\
\tgtb.2 & $<28$ & $<15$ & $<22$ & $<10$ & $<12$ & $<12$ & $<6$ & $<10$ \\
\tgtb.3 & $<29$ & $<13$ & $<23$ & $19\pm4$ & $13\pm4$ & $20\pm4$ & $17\pm2$ & $19\pm4$ \\
\tgtb.4 & $<32$ & $<14$ & $<24$ & $7\pm3$ & $12\pm4$ & $16\pm4$ & $15\pm2$ & $26\pm3$ \\
\tgtb.5 & $<34$ & $<15$ & $<26$ & $17\pm3$ & $14\pm4$ & $26\pm4$ & $28\pm2$ & $39\pm4$ \\
\tgtb\ (stack) & $<17$ & $<8$ & $28\pm7$ & $18\pm2$ & $14\pm3$ & $23\pm3$ & $24\pm2$ & $29\pm3$ \\
\hline

\hline\hline
ID & F210M & F277W & F300M & F322W2 & F356W & F410W & F444W \\
\hline
\tgta.1 & $94\pm5$ & $409\pm20$ & $417\pm21$ & $681\pm34$ & $966\pm48$ & $794\pm40$ & $897\pm45$ \\
\tgta.2 & $45\pm4$ & $140\pm7$ & $132\pm7$ & $253\pm13$ & $362\pm18$ & $303\pm15$ & $328\pm16$ \\
\tgta.3 & $58\pm5$ & $205\pm10$ & $205\pm10$ & $360\pm18$ & $524\pm26$ & $412\pm21$ & $451\pm23$ \\
\tgta.4 & $74\pm5$ & $260\pm13$ & $258\pm13$ & $437\pm22$ & $638\pm32$ & $516\pm26$ & $575\pm29$ \\
\tgta\ (stack) & $65\pm3$ & $192\pm10$ & $183\pm9$ & $341\pm17$ & $491\pm25$ & $403\pm20$ & $440\pm22$ \\
\tgtb.1 & $8\pm2$ & $19\pm4$ & $6\pm2$ & $30\pm2$ & $36\pm2$ & $21\pm2$ & $19\pm2$ \\
\tgtb.2 & $<10$ & $24\pm2$ & $8\pm2$ & $21\pm2$ & $30\pm2$ & $23\pm2$ & $21\pm2$ \\
\tgtb.3 & $14\pm4$ & $49\pm3$ & $36\pm4$ & $70\pm4$ & $89\pm4$ & $60\pm4$ & $60\pm4$ \\
\tgtb.4 & $18\pm4$ & $56\pm3$ & $39\pm4$ & $74\pm4$ & $95\pm5$ & $56\pm4$ & $62\pm4$ \\
\tgtb.5 & $38\pm4$ & $109\pm5$ & $80\pm4$ & $150\pm7$ & $197\pm10$ & $135\pm7$ & $149\pm7$ \\
\tgtb\ (stack) & $26\pm3$ & $64\pm3$ & $56\pm3$ & $84\pm4$ & $107\pm5$ & $81\pm4$ & $79\pm4$ \\
\hline
\end{tabular}
\end{table}

\clearpage

\begin{longtable}{ccccccc}
\caption{\label{tab:multi}\textbf{ List of multiple images used for strong lens mass modeling.}
The assumed positional error in units of arcsecond is indicated by $\sigma_{\mathrm{pos}}$, and $z$ and $\sigma_z$ refer to the source redshift and its error, respectively.
Photometric redshifts together with $1\sigma$ uncertainties are given with the absence of spectroscopic redshifts. 
If the redshift is not indicated, it is assumed to be a free parameter.
ID 1, 2, and 3 are presented in \cite{Cerny2018}, and ID 4, 5, 6. 7, 8, and 21 are presented in \cite{cerny25}. 
ID 9 and 10 corresponds to \tgta and \tgtb, respectively.
The spectroscopic redshifts of ID 14 and 32 are obtained through Keck/KCWI spectroscopy of \lya\ emission (private communication).
}\\
\hline\hline
ID & Redshift & RA [$^{\circ}$] & Dec [$^{\circ}$] & $\sigma_\mathrm{pos}$ [\arcsec] \\\hline
1.1  &  1.051         & 332.94000  & $-$3.82412 & 0.6 \\
 1.2  &                & 332.93302  & $-$3.82889 & 0.6 \\
 1.3  &                & 332.93848  & $-$3.83433 & 0.6 \\
 1.4  &                & 332.95233  & $-$3.83127 & 0.6 \\
 1.5  &                & 332.94080  & $-$3.82905 & 0.6 \\\hline
 2.1  &  1.9$\pm$0.2   & 332.93105  & $-$3.83665 & 0.6 \\
 2.2  &                & 332.93438  & $-$3.83923 & 0.6 \\
 2.3  &                & 332.95345  & $-$3.83825 & 0.6 \\\hline
 3.1  &  2.0$\pm$0.2   & 332.93148  & $-$3.83612 & 0.6 \\
 3.2  &                & 332.93403  & $-$3.83824 & 0.6 \\
 3.3  &                & 332.95434  & $-$3.83717 & 0.6 \\\hline
 4.1  &  4.4$\pm$0.2   & 332.93711  & $-$3.81924 & 0.6 \\
 4.2  &                & 332.92915  & $-$3.82778 & 0.6 \\
 4.3  &                & 332.93637  & $-$3.83717 & 0.6 \\
 4.4  &                & 332.95844  & $-$3.83086 & 0.6 \\
 4.5  &                & 332.94035  & $-$3.82940 & 0.6 \\\hline
 5.1  &  $\cdots$      & 332.94064  & $-$3.81186 & 0.6 \\
 5.2  &                & 332.93776  & $-$3.81210 & 0.6 \\
 5.3  &                & 332.95072  & $-$3.81538 & 0.6 \\\hline
 6.1  &  1.8$\pm$0.2   & 332.94164  & $-$3.84459 & 0.6 \\
 6.2  &                & 332.94318  & $-$3.84451 & 0.6 \\\hline
 7.1  &  $\cdots$      & 332.93139  & $-$3.83747 & 0.6 \\
 7.2  &                & 332.93342  & $-$3.83873 & 0.6 \\
 7.3  &                & 332.95470  & $-$3.83799 & 0.6 \\\hline
 8.1  &  3.2$\pm$0.2   & 332.93862  & $-$3.82232 & 0.6 \\
 8.2  &                & 332.93094  & $-$3.83324 & 0.6 \\
 8.3  &                & 332.93407  & $-$3.83684 & 0.6 \\
 8.4  &                & 332.95750  & $-$3.83386 & 0.6 \\\hline
 9.1  &  4.3$\pm$0.4   & 332.95386  & $-$3.83207 & 0.6 \\
 9.2  &                & 332.92441  & $-$3.82768 & 0.6 \\
 9.3  &                & 332.94059  & $-$3.84067 & 0.6 \\
 9.4  &                & 332.94436  & $-$3.81911 & 0.6 \\\hline
10.1  &  4.3$\pm$0.4   & 332.94047  & $-$3.83532 & 0.6 \\
10.2  &                & 332.94092  & $-$3.83182 & 0.6 \\
10.3  &                & 332.95469  & $-$3.82245 & 0.6 \\
10.4  &                & 332.92738  & $-$3.81977 & 0.6 \\
10.5  &                & 332.94432  & $-$3.81432 & 0.6 \\\hline
11.1  &  4.3$\pm$0.4   & 332.92728  & $-$3.82682 & 0.6 \\
11.2  &                & 332.94054  & $-$3.81775 & 0.6 \\
11.3  &                & 332.93839  & $-$3.83829 & 0.6 \\
11.4  &                & 332.95642  & $-$3.83043 & 0.2 \\
11.5  &                & 332.95646  & $-$3.82963 & 0.2 \\
11.6  &                & 332.95633  & $-$3.83006 & 0.2 \\\hline
12.1  &  2.6$\pm$0.2   & 332.94524  & $-$3.82534 & 0.6 \\
12.2  &                & 332.94318  & $-$3.82812 & 0.6 \\
12.3  &                & 332.92278  & $-$3.82890 & 0.6 \\\hline
13.1  &  1.8$\pm$0.2   & 332.93826  & $-$3.83611 & 0.6 \\
13.2  &                & 332.93000  & $-$3.82732 & 0.6 \\
13.3  &                & 332.95517  & $-$3.83043 & 0.6 \\
13.4  &                & 332.93961  & $-$3.82022 & 0.6 \\\hline
14.1  &  2.535         & 332.94171  & $-$3.81736 & 0.6 \\
14.2  &                & 332.92791  & $-$3.82434 & 0.6 \\
14.3  &                & 332.95511  & $-$3.82759 & 0.6 \\\hline
15.1  &  2.4$\pm$0.2   & 332.94504  & $-$3.83988 & 0.6 \\
15.2  &                & 332.94888  & $-$3.83764 & 0.6 \\
15.3  &                & 332.94454  & $-$3.82286 & 0.6 \\
15.4  &                & 332.92441  & $-$3.82929 & 0.6 \\\hline
16.1  &  4.4$\pm$0.2   & 332.92367  & $-$3.82770 & 0.6 \\
16.2  &                & 332.94260  & $-$3.84050 & 0.6 \\
16.3  &                & 332.95265  & $-$3.83281 & 0.6 \\\hline
17.1  &  4.4$\pm$0.2   & 332.93518  & $-$3.82096 & 0.6 \\
17.2  &                & 332.93546  & $-$3.83628 & 0.6 \\
17.3  &                & 332.93031  & $-$3.82849 & 0.6 \\
17.4  &                & 332.95919  & $-$3.83113 & 0.6 \\\hline
18.1  &  4.0$\pm$0.2   & 332.95375  & $-$3.83715 & 0.6 \\
18.2  &                & 332.92643  & $-$3.83227 & 0.6 \\
18.3  &                & 332.94177  & $-$3.82215 & 0.6 \\
18.4  &                & 332.93715  & $-$3.84224 & 0.6 \\\hline
19.1  &  3.6$\pm$0.2   & 332.94244  & $-$3.84389 & 0.6 \\
19.2  &                & 332.94604  & $-$3.84307 & 0.6 \\
19.3  &                & 332.92478  & $-$3.83394 & 0.6 \\\hline
20.1  &  2.3$\pm$0.2   & 332.92905  & $-$3.83857 & 0.6 \\
20.2  &                & 332.94886  & $-$3.84278 & 0.6 \\
20.3  &                & 332.93371  & $-$3.84257 & 0.6 \\\hline
21.1  &  6.4$\pm$0.2   & 332.93782  & $-$3.82639 & 0.2 \\
21.2  &                & 332.93826  & $-$3.82703 & 0.2 \\
21.3  &                & 332.93689  & $-$3.82410 & 0.6 \\\hline
22.1  &  4.4$\pm$0.2   & 332.93081  & $-$3.82963 & 0.6 \\
22.2  &                & 332.93537  & $-$3.82191 & 0.6 \\
22.3  &                & 332.93454  & $-$3.83573 & 0.6 \\
22.4  &                & 332.95944  & $-$3.83167 & 0.6 \\\hline
23.1  &  3.5$\pm$0.2   & 332.95137  & $-$3.83997 & 0.6 \\
23.2  &                & 332.93840  & $-$3.84343 & 0.6 \\
23.3  &                & 332.92597  & $-$3.83366 & 0.6 \\
23.4  &                & 332.94205  & $-$3.82462 & 0.6 \\\hline
24.1  &  3.2$\pm$0.2   & 332.94244  & $-$3.84333 & 0.6 \\
24.2  &                & 332.94670  & $-$3.84226 & 0.6 \\\hline
25.1  &  3.0$\pm$0.2   & 332.93535  & $-$3.84064 & 0.6 \\
25.2  &                & 332.94089  & $-$3.82417 & 0.6 \\
25.3  &                & 332.95449  & $-$3.83749 & 0.6 \\
25.4  &                & 332.92860  & $-$3.83446 & 0.6 \\\hline
26.1  &  6.4$\pm$0.2   & 332.94770  & $-$3.82331 & 0.6 \\
26.2  &                & 332.92109  & $-$3.82712 & 0.6 \\\hline
27.1  &  2.0$\pm$0.2   & 332.95153  & $-$3.81949 & 0.6 \\
27.2  &                & 332.94719  & $-$3.81613 & 0.6 \\\hline
28.1  &  3.1$\pm$0.2   & 332.93073  & $-$3.81751 & 0.6 \\
28.2  &                & 332.94329  & $-$3.81478 & 0.6 \\
28.3  &                & 332.95312  & $-$3.82092 & 0.6 \\\hline
29.1  &  3.6$\pm$0.2   & 332.92601  & $-$3.82254 & 0.6 \\
29.2  &                & 332.94624  & $-$3.81665 & 0.6 \\
29.3  &                & 332.95392  & $-$3.82433 & 0.6 \\
29.4  &                & 332.94080  & $-$3.83653 & 0.0 \\\hline
30.1  &  3.3$\pm$0.5   & 332.95153  & $-$3.82058 & 0.6 \\
30.2  &                & 332.94932  & $-$3.81832 & 0.6 \\\hline
31.1  &  7.3$\pm$0.2   & 332.94224  & $-$3.81121 & 0.6 \\
31.2  &                & 332.95513  & $-$3.81916 & 0.6 \\
31.3  &                & 332.93079  & $-$3.81439 & 0.6 \\\hline
32.1  &  2.647         & 332.95413  & $-$3.82471 & 0.6 \\
32.2  &                & 332.94399  & $-$3.81640 & 0.6 \\
32.3  &                & 332.92747  & $-$3.82210 & 0.6 \\\hline
33.1  &  2.2$\pm$0.2   & 332.93080  & $-$3.81982 & 0.6 \\
33.2  &                & 332.93993  & $-$3.81624 & 0.6 \\
33.3  &                & 332.95506  & $-$3.82470 & 0.6 \\\hline
34.1  &  2.2$\pm$0.2   & 332.93218  & $-$3.81853 & 0.6 \\
34.2  &                & 332.93911  & $-$3.81615 & 0.6 \\
34.3  &                & 332.95523  & $-$3.82433 & 0.6 \\\hline
35.1  &  6.4$\pm$0.2   & 332.92504  & $-$3.83190 & 0.6 \\
35.2  &                & 332.93862  & $-$3.84321 & 0.6 \\
35.3  &                & 332.95295  & $-$3.83817 & 0.6 \\\hline
36.1  &  5.7$\pm$0.2   & 332.95109  & $-$3.84068 & 0.6 \\
36.2  &                & 332.93955  & $-$3.84412 & 0.6 \\
36.3  &                & 332.92481  & $-$3.83336 & 0.6 \\
36.4  &                & 332.94240  & $-$3.82362 & 0.6 \\\hline
37.1  &  4.2$\pm$0.2   & 332.94913  & $-$3.84019 & 0.6 \\
37.2  &                & 332.94275  & $-$3.84264 & 0.6 \\
37.3  &                & 332.92416  & $-$3.83144 & 0.6 \\\hline
38.1  &  3.9$\pm$0.2   & 332.95586  & $-$3.83674 & 0.6 \\
38.2  &                & 332.92883  & $-$3.83442 & 0.6 \\
38.3  &                & 332.93474  & $-$3.84008 & 0.6 \\\hline
39.1  &  3.2$\pm$0.2   & 332.95363  & $-$3.82934 & 0.6 \\
39.2  &                & 332.94080  & $-$3.83846 & 0.6 \\
39.3  &                & 332.94479  & $-$3.81880 & 0.6 \\
39.4  &                & 332.92541  & $-$3.82612 & 0.6 \\\hline
40.1  &  2.3$\pm$0.2   & 332.94591  & $-$3.84386 & 0.6 \\
40.2  &                & 332.93978  & $-$3.84451 & 0.6 \\\hline
41.1  &  2.1$\pm$0.2   & 332.93915  & $-$3.82041 & 0.6 \\
41.2  &                & 332.93777  & $-$3.83636 & 0.6 \\
41.3  &                & 332.95566  & $-$3.83104 & 0.6 \\
41.4  &                & 332.93005  & $-$3.82783 & 0.6 \\\hline
42.1  &  2.0$\pm$0.5   & 332.94458  & $-$3.81519 & 0.6 \\
42.2  &                & 332.95279  & $-$3.82075 & 0.6 \\
42.3  &                & 332.92995  & $-$3.81822 & 0.6 \\\hline
43.1  &  $\cdots$      & 332.95395  & $-$3.83408 & 0.6 \\
43.2  &                & 332.93106  & $-$3.83193 & 0.6 \\
43.3  &                & 332.94032  & $-$3.82439 & 0.6 \\
43.4  &                & 332.93666  & $-$3.83707 & 0.6 \\
\hline
\end{longtable}

\end{document}